\documentclass[prd,a4paper,nofootinbib,reprint]{revtex4}
\pdfoutput=1 

\usepackage{graphicx}
\usepackage{amsfonts}
\usepackage{contractions}
\usepackage{soul}
\usepackage{amssymb}
\usepackage{amsmath}
\usepackage{slashed} 
\usepackage{enumerate}
\usepackage{color}
\usepackage{comment}
\usepackage{bbold}
\usepackage{tikz}
\usepackage{multirow}
\usepackage{hyperref}
\usepackage[utf8]{inputenc}
\usepackage{adjustbox}
\usepackage{erewhon}
\definecolor{VioletRed4}{rgb}{0, 0, 0}

\usepackage{hyperref}

\def\Im{\mbox{Im}\,}

\definecolor{oucrimsonred}{rgb}{0.6, 0.0, 0.0}
\definecolor{persianblue}{rgb}{0.11, 0.22, 0.73}
\definecolor{forestgreen}{rgb}{0.13,0.35,0.13}
\definecolor{lightgray}{rgb}{0.83, 0.83, 0.83}
 \hypersetup{colorlinks, citecolor=oucrimsonred, linkcolor=persianblue, urlcolor=oucrimsonred}
\def\hhref#1{\href{http://arxiv.org/abs/#1}{#1}} % in bibliography
\newcommand{\be}{\begin{equation}}
\newcommand{\ee}{\end{equation}}
\newcommand{\bea}{\begin{eqnarray}}
\newcommand{\eea}{\end{eqnarray}}
\newcommand{\nn}{\nonumber}

\definecolor{oucrimsonred}{rgb}{0.6, 0.0, 0.0}
\newcommand{\fd}[2]{\parbox{#1}{\includegraphics[width=#1]{figs/#2}}}

\definecolor{mtcolor}{rgb}{.8,.3,.1}

\definecolor{violachiaro}{rgb}{1,0.6,1}
 
\definecolor{verdechiaro}{rgb}{0.6,1,0.6}
\definecolor{giallochiaro}{rgb}{1,1,0.6}
\definecolor{bluscuro}{rgb}{0.15, 0.2, 0.9}
\definecolor{verdes}{rgb}{0.1, 0.5, 0.1}%
\definecolor{tangerineyellow}{rgb}{1.0, 0.8, 0.0}

\begin{document}
%%%%%%%%%%%%%%%%%%%%%%%%%%%%%%%%%%%%%%%%%%%%%%%%%%%%%%%%%%%  FRONT PAGE

\title[]{Non-gaussianities for primordial black hole formation}

\date{\today}
\author{Marco Taoso$^{a}$}
\author{Alfredo Urbano$^{b,c}$}
\affiliation{$^a$I.N.F.N. sezione di Torino, via P. Giuria 1, I-10125 Torino, Italy}
\affiliation{$^b$Dipartimento di Fisica, ``Sapienza'' Universit\`a di Roma, Piazzale Aldo Moro 5, 00185, Roma, Italy}
\affiliation{$^c$I.F.P.U., Institute  for  Fundamental Physics  of  the  Universe, via  Beirut  2, I-34014 Trieste, Italy.}

%%%%%%%%%%%%%%%%%%%%%%%%%%%%%%%%%%%%%%%%%%%%%%%%%%%%%%%%%%%%%%%%%%%%%
\begin{abstract}
\noindent  
We analyze primordial non-gaussianities in presence of an ultra-slow phase during the inflationary dynamics, focusing on scenarios relevant for the production of primordial black holes.
We compute the three-point correlation function of comoving curvature perturbations 
finding that non-gaussianities are sizable, and predominantly local.
In the context of threshold statistics, we analyze their impact for the abundance of primordial black holes,
 and their interplay with the non-gaussianities arising from the non-linear relation between density and curvature perturbations.
We find that non-gaussianities significantly modify the estimate of the primordial black holes abundance obtained with the gaussian approximation. 
However, we show that this effect can be compensated by a small change, of a factor $2\div3$ at most, of the amplitude of the primordial power spectrum of curvature perturbations.
This is obtained with a small tuning of the parameters of the inflationary model.
 \end{abstract}
%%%%%%%%%%%%%%%%%%%%%%%%%%%%%%%%%%%%%%%%%%%%%%%%%%%%%%%%%%%%%%%%%%%
\maketitle
%%%%%%%%%%%%%%%%%%%%%%%%%%%%%%%%%%%%%%%%%%%%%%%%%%%%%%%%%%%%%%%%%%%
 %%%%%%%%%%%%%%%%%%%%%%%%%%%%%%%%%%%%%%%%%%%%%%%%%%%%%%%%%%%%%%%%%%%
 
  \section{Motivations}\label{sec:Mot}
%%%%%%%%%%%%%%%%%%%%%%%%

The theory of inflation provides an elegant mechanism that explains the origin of structures in the Universe. 
In the inflationary picture, space-time fluctuates quantum mechanically around a background that is expanding exponentially fast. 
After the end of inflation, these fluctuations are transferred to the radiation field, creating slightly overdense and under-dense regions. These tiny ripples are the initial conditions that start the process of gravitational collapse which eventually 
forms the intricate architecture of galaxies that populate our observable Universe today.

If the amplitude of some of the overdense regions exceeds a  specific threshold value, 
the pull of gravity becomes unsustainable and  
the gravitational collapse directly forms primordial black holes (PBHs)\,\cite{Zelda}.
Interestingly, 
for masses in the range $10^{18} \lesssim M_{\rm PBH} [{\rm g}] \lesssim 10^{21}$, a population of
such PBHs may account for the totality of dark matter
observed in the Universe today, as envisaged by Hawking and Carr in their pioneering research\,\cite{Hawking:1971ei,Carr:1974nx}.
 
Because of their intrinsic quantum-mechanical origin,  the way in which 
quantum fluctuations lead to a classical pattern of perturbations can be described only in a probabilistic sense, and 
it is often assumed that primordial perturbations follow a gaussian statistics. 
It is important to stress that there is no fundamental reason to believe that this assumption is true, and it should be rather considered to be an approximation.
This approximation seems to work exquisitely well at the typical scales which are relevant for cosmic microwave background (CMB) anisotropy measurements, 
$0.005 \lesssim k\,[{\rm Mpc}^{-1}] \lesssim 0.2$, and this is because deviations from the gaussian pattern are expected to be 
suppressed by powers of the slow-roll parameters\,\cite{Maldacena:2002vr} which during conventional slow-roll dynamics take $O(\ll 1)$ values. 

As far as the formation of PBHs is concerned, however, the story might be drastically different. 
The process of PBH formation involves completely different scales compared to those probed by CMB observations.
Typically, in order to be compatible with observational bounds on PBH abundance in the present-day Universe, 
one needs to consider scales $10^{12} \lesssim k\,[{\rm Mpc}^{-1}]\lesssim 10^{14}$. At such small scales, 
the physics involved can be completely different compared to what expected at CMB scales.
Indeed, the formation of fluctuations large enough to trigger gravitational collapse into PBHs requires a departure from conventional slow-roll dynamics, and the slow-roll parameters can easily reach $O(1)$ values.
Consequently, the parametric suppression that limits the presence of non-gaussianities at CMB scales is not guaranteed anymore.

\begin{figure}[!htb!]
\begin{center}
\includegraphics[width=.93\textwidth]{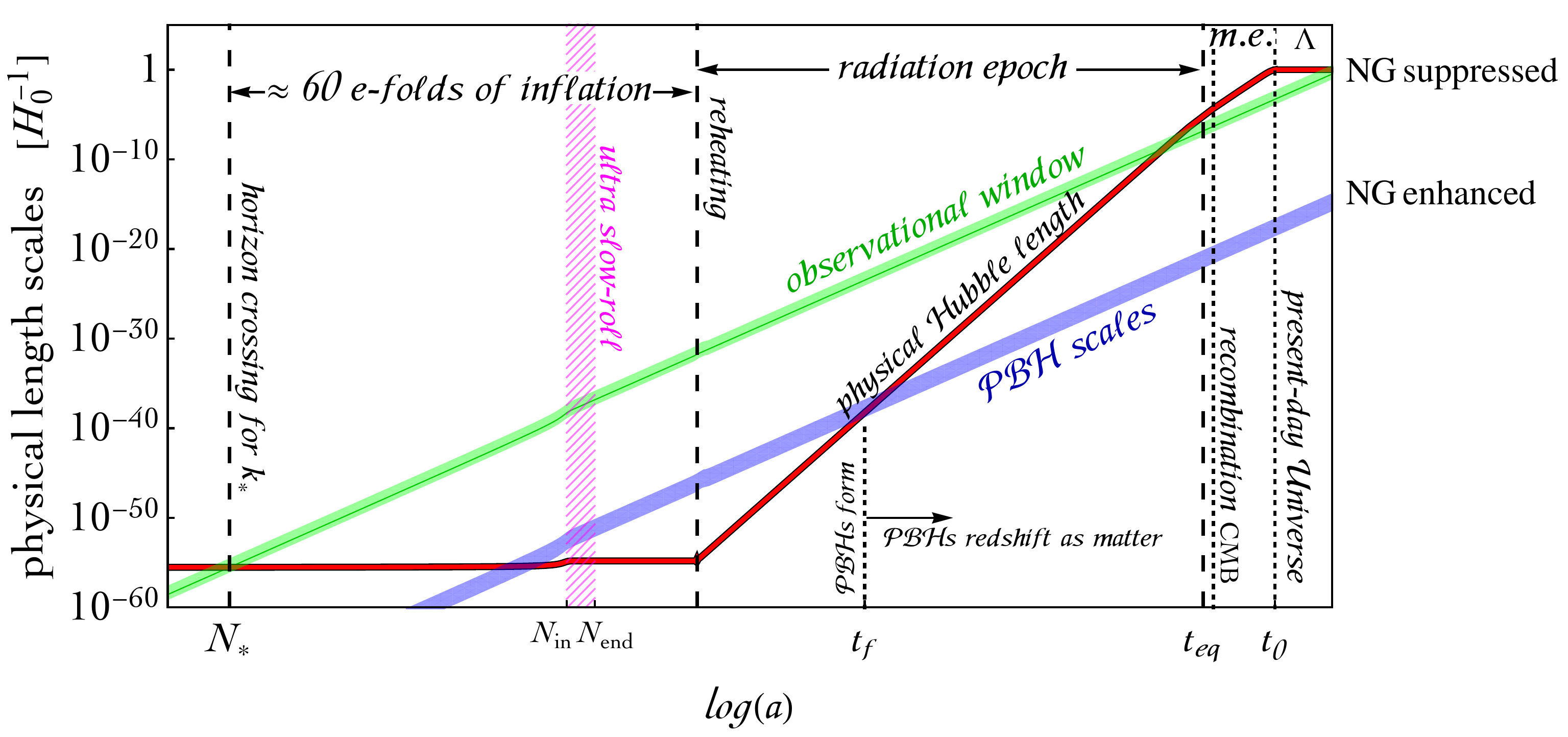}
\caption{\em \label{fig:Evol} 
Time-evolution of the physical Hubble length $1/H$ (red line), written in units of its present-day value
$1/H_0\simeq 4.5\times 10^3\,{\rm Mpc}$ (with $H_0 \simeq 10^{-42}$ GeV in natural units),
as function of the logarithm of the scale factor $a$
of the Friedmann-Lema\^{\i}tre-Robertson-Walker line element. 
We follow the evolution of $1/H$ throughout the history of our observable Universe. 
The inflationary phase is computed according to the model in ref.\,\cite{Ballesteros:2020qam}. 
After (instantaneous) reheating, the evolution of $1/H$ during the radiation epoch, matter epoch (m.e.) and the 
present-day epoch ($\Lambda$) is computed according to the standard $\Lambda$CDM model.
We superimpose the time-evolution of physical length scales $\lambda = a/k$, with $k$ the comoving wavenumber, 
that are relevant for CMB observations (the green band labelled ``observational window''
with $0.005 \lesssim k\,[{\rm Mpc}^{-1}] \lesssim 0.2$ and $k_* = 0.05$ Mpc$^{-1}$). 
We also show the time-evolution  of physical length scales associated to perturbations that are responsible for PBH formation (the blue band labelled ``PBH scales'' with $10^{12} \lesssim k\,[{\rm Mpc}^{-1}]\lesssim 10^{14}$).
  Scales that cross the ultra slow-roll phase when they are deep in the super-Hubble regime are not sizably 
  affected by the presence of negative friction\,\cite{Ballesteros:2020qam}. For these scales, non-gaussianities are slow-roll suppressed.
  This is the case of scales relevant for CMB observations.
  On the contrary, scales that cross the Hubble radius close to the ultra slow-roll regime can be 
  exponentially enhanced\,\cite{Ballesteros:2020qam}.
  The interval of $e$-folds during which the 
 ultra slow-roll phase takes place is indicated with a vertical strip. At these scales, 
 non-gaussianities are enhanced.  
 PBH formation takes place at time $t_f$ after the PBH scales re-enter the Hubble radius during the radiation epoch.
 Once formed, we consider the simple scenario in which the PBH energy density redshifts as matter.
 }
\end{center}
\end{figure}

For illustration, we summarize the time-evolution of physical scales probed by CMB observations and scales relevant for PBH formation in fig.\,\ref{fig:Evol}.

The introductory discussion above makes clear that the possible presence of non-gaussianities plays a fundamental role in the precise computation of the PBH abundance. The formation of PBHs requires perturbations above some threshold, located in the tail of their 
probability distribution. Small deviations from the gaussian shape of the tail, therefore, may produce exponentially different results.

In this work we investigate the role of non-gaussianities in the context of inflationary models featuring an ultra slow roll (USR) phase\,\cite{Leach:2000yw,Leach:2001zf,Tsamis:2003px,Kinney:2005vj,Kinney:1997ne}. This part of the inflationary dynamics has been invoked for the production of PBHs 
(see refs.\,\cite{Ivanov:1994pa,Saito:2008em} for the earliest proposal in this direction\footnote{Ref.\cite{Ivanov:1994pa} investigates a potential with a plateau and two breaks. The inflationary dynamics in presence of such features was studied in\,\cite{Starobinsky:1992ts}.}).
This is because,
in these scenarios, an USR phase allows to boost the primordial fluctuations at small scales, leading to a sizable population of PBHs.
Crucially, conventional slow-roll dynamics is violated during such a period, potentially enabling the generation of large non-gaussianities.
In our analysis, we focus on single-field inflation models in which the USR phase arises when the inflaton, rolling down its potential, crosses an approximate stationary inflection point before the end of inflation.
Specifically, we consider both a toy model, which allows an analytical exposition, and a more realistic parametrization of the inflaton potential.

There exist a number of preceding studies devoted to the understanding of non-gaussianities in the context of PBH formation, and we will properly acknowledge them in the course of our work.
Before diving into the details of our analysis, let us sketch briefly the structure of this paper.

In section\,\ref{sec:Abundance}, we will frame in a more formal way the computation of the abundance of PBHs and 
the connection with inflationary dynamics. 
In section\,\ref{sec:SkewnessExact}, we will compute the so-called bispectrum of comoving curvature perturbations which
controls the first non-trivial (i.e. connected) deviation from exact gaussianity. 
The bispectrum vanishes identically for a gaussian 
distribution, and a large non-zero result will introduce sizable deviations from the gaussian statistics. 
In section\,\ref{sec:Kurtosis}, we will move to consider the connected part of the trispectrum  of comoving curvature perturbations
which
controls deviation from exact gaussianity at the level of fourth-order statistical correlators.
We will not compute the full connected trispectrum but we will argue that the non-zero contribution which is 
generated by cubic interactions plays a crucial role in the computation of the PBH abundance. 
In section\,\ref{sec:Local} we discuss the computation of PBH abundance in the presence of 
non-gaussianities.
We will show that their impact is not dramatic. 
We find that the size of non-gaussianities is large enough to 
significantly affect the PBH abundance. On the other hand, it is enough to
change the amplitude of the power spectrum of comoving curvature perturbations of a factor $2\div3$
to obtain the same abundance obtained with the gaussian approximation.

We present our conclusions in section\,\ref{sec:Conclusions}.
Appendix\,\ref{secAPP} complements the computations illustrated in the main text with further technical details. 

\section{On the abundance of primordial black holes: from theory to observations and back}\label{sec:Abundance}
%%%%%%%%%%%%%%%%%%%%%%%%
In cosmology and in the theory of structure formation, we assume that fluctuations in the energy density or in the
gravitational potential 
form a three-dimensional random field. 
A three-dimensional random field $\delta(\vec{x})$ is a set of random variables, one for each point
in the three-dimensional real space, defined by a probability functional $P[\delta(\vec{x})]$ which specifies the probability for the occurrence of a particular
realization of the field over the ensemble. 
In cosmology, the random field $\delta(\vec{x})$ is the three-dimensional mass overdensity field
which gives the fractional overdensity in a
given region of space (and at a given cosmic time) with respect to the unperturbed background (a.k.a. density contrast).
A random field is fully specified by the entire hierarchy of 
its correlation functions, $\langle \delta(\vec{x}_1)\delta(\vec{x}_2)\dots\delta(\vec{x}_n)\rangle$.
 The simplest type of random field is the gaussian random field for which the probability functional $P$ is a gaussian 
distribution.
The statistical properties of a gaussian (subscript $_{\mathcal{G}}$ hereafter) random field are completely specified by the two-point
correlation function or, working in Fourier space, by the power spectrum.
The $n$-point correlation functions either vanish (for odd $n$) or can be expressed in terms of the power spectrum as a consequence of the Isserlis' theorem (for even $n$).  
In Fourier space, we have (assuming a vanishing mean value as well as homogeneity and isotropy of space)
\begin{align}
\langle \delta_{k}\rangle & = 0\,,\\
\langle \delta_{k}\delta_{k^{\prime}} \rangle & = (2\pi)^3\delta^{(3)}(\vec{k} + \vec{k}^\prime)\Delta_{\delta}(k)\,,\label{eq:Gaussian2pt}\\
\langle \delta_{k_1}\delta_{k_2}\delta_{k_3} \rangle_{\mathcal{G}} & = 0\,,\label{eq:Skew}\\
\langle \delta_{k_1}\delta_{k_2}\delta_{k_3}\delta_{k_4} \rangle_{\mathcal{G}} & =
\langle\delta_{k_1}\delta_{k_2}\rangle\langle\delta_{k_3}\delta_{k_4}\rangle +
\langle\delta_{k_1}\delta_{k_3}\rangle\langle\delta_{k_2}\delta_{k_4}\rangle +
\langle\delta_{k_1}\delta_{k_4}\rangle\langle\delta_{k_2}\delta_{k_3}\rangle \nonumber \\
 & = (2\pi)^6\delta^{(3)}(\vec{k}_1 + \vec{k}_2)
 \delta^{(3)}(\vec{k}_3 + \vec{k}_4)\Delta_{\delta}(k_1)\Delta_{\delta}(k_3) + {\rm two\,permutations}\,,\label{eq:Gaussian4pt}
\end{align}
and so on.
 The variance of the gaussian distribution is
\begin{align}\label{eq:NaiveVariance}
\sigma^2 = 
\lim_{\vec{y}\to \vec{x}}\langle \delta(\vec{x})\delta(\vec{y})\rangle =
 \langle \delta(\vec{x})\delta(\vec{x})\rangle =  
\int\frac{dk}{k} \mathcal{P}_{\delta}(k)\,,~~~~~\Delta_{\delta}(k) = \frac{2\pi^2}{k^3}\mathcal{P}_{\delta}(k)\,,
\end{align}
from which we see that the power spectrum of density fluctuations $\mathcal{P}_{\delta}(k)$ is the contribution to
the variance of the field per unit logarithmic interval of $k$, and saying that $\mathcal{P}_{\delta}(k) \sim 1$ means that
the Fourier modes in a unit logarithmic bin around wavenumber $k$ generate fluctuations $\delta\rho/\rho$ of order unity.
Notice that $\sigma^2$ 
does not depend on the position, by virtue of the delta function eq.\,(\ref{eq:Gaussian2pt}) which is indeed a consequence of
homogeneity.  
We are interested in the situation in which there is a sizable probability to form 
in a given region of typical size $R$
density 
fluctuations $\delta\rho/\rho$ large enough to collapse and form black holes under the pull of their own gravitational weight.
In order to study PBH formation and make contact with the power spectrum of comoving curvature perturbations, eq.\,(\ref{eq:NaiveVariance}) has to be modified in two ways. 
First, we convolve the random field $\delta(\vec{x})$ with some window function $W(|\vec{x}-\vec{y}|,R)$ to obtain
 $\delta_R(\vec{x}) = \int d^3\vec{y}\,W(|\vec{x}-\vec{y}|,R)\,\delta(\vec{y})$
so that, at every point $\vec{x}$, the smoothed field represents the weighted average
of $\delta$ over a spherical region of characteristic dimension $R$ centred in $\vec{x}$.
Second, we relate the density contrast to the comoving curvature perturbation $\mathcal{R}$ by means of the linear-order relation (in Fourier space)\,\cite{Green:2004wb} 
\begin{align}\label{eq:DensityCurvature}
\delta_k = \frac{2(1+\omega)}{(5+3\omega)}\left(\frac{k}{aH}\right)^2\mathcal{R}_k\,,
\end{align}
where $\omega = p/\rho$ defines the equation of state, and we shall take $\omega=1/3$ for radiation domination. 
It is important to remark that eq.\,(\ref{eq:DensityCurvature}) is not an exact relation but it is only valid at the linear order, and 
non-linear corrections to eq.\,(\ref{eq:DensityCurvature}) unavoidably source non-gaussianities, see \,\cite{DeLuca:2019qsy,Young:2019yug,Yoo:2019pma,Kehagias:2019eil,Yoo:2018kvb,Kawasaki:2019mbl,Germani:2019zez,Yoo:2020dkz,Musco:2020jjb} for related studies.
We will come back to this point in section\,\ref{sec:Local}. For the moment, we shall work assuming the linear relation in eq.\,(\ref{eq:DensityCurvature}).

Eq.\,(\ref{eq:NaiveVariance}) becomes 
\begin{align}\label{eq:SmoothedVariance}
\langle \delta^2\rangle_R = \sigma_R^2 \equiv  \frac{4(1+\omega)^2}{(5+3\omega)^2}
\int\frac{dk}{k}\,W(kR)^2\left(\frac{k}{aH}\right)^4\mathcal{P}_{\mathcal{R}}(k)\,,
\end{align}
where $\mathcal{P}_{\mathcal{R}}(k)$ is the power spectrum of the the comoving curvature perturbation.
$W(kR)$ is the Fourier transform of the window function introduced before, and we shall use the 
volume-normalized gaussian window function $W(kR) = \exp(-k^2R^2/2)$\footnote{See\,\cite{Young:2019osy} for a study of the role of the window function for the PBHs formation.}.
Assuming that PHBs are  formed  when  the  density  perturbation  exceeds the threshold value $\delta_{\rm th}$,  the  fraction  of  the
Universe ending up in PHBs is given by the tail of the density contrast distribution 
$\beta(M_{\rm PBH}) = \gamma\int_{\delta_{\rm th}}^{\infty}P(\delta)d\delta$,
and in the gaussian approximation we have 
\begin{align}\label{eq:FundBeta}
\beta_{\mathcal{G}}(M_{\rm PBH}) = \gamma\int_{\delta_{\rm th}}^{\infty}\frac{d\delta}{\sqrt{2\pi}\sigma_R(M_{\rm PBH})}
\exp\left[-\frac{\delta^2}{2\sigma_R(M_{\rm PBH})^2}\right]\,,
\end{align}
where we convert the smoothed variance $\sigma_R$ to a function of the PBH mass taking into account that the size $R$ is related to the mass $M_{\rm PBH}$ as $R\approx 2G_N M_{\rm PBH}/\gamma a_{\rm form}$.
The parameter $\gamma$ depends on the details of the gravitational collapse and $a_{\rm form}$ is the scale factor at the time at which the PBH is formed.

The distribution of PHBs as function of their mass is strongly peaked at the value that maximizes $\sigma(M_{\rm PBH})$, in view of the exponential factor in eq.\,(\ref{eq:FundBeta}). 
In terms of the power spectrum $\mathcal{P}_{\mathcal{R}}$ this means that the PBH abundance roughly scales as 
$e^{-1/\mathcal{P}_{\mathcal{R}}}$. 
Finally, the present-day fractional abundance of dark matter in the form of PBH is given by
\begin{align}\label{eq:PresentDayAbundance}
f_{\rm PBH} = 
\left(\frac{\gamma}{0.2}\right)^{1/2}\left[\frac{\beta(M_{\rm PBH})}{1.6\times 10^{-16}}\right]
\left[\frac{g_{*}(T_{\rm form})}{106.75}\right]^{-1/4}\left(\frac{M_{\rm PBH}}{10^{18}\,{\rm g}}\right)^{-1/2}\,.
\end{align}

The computation of $\beta$ by means of eq.\,(\ref{eq:FundBeta}) is known as threshold statistics.
The value of $\delta_{\rm th}$ is subject to some uncertainty (varying in the range 
$0.4 \lesssim \delta_{\rm th} \lesssim 2/3$), and  also depends on the specific shape of the power spectrum. 
In our analysis, for simplicity, we adopt the fixed values $\delta_{\rm th} = 0.45$ and $\gamma = 0.2$\,\cite{Musco:2004ak}.

An alternative way of computing the PBH abundance relies on the use of peak theory\,\cite{Young:2014ana}, in which one computes 
the number density of peaks of the overdensity field $\delta$ (which in general contains many peaks, i.e. local maxima,
 and valleys, i.e. local minima) and integrate above some threshold $\delta_c$\,\cite{Bardeen:1985tr}. The latter in general differs from 
$\delta_{\rm th}$ and depends on the shape of the power spectrum\,\cite{Germani:2018jgr} (see also ref.\,\cite{Germani:2019zez} in which the full non-linear relation between density and curvature perturbation is explored). 
Using again the gaussian approximation, one finds\,\cite{Bardeen:1985tr,Young:2014ana,DeLuca:2019qsy}
\begin{align}\label{eq:Peak}
\beta_{\mathcal{G}}^{\rm pk}(M_{\rm PBH}) \simeq \frac{\gamma}{3\pi}\left(\frac{\langle k^2\rangle}{3}\right)^{3/2}r_m^3
\left(\frac{\delta_{c}^2}{\sigma^2}\right)e^{-\delta_{c}^2/2\sigma^2}\,,~~~~~{\rm with}\,\, \langle k^2\rangle = 
\frac{1}{\sigma^2}\int\frac{d^3\vec{k}}{(2\pi)^3}k^2\Delta_{\delta}(k)\,,
\end{align}
where $\delta_c = \delta_{\rm th}/3\psi(r_m)$ with $\psi$ the average density profile and $r_m$ is the comoving distance where the so-called compaction function is maximized (we refer to ref.\,\cite{Germani:2018jgr,Musco:2018rwt} for a comprehensive discussion and more detailed definitions).\footnote{
Eq.\,(\ref{eq:Peak}) is obtained as follows. Introducing $\nu\equiv\delta/\sigma,$ the comoving number density of peaks is\,\cite{Bardeen:1985tr}
\begin{align}\label{eq:PeakNumberDensity}
n_{\mathcal{G},{\rm com}}^{\rm pk} = \frac{1}{(2\pi)^2}\left(\frac{\langle k^2\rangle}{3}\right)^{3/2}\,\nu^2\,e^{-\nu^2}\,,
\end{align}
and in physical space one has $n_{\mathcal{G},{\rm phys}}^{\rm pk} = n_{\mathcal{G},{\rm com}}^{\rm pk}/a_f^3,$ with $a_f$ is the scale factor at the formation time $t_f.$ The PBH mass reads 
$\mathcal{M}_{\rm PBH}(\nu) = \mathcal{K}\,M_{\rm PBH}(t_m)\,\sigma^{\tilde{\gamma}}\,(\nu-\nu_c)^{\tilde{\gamma}}$ 
where $\tilde{\gamma}$ is a critical exponent that depends on the equation of state in the formation era (with $\tilde{\gamma} \simeq 0.36$ for radiation\,\cite{Neilsen:1998qc}) and $\mathcal{K} \sim O(1)$. $M_{\rm PBH}(t_m)$ is the horizon mass at horizon-crossing time $t_m$ defined implicitly by the relation $a(t_m)H(t_m)\,r_m=1.$ 
One can write $\beta_{\mathcal{G}}^{\rm pk}=\int_{\nu_c}^{\infty}d\nu\,\mathcal{M}_{\rm PBH}(\nu)\,n_{\mathcal{G},{\rm phys}}^{\rm pk}(\nu)/\rho_f$ where $\rho_f$ is the background density at the formation time.
Using the result from numerical simulation $a_f/a_m \sim 3$\,\cite{Musco:2018rwt,Kalaja:2019uju} and in the limit $\nu_c\equiv\delta_c/\sigma\gg1$ one  obtains eq.\,(\ref{eq:Peak}) with $\gamma = 3\mathcal{K}\Gamma(1+\tilde{\gamma})\nu_c^{-\tilde{\gamma}}\sigma^{\tilde{\gamma}}.$
}

Notice also that, following refs.\,\cite{Germani:2018jgr,Musco:2018rwt,DeLuca:2019qsy}, we do not use the smoothing in eq.\,(\ref{eq:Peak}) when dealing with peak theory. The reason is that when applying peak theory, and in particular when computing the threshold for collapse, the scale $r_m$ in position space associated to the peak of the power spectrum arises naturally.

On general ground, the abundance obtained by means of peak theory is larger than the one provided by threshold statistics\,\cite{Young:2014ana,Germani:2018jgr}.
In this paper, however, our interest is not a comparison between the two approaches. Instead, the issue that we want to explore is whether the gaussian approximation in eqs.\,(\ref{eq:FundBeta},\,\ref{eq:Peak}) is justified.
It is clear that the formation of PBH is a tail effect and, as such, strongly depends on possible deviations from the gaussian approximation. 
Investigating the possible presence of non-gaussianities in the context of realistic inflationary models in which one generates a sizable abundance of PBH is precisely the main goal of this work.
After clarifying whether non-gaussianities are present and what are their main properties, we will quantify their impact on the PBH abundance focussing on threshold statistics.
We will consider the case of peak theory and present a comparison with threshold statistics in a forthcoming work\,\cite{Riccardi:2021rlf}.

Let us, therefore, illustrate in more details what are the conditions that the inflationary dynamics should possess in order to generate the large density fluctuations needed for PBH formation.

\begin{figure}[!htb!]
\begin{center}
$$\includegraphics[width=.42\textwidth]{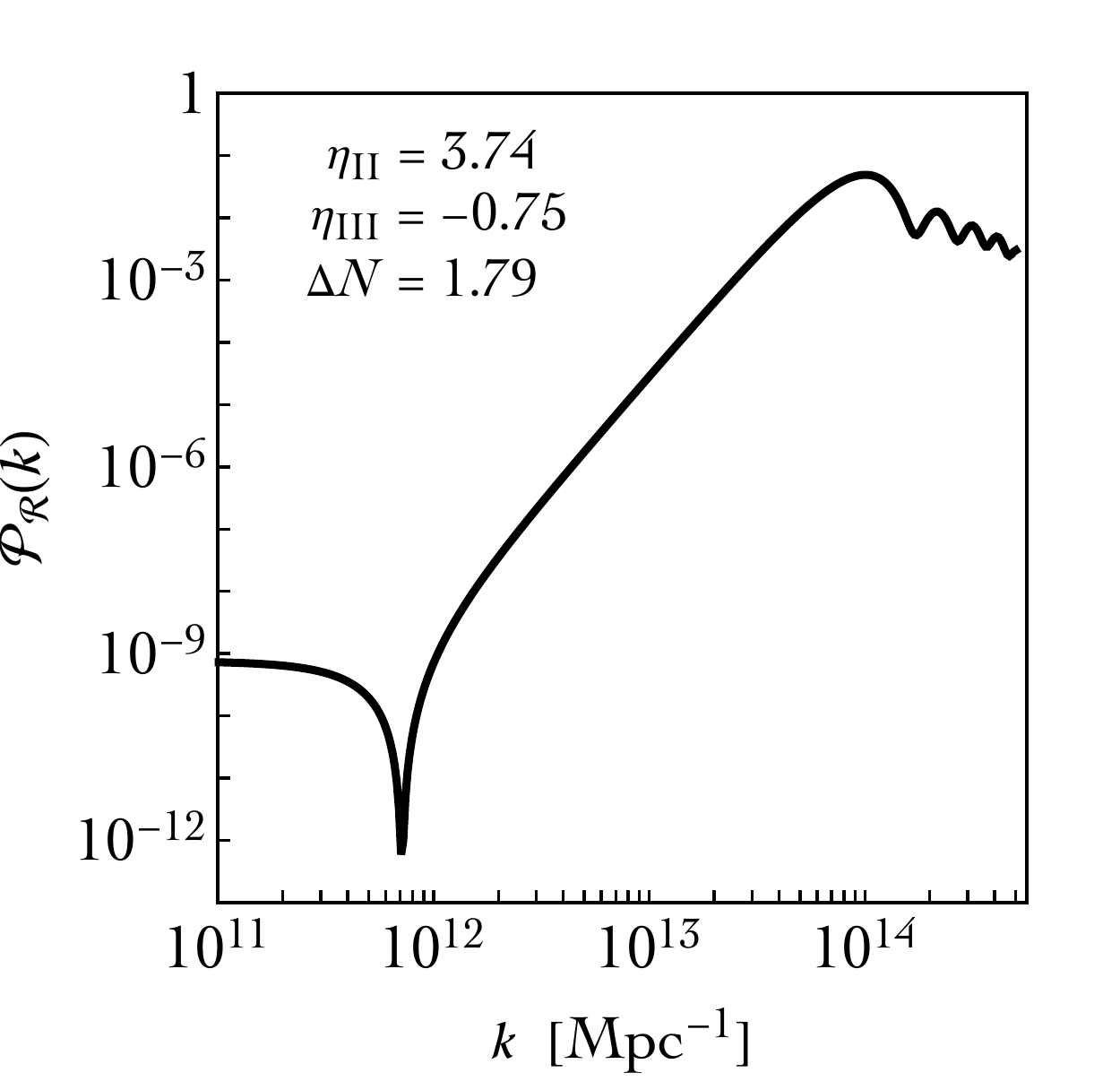}
\qquad\includegraphics[width=.42\textwidth]{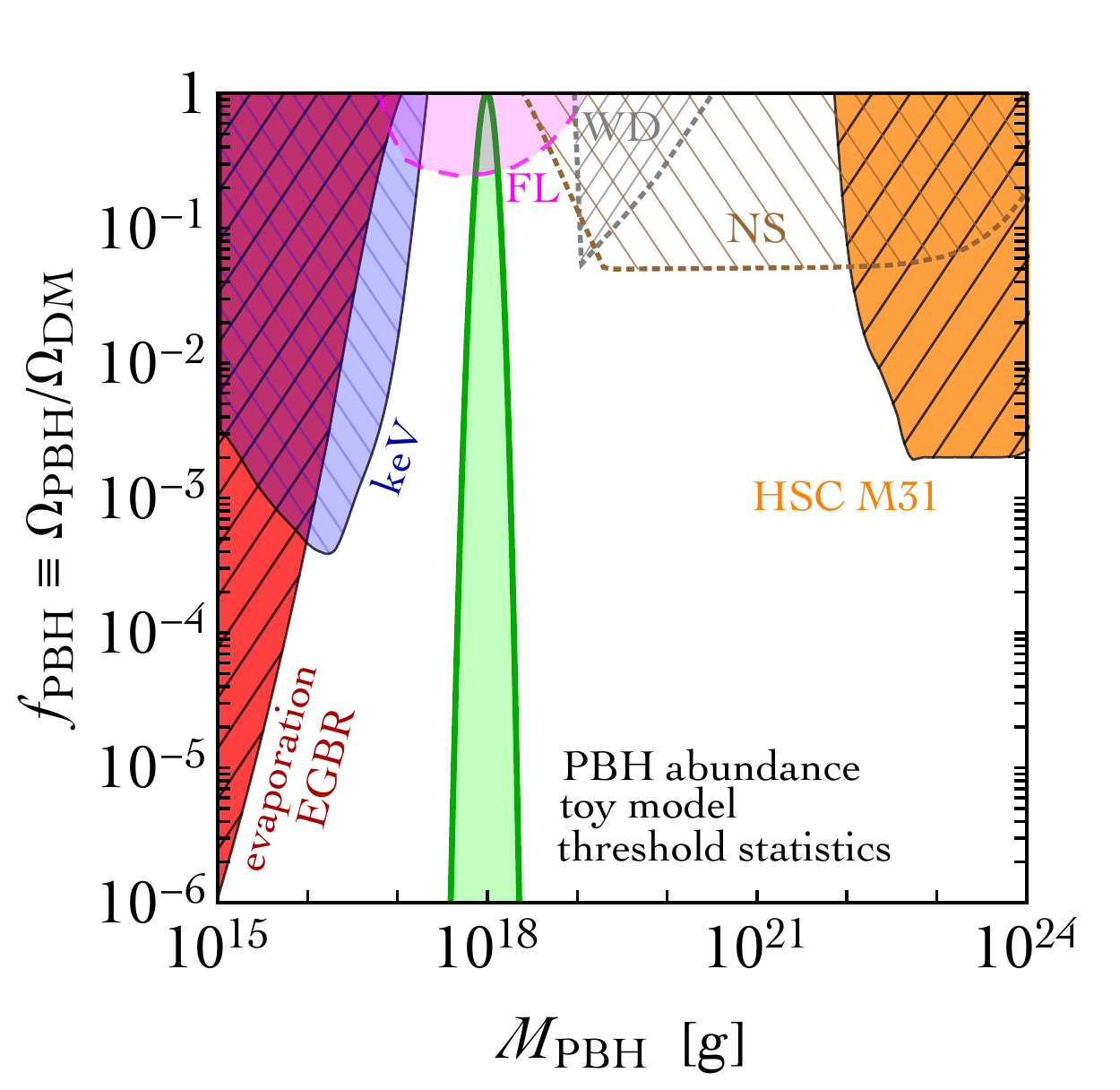}$$
\caption{\em \label{fig:PhysSpectrum}
Power spectrum of comoving curvature perturbations and PBH abundance in the gaussian approximation.
Both quantities refer to the toy analytical model.
The abundance is computed by means of threshold statistics, eq.\,(\ref{eq:PresentDayAbundance}). 
We compare the abundance of PBH with existing bound from Hawking evaporation using both the extragalactic background radiation (EGBR, ref.\,\cite{Carr:2009jm}) and the 511 keV gamma-ray line (keV, ref.\,\cite{Laha:2019ssq})
and micro-lensing constraints from observation of the Andromeda galaxy M31 (HSC\,M31, ref.\,\cite{Niikura:2017zjd}).
We also show future detection prospects using femto-lensing  of gamma-ray bursts (FL, ref.\,\cite{Katz:2018zrn}).
For completeness, we also show bounds from neutron star disruption (NS, ref.\,\cite{Capela:2013yf}; see also refs.\,\cite{Pani:2014rca,Capela:2014qea,Defillon:2014wla,Montero-Camacho:2019jte} for a more detailed discussion), and white dwarf explosions (WD, ref.\,\cite{Montero-Camacho:2019jte,Graham:2015apa}).
 }
\end{center}
\end{figure}

In the gaussian approximation, a sizable PBH abundance is obtained for $\mathcal{P}_{\mathcal{R}}\simeq 10^{-1}\textrm{--}10^{-2}$ at the PBH scales, requiring therefore a huge enhancement of $\mathcal{P}_{\mathcal{R}}$ with respect to its value at the CMB scales, $\mathcal{P}_{\mathcal{R}}\simeq 10^{-9}$.
This can happen in presence of an USR phase during inflation.
In fact, during such a transitory part of the inflationary dynamics, the amplitude of the comoving curvature perturbations can be exponentially enhanced. Let us introduce the Hubble parameters $\epsilon$ and $\eta:$
 \begin{align}\label{eq:HubbleEps}
\epsilon \equiv - \frac{\dot{H}}{H^2}\,~~~~~~~~\eta \equiv -\frac{\ddot{H}}{2H\dot{H}}\,,
\end{align}
where $H$ is the Hubble function and dot is the derivative with respect to time. 
During standard slow roll inflation $|\epsilon |\ll1,$ $|\eta|\ll1$ and, in the super-horizon limit, the evolution of $\mathcal{R}_k$ is a linear combination of a constant mode and an exponentially decaying one. The latter becomes quickly negligible and the amplitude of $\mathcal{R}_k$ freezes to a constant value until the end of inflation. Instead, if $3+\epsilon-2\eta\simeq 3-2\eta<1,$ a condition which we take as a definition of the USR phase, the decaying mode becomes a growing one. % 
In practice $\mathcal{R}_k$ evolves as dumped harmonic oscillator, with the sign of the friction term controlled by the 
combination $3+\epsilon-2\eta.$ During the USR phase this sign flips and the friction term becomes a driving force.
This is responsible for the exponential growth of $\mathcal{R}_k$ and the enhancement of $\mathcal{P}_{\mathcal{R}}$ at small scales. Notice however that for certain comoving wavelengths the amplitude of the growing mode (once it freezes to a constant value after the USR phase) is opposite to the one of the constant mode. This cancellation produces a dip in the power spectrum.
The outcome of this intricate dynamics can produce a power spectrum of comoving curvature perturbations as the one in fig.\,\ref{fig:PhysSpectrum}.

\begin{figure}[!htb!]
\begin{center}
$$
\qquad\includegraphics[width=.48\textwidth]{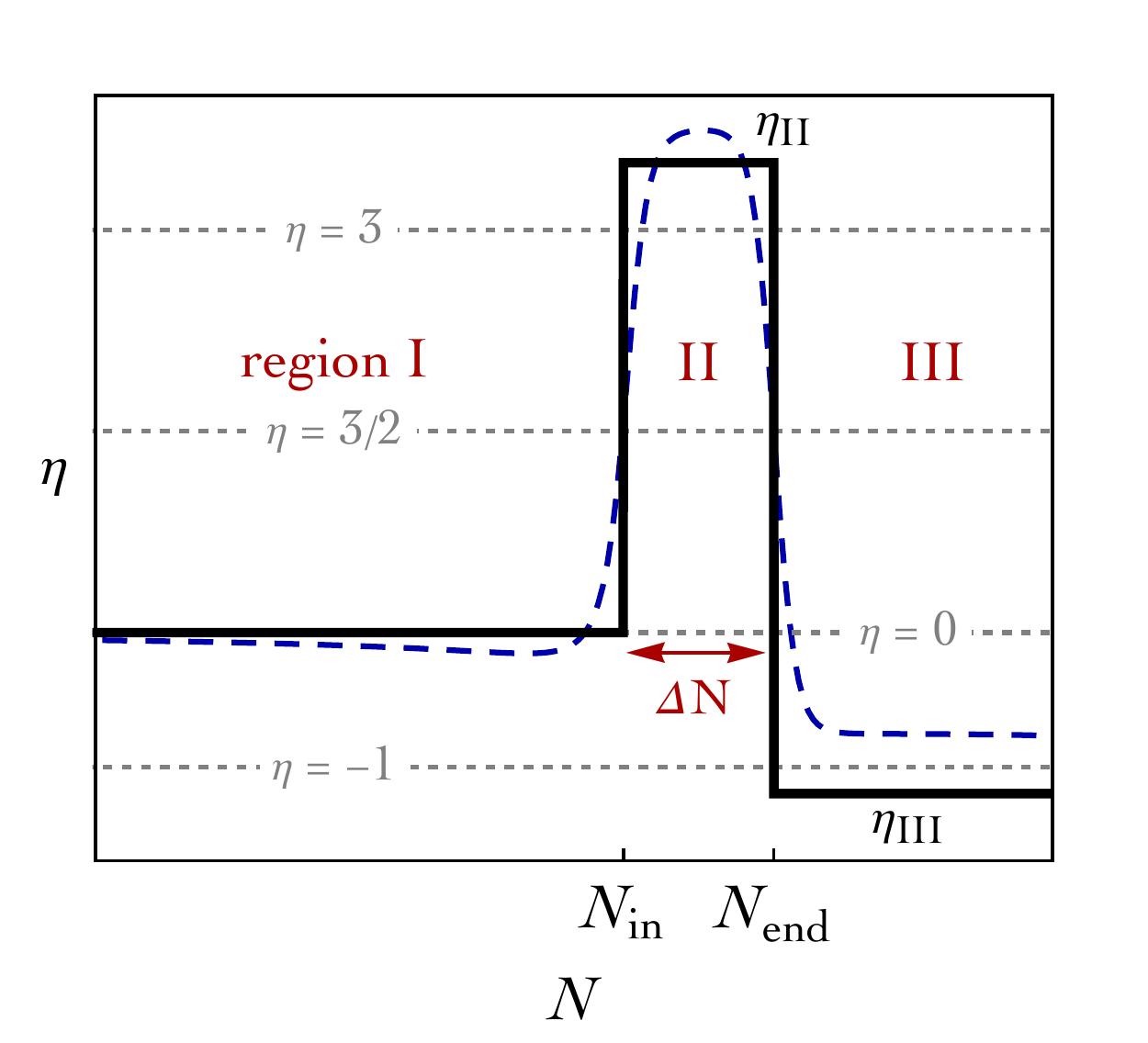}$$
\caption{\em \label{fig:Pot} 
The blue dashed line shows the evolution of $\eta$ as function of the number of $e$-folds in the model of ref.\,\cite{Ballesteros:2020qam}. Specifically, within this solution (depicted with a cyan star in\,\cite{Ballesteros:2020qam}) all the dark matter is in form of PBHs.
The black line shows the behaviour of $\eta$ in our toy-model.  
 }
\end{center}
\end{figure}

To make a quantitative analysis we shall consider an explicit inflationary model where the USR phase is realized. We focus on the single field model discussed in~\cite{Ballesteros:2020qam}, described by the following action in the Jordan frame 
\begin{align} \label{act}
\mathcal{S}   = \int d^4x\, \sqrt{-g}\, \left[  - \frac{1}{2}\left(M_P^2+\xi\,\phi^2\right) R + \frac{1}{2} g_{\mu\nu}\,\partial^{\mu}\phi\,\partial^{\nu}\phi -V(\phi)\right]\,.
\end{align}
where $\phi$ is the inflaton field, $M_P$ is the reduced Planck mass and $\xi$ the non-minimal coupling to gravity.
The potential is a polyonomial
\begin{align}\label{eq:TempHDO1}
V(\phi) = a_2\phi^2 + a_3\phi^3 + a_4\phi^4 + \sum_{n=5}^{\mathcal{N}}a_n\phi^n.
\end{align}  
and in full generality contains also higher order operators ($n>4$). These are not mandatory for the production of PBH but, as noted in~\cite{Ballesteros:2020qam}, they improve the compatibility of the model with the CMB observables.
After a judicious tuning of the coefficients of the quadratic and cubic terms, the potential presents an approximate stationary inflection point. 
This key features induces an USR phase in the inflationary dynamics. The dashed blue line in fig.\,\ref{fig:Pot} shows the evolution of $\eta$ as a function of the $e$-fold time for a specific choice of the parameters of the model.
As it is evident, the condition $\eta>3/2$, and therefore the USR phase, is realized for a while.
This solution, depicted with a cyan star in~\cite{Ballesteros:2020qam}, is in agreement with CMB observables and present 100\% of DM in the form of PBHs (the abundance is computed with eq.\,(\ref{eq:FundBeta})).  
For concreteness, in the rest of the paper, we focus on this benchmark point for our numerical analysis. We refer to it as numerical model in the following.
However, let us stress that our results will not be limited to this specific solution. On the contrary, as we will see, our conclusion will be of general validity for all the single-field inflationary models in which PBH production is induced by the presence of an approximate stationary inflection point before the end of inflation.

As shown by the black line in fig.\,\ref{fig:Pot}, the evolution of $\eta$ can be approximated, at least at the qualitative level, by  a piecewise function. This observation has prompted a simple analytical model to describe the inflationary dynamics~\cite{Ballesteros:2020qam,Ballesteros:2020sre}.
The inflationary phase is splitted in the three periods, each of them with constant values of $\eta$ (and $\epsilon\ll1$).
In this way one can obtain an analytical solution for the evolution of $\mathcal{R}_k$ and consequently its power spectrum.  The advantage is that one can gain more insight into the features of the USR dynamics, as the enhancement or decrease of $\mathcal{R}_k$ for certain comoving wavelenghts (see~\cite{Ballesteros:2020qam} for an in-depth discussion). Moreover, the analytical model does not rely on a specific inflationary model, rather, changing its parameters, it can describe different USR phases.
The drawback is that it is not realistic enough for a detailed comparison with observables.
In the rest of the paper we will compare the results based on the analytical model with the outcome of a numerical analysis performed in the context of the inflationary model described above.

Let us briefly summarize the main features of the analytical model. The three regions discussed above are as follows.
\begin{itemize}
\item [$\circ$] \textbf{Region\,I}. It corresponds to a standard slow roll phase. We assume $\eta_I=0$ and therefore we have a constant $\epsilon_{\rm I}\ll1.$
\item [$\circ$] \textbf{Region\,II}. It corresponds to the USR phase, which extends from the the $e$-fold time $N_{\rm in}$ to $N_{\rm end}$. We define $\Delta N\equiv N_{\rm end}-N_{\rm in}$ and take $\eta_{\rm II}\gg3/2.$ 
Considering $\epsilon \ll\eta, $ its evolution is $\epsilon_{\rm II}(N) = \epsilon_{\rm I}e^{-2\eta_{\rm II}(N-N_{\rm in})}.$
\item [$\circ$] \textbf{Region\,III}. It corresponds to the last part of the inflationary dynamics. We assume a negative value for $\eta_{\rm III}.$ $\epsilon$ evolves according to $\epsilon_{\rm III}(N) = \epsilon_{\rm I}e^{-2\eta_{\rm II}\Delta N}e^{-2\eta_{\rm III}(N-N_{\rm end})}.$
\end{itemize}
In each region the evolution of the $\mathcal{R}_k$ can be obtained analytically by solving the Mukhanov-Sasaki equation. Switching to conformal time $\tau$ (we remind that $dN_{e}/d\tau=a\,H$), for each region $\rm (i)$ we have 
\begin{align}\label{eq:HankelGen}
\mathcal{R}^{\rm (i)}_{k}(\tau) = \frac{H\,\sqrt{\pi}}{2\,\sqrt{2\,\epsilon^{\rm (i)}(\tau)}} (-\tau)^{3/2}\left[ \alpha^{\rm (i)}_k\,H_{\nu_{\rm i}}^{(1)}(-k\tau)\, e^{i(\nu_{\rm i} + 1/2)\pi/2}\,+ \beta^{\rm (i)}_k\,H_{\nu_{\rm i}}^{(2)}(-k\tau)\, e^{-i(\nu_{\rm i} + 1/2)\pi/2}\, \right],
\end{align}
where $H$ is taken constant since we always have $\epsilon\ll 1$. $H_{\nu_{\rm i}}^{(1)}$ and $H_{\nu_{\rm i}}^{(2)}$ are the Hankel function of first and second kind respectively and $\nu_{\rm i}=\sqrt{9/4 -\eta_{\rm i}\left(3-\eta_{\rm i}\right)}.$
Imposing Bunch-Davies initial conditions, one has $\alpha^{\rm I}_k=1$ and $\beta^{\rm I}_k=0$  in the first region.
The complex coefficients $\alpha^{\rm (i)}_k$ and $\beta^{\rm (i)}_k$ in the regions II and III are obtained requiring continuity of $\mathcal{R}_k$ and its first derivative across the three regions.
Finally, the power spectrum $\mathcal{P}_{\mathcal{R}}(k)$ is computed at super-Hubble scales 
as $\mathcal{P}_{\mathcal{R}}(k) \equiv \lim_{k\tau\to 0^-} \frac{k^3}{2\pi^2} |\mathcal{R}_{k}(\tau)|^2.$ 

We can now try to relate the parameters of the model to observable quantities (see table\,\ref{eq:ModelTab}). 
For this purpose in the left panel of fig.\,\ref{fig:PhysSpectrum}  we show the power spectrum of comoving curvature perturbations as a function of the comoving scale $k$ for a choice of the parameters of the analytical model.
The large enhancement at small scales allows for a sizable PBH production. As anticipated, the analytical model reproduces, at least qualitatively, the main features obtained with a more accurate numerical calculation, namely the presence of a peak and a dip in the power spectrum.
The correspondence between what we compute and what we observe can be explained with the following points.
\begin{table}[htp]
\begin{center}
\begin{tabular}{||c||c|c|c|c|c||}
\hline\hline
 \multirow{2}{*}{\textbf{Model parameter}} & \multirow{2}{*}{$H^2/8\pi^2\epsilon_{\rm I}$} & \multirow{2}{*}{$N_{\rm in}$} & \multirow{2}{*}{$\Delta N$} & \multirow{2}{*}{$\eta_{\rm II}$} & \multirow{2}{*}{$\eta_{\rm III}$} \\ 
 & & & & & \\ \hline
\multirow{2}{*}{\textbf{Description}} & \multirow{2}{*}{$\mathcal{P}_{\mathcal{R}}$ in region\,I}  & \multirow{2}{*}{beginning of USR} & \multirow{2}{*}{duration of USR} & 
\multirow{2}{*}{value of $\eta$ during USR} & 
\multirow{2}{*}{value of $\eta$ after USR}   \\ 
 & & & & &  \\ \hline
\textbf{Features in the} & normalization of $\mathcal{P}_{\mathcal{R}}$ & \multirow{2}{*}{$k_{\rm peak}$} & 
\multirow{2}{*}{$k_{\rm dip}$,\,$\mathcal{P}_{\mathcal{R}}(k_{\rm peak})$} & \multirow{2}{*}{$k_{\rm dip}$,\,$\mathcal{P}_{\mathcal{R}}(k_{\rm peak})$} & \multirow{2}{*}{$\mathcal{P}_{\mathcal{R}}(k)$}   \\ 
\textbf{power spectrum} & at large scales & & & &  \\ \hline
 \multirow{2}{*}{\textbf{Observable}} & \multirow{2}{*}{$A_s$} & 
 \multirow{2}{*}{$M_{\rm PBH}$} & \multirow{2}{*}{$n_s$} & \multirow{2}{*}{$f_{\rm PBH}$} & 
 \multirow{2}{*}{total number of $e$-folds}   \\ 
 & & & & & \\ \hline\hline
\end{tabular}
\end{center}\vspace{-0.35cm}
\caption{{\it 
Correspondence between the parameters of the model introduced in section\,\ref{sec:Abundance} and the cosmological observables relevant for PBH formation.
}}\label{eq:ModelTab}
\end{table}
\begin{itemize}
\item [{\it i)}] We fix the position of the peak of $\mathcal{P}_{\mathcal R}(k)$ at $k_{\rm peak} = 10^{14}$ Mpc$^{-1}$. 
This choice, in turn, gives a PBH distribution peaked around $M_{\rm PBH} = 10^{18}$ g. 
These values are motivated by the fact that such population of PBHs can successfully account for the totality of dark matter in the present-day Universe. 
Furthermore, from the value $M_{\rm PBH} = 10^{18}$ g we know that the comoving wavenumber 
$k_{\rm peak}$ crosses the horizon approximately $36$ $e$-folds after the perturbation associated to the
 pivot wavenumber associated with the CMB scale $k_* = 0.05$ Mpc$^{-1}$ does it. From that, we infer the value of $N_{\rm peak}$ (and $N_{\rm in}$).
\item [{\it ii)}] For a given value of $\Delta N$ and $\eta_{\rm II}$, we can compute the position of the dip and, consequently, the number of $e$-folds between horizon crossing of the pivot scale $k_*$ and $k_{\rm dip}$. 
We can, therefore, compute the normalization of the power spectrum $H^2/8\pi^2\epsilon_{\rm I}$ by rescaling 
the observed value at the pivot scale, namely $A_s \simeq 2\times 10^{-9}$, up to $k_{\rm dip}$ by means of the slow-roll evolution $\mathcal{P}_{\mathcal R}(k) = A_s(k/aH)^{n_s - 1}$ where we take $n_s = 0.965$.
\item [{\it iii)}] Finally, the values of $\Delta N$, $\eta_{\rm II}$ and $\eta_{\rm III}$ are readjusted--recomputing for each choice the normalization described at point {\it ii)}--by requiring $f_{\rm PBH} = 1$. The value of $\eta_{\rm III}$ mostly controls the dynamics 
after the USR phase (although it also enters in the computation of the power spectrum), and its value can be gauged in order to guarantee that inflation lasts approximately 60 $e$-folds.

\end{itemize}
Typical values of $\Delta N$, $\eta_{\rm II}$ and $\eta_{\rm III}$ which generate the correct PBH abundance of dark matter 
are shown in fig.\,\ref{fig:PhysSpectrum}.
Despite the simplicity of the model, it is nice to see that these numbers are in full agreement with those obtained, by means of a careful numerical analysis, in the context of more complete models\,\cite{Ballesteros:2020qam}.

However, this discussion makes also clear that the picture risks being 
grossly incomplete without including the effect of 
non-gaussianities. This is mostly because the integral in eq.\,(\ref{eq:FundBeta}) is dominated by the tail of the distribution where deviation from the gaussian approximation may give a sizable effect.

On general ground, deviations from the gaussian approximation are expected in realistic models of inflation. 
At CMB scales, non-gaussianities are typically negligible since suppressed by the small values of the 
Hubble parameters $\epsilon$ and $\eta$ (see ref.\,\cite{Akrami:2019izv} for the latest experimental 
constraints on primordial non-gaussianities at the scales probed by the Planck experiment). 
However, we are interested in the opposite situation in which slow-roll is violated, and 
in particular $\eta$ takes sizable values. It is, therefore, plausible to expect sizable  non-gaussianities
at scales much smaller that those probed by CMB measurements where the USR dynamics might take place.

Deviation from normality can be explored by considering skewness and kurtosis.
Skewness ($\mathcal{S}$) and kurtosis ($\mathcal{K}$) measure deviations with respect to a gaussian  normal distribution. 
Gaussian distributions are symmetric around their mean while, for a generic distribution, skewness measures the degree of departure from the symmetric shape. 
Kurtosis, on the contrary, is a measure of how differently shaped are the tails of a distribution 
with respect to the tails of the gaussian distribution. 
In the presence of non-gaussianities, eq.\,(\ref{eq:FundBeta}) is modified as\,\cite{Matarrese:1986et,Franciolini:2018vbk} 
\begin{align}\label{eq:FundBetaNG}
\beta_{\mathcal{NG}}(M_{\rm PBH}) = \beta_{\mathcal{G}}(M_{\rm PBH}) + 
\gamma \frac{e^{-\delta_{\rm th}^2/2\sigma_R^2}}{\sqrt{\pi}}\left[
\frac{\mathcal{S}_R}{2^{3/2}3!}H_2\left(\frac{\delta_{\rm th}}{\sqrt{2}\sigma_R}\right) +
\frac{\mathcal{K}_R}{2^{2}4!}H_3\left(\frac{\delta_{\rm th}}{\sqrt{2}\sigma_R}\right) + \dots
\right]\,,
\end{align}
which can be considered as an application of the Gram-Charlier series, 
that is the expansion of the probability density function (PDF) in terms of its cumulant 
(sometimes also called Edgeworth expansion, the latter being the same series but with a different truncation and ordering of terms). 
In eq.\,(\ref{eq:FundBetaNG}), $\mathcal{S}_R$ and $\mathcal{K}_R$ are the skewness and the kurtosis smoothed on the scale $R$ (in analogy with 
eq.\,(\ref{eq:SmoothedVariance})) and their explicit definitions will be discussed in the next sections. 
In short, skewness is related to the three-point correlation function which vanishes in the gaussian limit (see, e.g., eq.\,(\ref{eq:Skew})) 
while kurtosis is related to the connected part of the four-point correlation function (that is the part of the four-point correlator that remains after subtracting the gaussian contribution, see, e.g., eq.\,(\ref{eq:Gaussian4pt})).
The dots in eq.\,(\ref{eq:FundBeta}) indicate contributions from higher order correlators ($n\geqslant 5$) and 
$H_n$ are the Hermite polynomials.\footnote{Notice that we use the so-called ``physicists' Hermite polynomials'' which are defined by $H_n(x) = (-1)^n e^{x^2} d^ne^{-x^2}/dx^n$ instead of the so-called ``probabilists' Hermite polynomials'' $He_n(x)$ often preferred in statistics theory. The relation among the two is $He_n(x) = 2^{-n/2}H_n(x)$.}
In the following, we shall proceed with the discussion of non-gaussianities assuming eq.\,(\ref{eq:FundBetaNG}).
In the next section we will compute the skewness and in section\,\ref{sec:Kurtosis} we will focus on the kurtosis.
With these results, in section\,\ref{sec:Local} we will investigate the impact of non-gaussianities on the abundance of PBHs.

\section{The bispectrum and the skewness}\label{sec:SkewnessExact}

The three-point correlation function in momentum space is
\begin{align}\label{eq:Bispectrum}
\langle \mathcal{R}_{k_1}\mathcal{R}_{k_2}\mathcal{R}_{k_3} \rangle = (2\pi)^3\delta^{(3)}(\vec{k}_1+\vec{k}_2+\vec{k}_3)
B_{\mathcal{R}}(k_1,k_2,k_3)\,,~~~~~~~~~~
\raisebox{-7mm}{\begin{tikzpicture}
	\draw [->,thick,draw=oucrimsonred]  (0,0.95)--(1.45,0.15);
	\draw [->,thick,draw=verdes]  (1.45,0.27)--(1.6,1.55);
	\draw [->,thick,draw=persianblue]  (1.55,1.55)--(0,1);
	\node[anchor=west] at (.5,1.6) {\scalebox{0.85}{{\color{persianblue}{$\vec{k}_1$}}}};
	\node[anchor=west] at (0,.5) {\scalebox{0.85}{{\color{oucrimsonred}{$\vec{k}_2$}}}};
	\node[anchor=west] at (1.5,1.) {\scalebox{0.85}{{\color{verdes}{$\vec{k}_3$}}}};
	\end{tikzpicture}}
\end{align}
where $B_{\mathcal{R}}(k_1,k_2,k_3)$ defines the bispectrum. 
In the gaussian approximation, the bispectrum vanishes.
While the power spectrum is a function of $k$ only, the bispectrum is a three-dimensional function defined
on a tetrahedral region for which $\vec{k}_1$, $\vec{k}_2$ and $\vec{k}_3$ satisfy the triangle condition enforced--as illustrated in the schematic picture above--by
the delta function in eq.\,(\ref{eq:Bispectrum}).\footnote{The delta function corresponds to invariance under translations, and the fact that the bispectrum depends only on the lengths of the three sides of the triangle formed by the momenta corresponds to invariance under rotations. Homogeneity and isotropy, therefore, reduce the nine parameters 
of the triad $\vec{k}_{1,2,3}$ down to three.}
It is convenient to define the reduced bispectrum $\mathcal{B}_{\mathcal{R}}(k_1,k_2,k_3)$ by means of
\begin{align}\label{eq:ReducedBi}
B_{\mathcal{R}}(k_1,k_2,k_3) = \mathcal{B}_{\mathcal{R}}(k_1,k_2,k_3)\left[
\Delta_{\mathcal{R}}(k_1)\Delta_{\mathcal{R}}(k_2) + 
\Delta_{\mathcal{R}}(k_1)\Delta_{\mathcal{R}}(k_3)+
\Delta_{\mathcal{R}}(k_2)\Delta_{\mathcal{R}}(k_2)
\right]\,.
\end{align}
This is a useful definition since the reduced bispectrum $\mathcal{B}_{\mathcal{R}}$ 
is dimensionless (roughly speaking, the scaling $B_{\mathcal{R}} \sim 1/k^6$ is absorbed by $\Delta_{\mathcal{R}}^2 \sim 1/k^6$).
In position space, the skewness is
\begin{align}\label{eq:skewness}
\mathcal{S} = \frac{1}{(\sigma^2)^{3/2}}\int\frac{d^3\vec{k}_1}{(2\pi)^3}\frac{d^3\vec{k}_2}{(2\pi)^3}
\frac{d^3\vec{k}_3}{(2\pi)^3}\langle \mathcal{R}_{k_1}\mathcal{R}_{k_2}\mathcal{R}_{k_3} \rangle 
= \left.\frac{1}{(\sigma^2)^{3/2}}\int\frac{d^3\vec{k}_1}{(2\pi)^3}
\frac{d^3\vec{k}_2}{(2\pi)^3}B_{\mathcal{R}}(k_1,k_2,k_3)\right|_{\vec{k}_1+\vec{k}_2+\vec{k}_3 = 0}\,.
\end{align}
We take this opportunity to give a more general definition. We define the $n$-th cumulant
\begin{align}\label{eq:Cumulants}
\mathcal{C}^{(n)} \equiv \frac{\langle\overbrace{\mathcal{R}(\vec{x})\dots\mathcal{R}(\vec{x})}^{n\,{\rm times}} \rangle}{(\sigma^2)^{n/2}} = \frac{\langle \mathcal{R}^n \rangle}{\sigma^n}\,,
\end{align}
as the connected part of the $n$-point correlator (evaluated at the same point) normalized by the $n$-th power of the standard deviation. The connected part of the $n$-point correlator is obtained by subtracting from the full expression of the 
$n$-point correlator the gaussian contribution (which is zero for the 3-point correlator but, as we shall see in the next section, does not vanish for $n\geqslant 4$, see e.g. eq.\,(\ref{eq:Gaussian4pt})). Given the definition in eq.\,(\ref{eq:Cumulants}), 
we have $\mathcal{C}^{(3)} = \mathcal{S}$.

The integration in $\mathbb{R}^3$ is limited by the tetrahedral region illustrated below (with semi-perimeter $K\equiv (k_1+k_2+k_3)/2$)
  \begin{align}\label{eq:IntegrationBispectrum}
\mathcal{V}_{\mathcal{B}}\equiv \left\{
\begin{array}{c}
 k_1 = K(1-\alpha+\beta)/2  \\
 k_2  =  K(1+\alpha+\beta)/2  \\
 k_3 = K(1-\beta)\hspace{10mm}
\end{array}
\right.~~
\resizebox{30mm}{!}{
\parbox{35mm}{
\begin{tikzpicture}[]
\node (label) at (0,0)[draw=white]{ 
       {\fd{3.5cm}{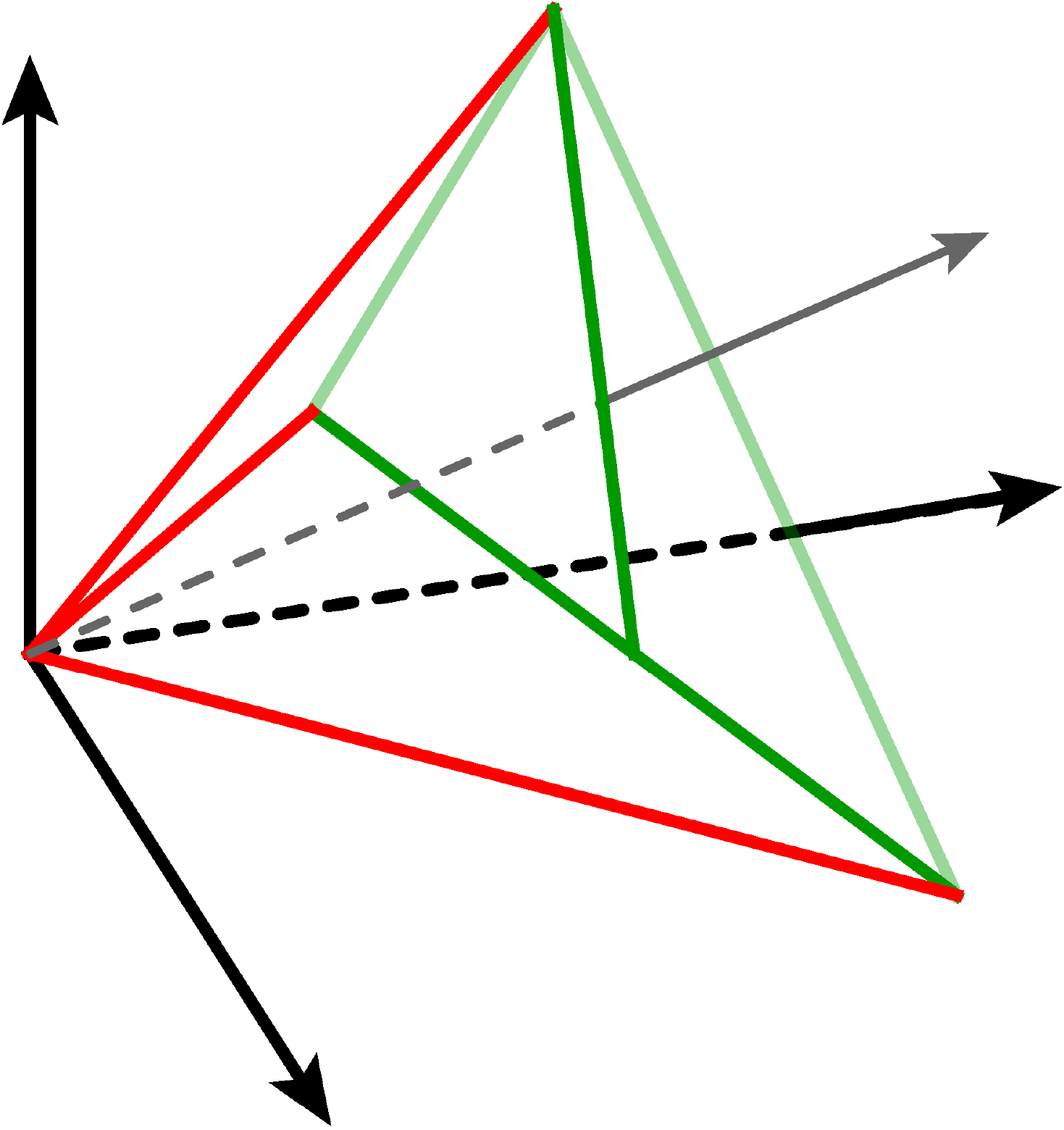}} 
      };
      \node[anchor=north] at (-2,1.75) {\scalebox{0.85}{$\vec{k}_1$}};
      \node[anchor=north] at (2,.75) {\scalebox{0.85}{$\vec{k}_2$}}; 
      \node[anchor=north] at (-1.4,-1.2) {\scalebox{0.85}{$\vec{k}_3$}};
      \node[anchor=north] at (1.7,1.6) {\scalebox{0.85}{$K$}};
      \node[anchor=north,rotate=-12] at (.7,-.9){\scalebox{0.85}{{\color{red}{$k_2=k_3,\,k_1=0$}}}}; 
      \node[anchor=north,rotate=52] at (-0.95,1.3){\scalebox{0.85}{{\color{red}{$k_1=k_2,\,k_3=0$}}}}; 
      \node[anchor=north,] at (.5,-.5){\scalebox{0.85}{{\color{verdes}{$\alpha$}}}}; 
      \node[anchor=north,] at (.45,0.6){\scalebox{0.85}{{\color{verdes}{$\beta$}}}}; 
\end{tikzpicture}
}}\hspace{1.5cm}{\rm with\,slices\,}\hspace{-.5cm}
\resizebox{30mm}{!}{
\parbox{35mm}{
\begin{tikzpicture}[]
\node (label) at (0,0)[draw=white]{ 
       {\fd{3.2cm}{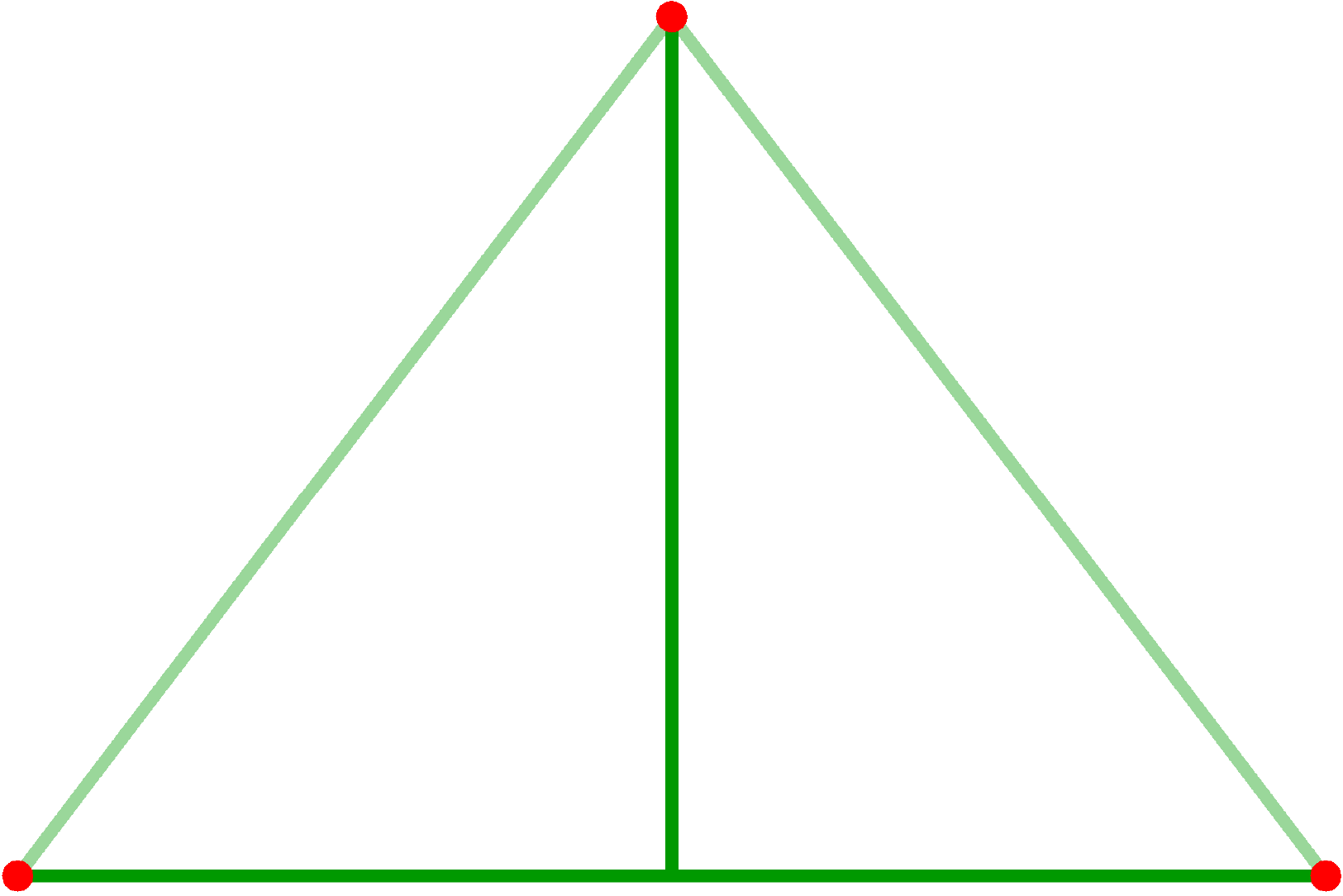}}        
      };
      \node[anchor=north,] at (0,1.5){\scalebox{0.85}{{\color{verdes}{$\beta = 1$}}}};
      \node[anchor=north,] at (0,-1.1){\scalebox{0.85}{{\color{verdes}{$\beta = \alpha = 0$}}}};
      \node[anchor=north,] at (-1.8,-1.1){\scalebox{0.85}{{\color{verdes}{$\alpha = -1$}}}};
      \node[anchor=north,] at (1.8,-1.1){\scalebox{0.85}{{\color{verdes}{$\alpha = 1$}}}};
      \node[anchor=north,] at (0,1.9){\scalebox{0.85}{{\color{black}{$k_1=k_2=K$,\,$k_3=0$}}}}; 
      \node[anchor=north,] at (-1.8,-1.4){\scalebox{0.85}{{\color{black}{$k_1=k_3=K$}}}};
      \node[anchor=north,] at (-2.15,-1.7){\scalebox{0.85}{{\color{black}{$k_2=0$}}}};
      \node[anchor=north,] at (1.8,-1.4){\scalebox{0.85}{{\color{black}{$k_2=k_3=K$}}}}; 
      \node[anchor=north,] at (1.45,-1.7){\scalebox{0.85}{{\color{black}{$k_1=0$}}}}; 
      \node[anchor=north,] at (1.45,0.95){\scalebox{0.85}{{\color{black}{constant $K$}}}}; 
\end{tikzpicture}
}}
\end{align}
and ranges $K\in[0,\infty)$, $\alpha\in [-(1-\beta),1-\beta]$ and $\beta\in [0,1]$.
Qualitatively, if we write (assuming $\mathcal{P}_{\mathcal{R}}$ and $\mathcal{B}_{\mathcal{R}}$ are 
scale-independent) $(\sigma^2)^{3/2} \sim \mathcal{P}_{\mathcal{R}}^{3/2}$ and 
$B_{\mathcal{R}}\sim \mathcal{B}_{\mathcal{R}}\mathcal{P}_{\mathcal{R}}^{2}$, 
we see that one expects $\mathcal{S} \sim \mathcal{B}_{\mathcal{R}}\mathcal{P}_{\mathcal{R}}^{1/2}$.
As a rule of thumb, we expect an almost gaussian perturbation if 
$\mathcal{B}_{\mathcal{R}} \ll \mathcal{P}_{\mathcal{R}}^{-1/2}$. 
This means that $\mathcal{B}_{\mathcal{R}}$ is an interesting quantity to compute to have at least a first idea about 
the impact of the three-point correlator on non-gaussianities. 
To perform the integral in eq.\,(\ref{eq:skewness}) some care is needed for the angular part since the triangular 
condition 
enforced by the delta function involves also angles. 
We use the exponential integral form of the delta function, by means of which we write
\begin{align}
\int\frac{d^3\vec{k}_1}{(2\pi)^3}\frac{d^3\vec{k}_2}{(2\pi)^3}
\frac{d^3\vec{k}_3}{(2\pi)^3}
(2\pi)^3\delta^{(3)}(\vec{k}_1+\vec{k}_2+\vec{k}_3) & =
\frac{1}{2\pi^5}\int 
d\log x \int dk_1 dk_2 dk_3 (k_1k_2k_3)\sin(k_1 x)\sin(k_2 x)\sin(k_3 x)\nonumber \\
& = \frac{1}{8\pi^4}\int_{\mathcal{V}_{\mathcal{B}}} dk_1 dk_2 dk_3 (k_1k_2k_3)\,,\label{eq:AngularRewriting}
\end{align} 
where, crucially, the second equality in which we performed the integral over $x$ is valid only 
inside the tetrahedron defined in eq.\,(\ref{eq:IntegrationBispectrum}) (as one can check by using the
parametrization of $k_{1,2,3}$ given in eq.\,(\ref{eq:IntegrationBispectrum}) and performing the integral over $x$ after imposing the 
correct constraints for $\alpha$ and $\beta$). Of course, this way of rewriting the angular integral works only 
because we are assuming isotropy for the bispectrum (i.e. the omitted integrand in eq.\,(\ref{eq:AngularRewriting}) does not depend on angles). 

In analogy with eq.\,(\ref{eq:SmoothedVariance}), we define the smoothed skewness as
\begin{align}
\mathcal{S}_R &=  \frac{1}{(\sigma_R^2)^{3/2}}\int\frac{d^3\vec{k}_1}{(2\pi)^3}\frac{d^3\vec{k}_2}{(2\pi)^3}
\frac{d^3\vec{k}_3}{(2\pi)^3}
W(k_1R)W(k_2R)W(k_3R)\frac{8(1+\omega)^3}{(5+3\omega)^3}
\left(\frac{k_1}{aH}\right)^2\left(\frac{k_2}{aH}\right)^2\left(\frac{k_3}{aH}\right)^2
\langle \mathcal{R}_{k_1}\mathcal{R}_{k_2}\mathcal{R}_{k_3} \rangle \nonumber \\
&= \frac{1}{(\sigma_R^2)^{3/2}}
\underbrace{\frac{8}{729\pi^4}\int_{\mathcal{V}_{\mathcal{B}}} dk_1dk_2dk_3(k_1k_2k_3)W(k_1R)W(k_2R)W(k_3R)
\left(\frac{k_1}{aH}\right)^2\left(\frac{k_2}{aH}\right)^2\left(\frac{k_3}{aH}\right)^2
B_{\mathcal{R}}(k_1,k_2,k_3)}_{\equiv \langle \delta^3 \rangle_R}\,,\label{eq:SmoothedSkew}
\end{align}
computed in the region defined by eq.\,(\ref{eq:IntegrationBispectrum}). 
We define the $n$-th cumulant (in analogy with eq.\,(\ref{eq:Cumulants}) but now in terms of $\delta$, and including the smoothing)
\begin{align}\label{eq:CumulantsR}
\mathcal{C}_R^{(n)} \equiv \frac{\langle\overbrace{\delta(\vec{x})\dots\delta(\vec{x})}^{n\,{\rm times}} \rangle}{(\sigma_R^2)^{n/2}} = \frac{\langle \delta^n \rangle_R}{\sigma_R^n}\,.
\end{align}
By using eq.\,(\ref{eq:SmoothedSkew}), the goal is to compute the skewness as function of the PBH mass.
To this end, we need to compute the three-point function 
$\langle \mathcal{R}_{k_1}\mathcal{R}_{k_2}\mathcal{R}_{k_3} \rangle$ in Fourier space. 
Notice that, in order to compute the integral in eq.\,(\ref{eq:SmoothedSkew}) without any approximation, 
we would like to keep the full shape in momentum space of $B_{\mathcal{R}}(k_1,k_2,k_3)$ into account.

We compute $\langle \mathcal{R}_{k_1}\mathcal{R}_{k_2}\mathcal{R}_{k_3} \rangle$ by means of the 
standard background+perturbations splitting and using the standard {\it in-in} formalism. We report directly the results and refer to appendix\,\ref{app:ThreePoints} for details.
One can isolate two dominant contributions in the calculation of three-point function. The first one (the sum of expressions {\it in} and {\it end} below) originates from the following operator in the cubic action of $\mathcal{R}:$ 
$\mathcal{S}_3  \supset -\int d\tau\,d^3x\, a^2\,\epsilon\,\eta^{\prime} \mathcal{R}^2\mathcal{R}^{\prime}.$
The second contribution (dubbed {\it red}) originates by a convenient redefinition of $\mathcal{R},$ see eq.\,(\ref{eq:Red1}).

We consider first the analytical model discussed in section\,\ref{sec:Abundance}.
Our result for $B_{\mathcal{R}}(k_1,k_2,k_3)$ is 
\begin{align}\label{eq:BispectrumStructure}
B_{\mathcal{R}}(k_1,k_2,k_3) = B^{({\rm in})}_{\mathcal{R}}(k_1,k_2,k_3) + 
B^{({\rm end})}_{\mathcal{R}}(k_1,k_2,k_3) + 
\underbrace{B^{({\rm red})}_{\mathcal{R}}(k_1,k_2,k_3)}_{\rm local}\,,
\end{align}
where
\begin{align}
k_{\rm in}^6 B^{({\rm in})}_{\mathcal{R}} & = -\Im\frac{H^4}{16\epsilon_{\rm I}^2}
\frac{\Gamma(\nu_{\rm III})^3}{\pi^{3/2}}\frac{e^{3(\eta_{\rm II}-\eta_{\rm III})\Delta N}}
{2^{3/2-3\nu_{\rm III}}}
(x_1 x_2 x_3)^{\eta_{\rm III}-\frac{3}{2}}\eta_{\rm II}\,\mathcal{R}^3_{\rm III}\,
\mathcal{F}_{\rm in}(x_1,x_2,x_3)\,,\label{eq:Bi1}\\
\mathcal{R}^3_{\rm III} & \equiv \prod_{i=1}^{3}\left[
\alpha_{k_i}^{\rm III}e^{\frac{i\pi}{2}(\nu_{\rm III}+\frac{1}{2})}-
\beta_{k_i}^{\rm III}e^{-\frac{i\pi}{2}(\nu_{\rm III}+\frac{1}{2})}
\right]\,,\\
k_{\rm in}^6 B^{({\rm end})}_{\mathcal{R}}  & = -\Im\frac{H^4}{16\epsilon_{\rm I}^2}
\frac{\Gamma(\nu_{\rm III})^3}{\pi^{3/2}}
\frac{e^{3(\eta_{\rm II}-\eta_{\rm III})\Delta N}}
{2^{3/2-3\nu_{\rm III}}}(x_1 x_2 x_3)^{\eta_{\rm III}-\frac{3}{2}}
(-\eta_{\rm II}+\eta_{\rm III})e^{2\Delta N(1-\eta_{\rm II})}\,\mathcal{R}^3_{\rm III}\,\mathcal{F}_{\rm end}(x_1,x_2,x_3)\,,\label{eq:Bi2}\\
k_{\rm in}^6 B^{({\rm red})}_{\mathcal{R}}  & = 
-\frac{4\pi^4 \eta_{\rm III}}{(x_1 x_2 x_3)^3}\left[
x_3^3\mathcal{P}_{\mathcal{R}}(k_1)\mathcal{P}_{\mathcal{R}}(k_2) 
+ x_1^3\mathcal{P}_{\mathcal{R}}(k_2)\mathcal{P}_{\mathcal{R}}(k_3)+
x_2^3\mathcal{P}_{\mathcal{R}}(k_1)\mathcal{P}_{\mathcal{R}}(k_3)  
\right]\,.\label{eq:Bi3}
\end{align}
In the expressions above we defined $x_{\rm i}\equiv k_{\rm i}/k_{\rm in},$ with $k_{\rm in}$ the comoving wavenumber associated to the start of the USR phase at conformal time $\tau_{\rm in}$ by the relation $k_{\rm in}\,\tau_{\rm in}=-1$.
The two functions $\mathcal{F}_{\rm in}(x_1,x_2,x_3)$ and $\mathcal{F}_{\rm end}(x_1,x_2,x_3)$ are defined by the 
general expression
\begin{align}
\mathcal{F}_{i}(x_1,x_2,x_3) \equiv 
\frac{i8\epsilon_{\rm I}^{3/2}k_{\rm in}^{7/2}}{H^3}\left.\left(
\mathcal{R}_{k_1}^*\mathcal{R}_{k_2}^*\frac{d\mathcal{R}_{k_3}^*}{d\tau} +
\mathcal{R}_{k_1}^*\mathcal{R}_{k_3}^*\frac{d\mathcal{R}_{k_2}^*}{d\tau} + 
\mathcal{R}_{k_2}^*\mathcal{R}_{k_3}^*\frac{d\mathcal{R}_{k_1}^*}{d\tau}\right)
\right|_{\tau = \tau_i}\,,
\end{align}
and we have
\begin{align}
\mathcal{F}_{\rm in}(x_1,x_2,x_3) = \frac{e^{-i(x_1+x_2+x_3)}}{(x_1x_2x_3)^{3/2}}\left[
x_3^{2}(x_1x_2 -ix_1 -ix_2 -1)
+ x_1^{2}(x_2x_3 -ix_2 -ix_3 -1) + 
x_2^{2}(x_1x_3 -ix_1 -ix_3 -1)
\right]\,,
\end{align}
while the corresponding expression for $\mathcal{F}_{\rm end}(x_1,x_2,x_3)$ is more complicated and must be evaluated 
numerically for non-integers $\eta_{\rm II,III}$. 
Armed with these expressions, we are ready to compute the reduced bispectrum in eq.\,(\ref{eq:ReducedBi}) 
and tackle the integral in eq.\,(\ref{eq:SmoothedSkew}).

\begin{figure}[!htb!]
\begin{center}
$$\includegraphics[width=.49\textwidth]{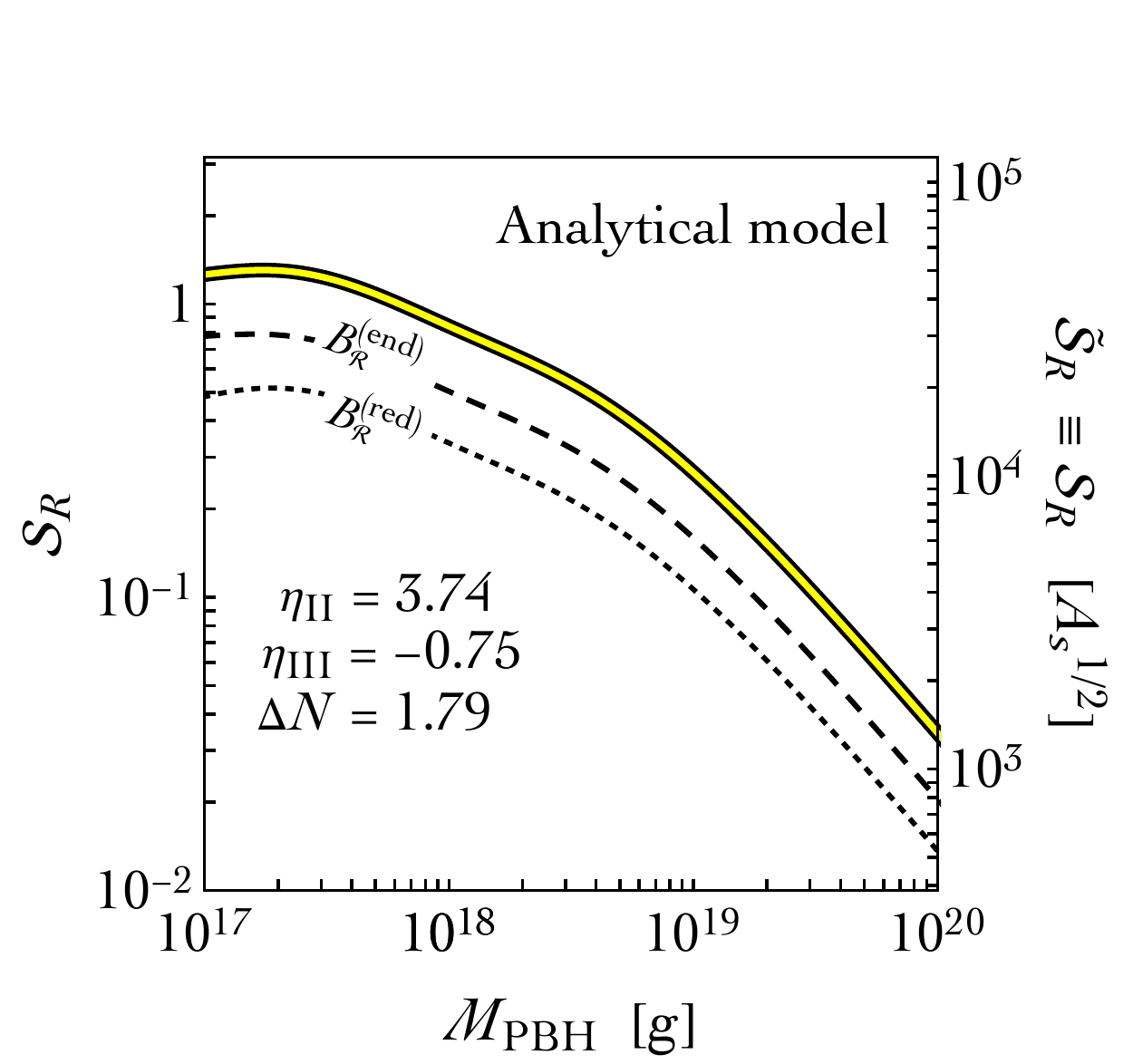}
\qquad\includegraphics[width=.49\textwidth]{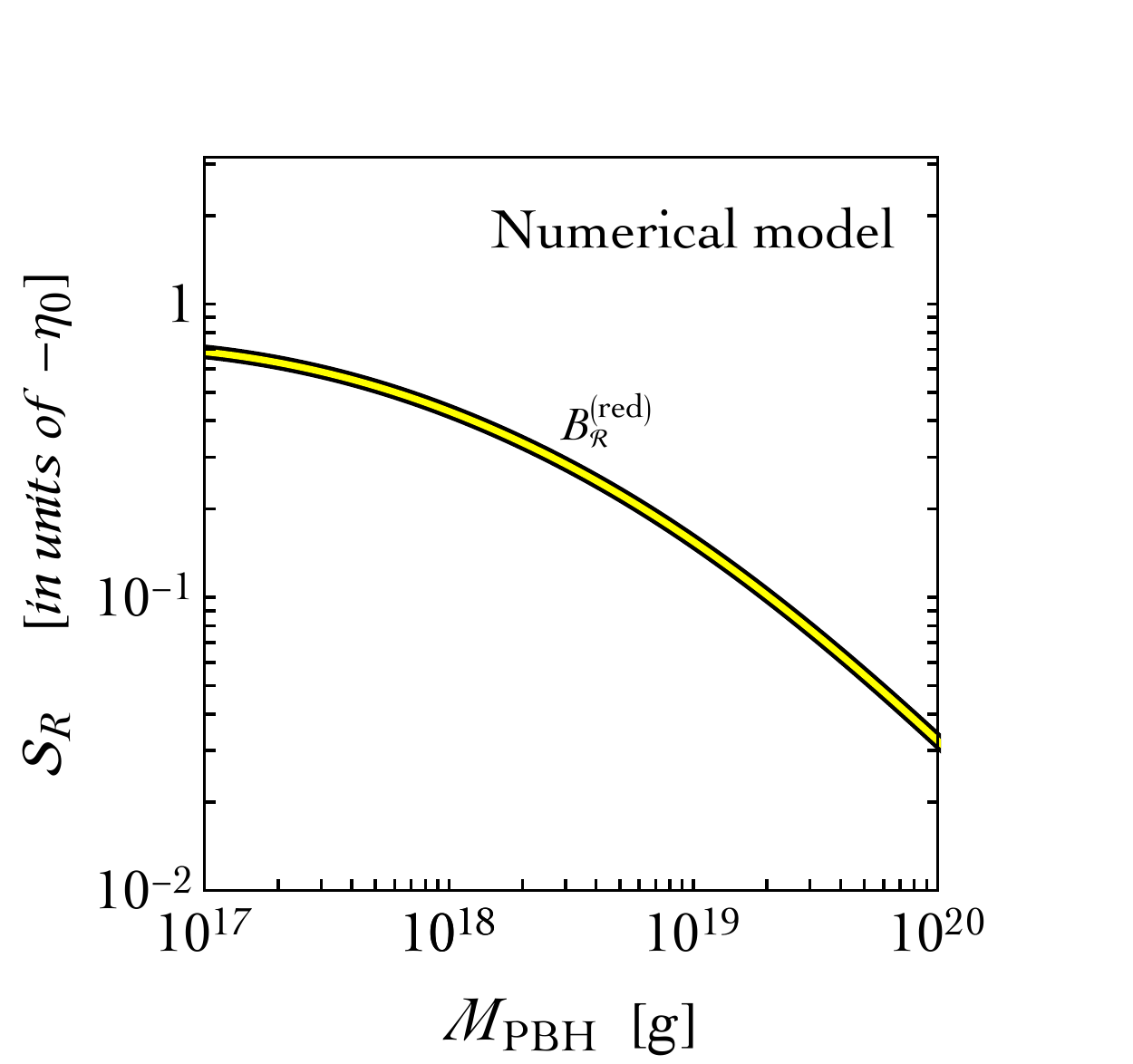}$$
\caption{\em \label{fig:Skewness}  
Smoothed skewness in eq.\,(\ref{eq:SmoothedSkew}) as function of the PBH mass computed for the model
in which $f_{\rm PBH} = 1$ in the gaussian approximation (see fig.\,\ref{fig:PhysSpectrum}). 
The yellow line represents the sum of the three contributions in eq.\,(\ref{eq:BispectrumStructure}).
Separately, we also show the contribution from $B_{\mathcal{R}}^{({\rm end})}$ and 
$B_{\mathcal{R}}^{({\rm red})}$ (that is the local contribution); $B_{\mathcal{R}}^{({\rm in})}$ gives a negligible contribution not shown in the plot. 
The right-side of the $y$-axis shows the value of $\mathcal{S}_R$ in units of $(A_s)^{1/2}$ with $A_s = H^2/8\pi^2\epsilon_I$. 
On the left-side $y$-axis, $A_s$ is fixed by the normalization chosen in fig.\,\ref{fig:PhysSpectrum}.
Right panel.
 Smoothed skewness in eq.\,(\ref{eq:SmoothedSkew}) as function of the PBH mass computed for the numerical model in 
 ref.\,\cite{Ballesteros:2020qam}. 
 In the numerical model, only the contribution $B_{\mathcal{R}}^{({\rm red})}$ survives (solid yellow line).
 }
\end{center}
\end{figure}

In fig.\,\ref{fig:Skewness} we show the bispectrum as function of the PBH mass.
The results obtained in the context of  the analytical model are shown in the left panel of fig.\,\ref{fig:Skewness}.
Notice that the ratio $\mathcal{S}_R = \langle \delta^3 \rangle_R/\sigma_R^3$ is proportional to the factor 
$A_s^{1/2} = H/\sqrt{8\pi^2\epsilon_I}$.  
On the left-side of the $y$-axis, we fix this value by means of the procedure outlined in the introduction of section\,\ref{sec:Abundance} (see table\,\ref{eq:ModelTab} and related discussions). 
On the right-side $y$-axis, on the contrary, we indicate the size of the ratio $\mathcal{S}_R/A_s^{1/2}$ such that this quantity is not affected by the details of the fit of the power spectrum at large scales (where our analytical model is not accurate since $\epsilon_{\rm I}$ is taken to be constant).

We find that $B^{({\rm in})}_{\mathcal{R}}$ is negligible while $B^{({\rm end})}_{\mathcal{R}}$ gives the dominant contribution compared to the local term $B_{\mathcal{R}}^{({\rm red})}$.  
In the analytical model, the two contributions $B^{({\rm in})}_{\mathcal{R}}$ and 
$B^{({\rm end})}_{\mathcal{R}}$ come, respectively, from the two delta function transition in the time evolution of $\eta$
\begin{align}
\frac{d\eta}{d\tau} = \eta_{\rm II}\delta(\tau-\tau_{\rm in}) + (-\eta_{\rm II} + \eta_{\rm III})\delta(\tau-\tau_{\rm end})\,.
\end{align}
Modeling the transition at $\tau = \tau_{\rm end}$ as a sharp step function can be misleading.  
This was already noticed in ref.\,\cite{Cai:2017bxr} (even though in the context of non-attractor inflation models) 
and further analyzed in ref.\,\cite{Passaglia:2018ixg}.
The outcome of the analysis carried out in ref.\,\cite{Cai:2017bxr} is that if one takes---in place of the sharp step function---a smooth transition at $\tau_{\rm end}$, 
the contribution $B^{({\rm end})}_{\mathcal{R}}$ gets drastically reduced. 

The evolution of $\eta$ was already shown in fig.\,\ref{fig:Pot} where it is indeed clear that  the step-function approximation at $\tau_{\rm end}$ is the crude approximation of a smoother transition.  
When the smooth transition is implemented, we indeed find that $B^{({\rm end})}_{\mathcal{R}}$ is completely negligible in the computation of the bispectrum. 
In short, this is because in the presence of a smooth transition $B^{({\rm end})}_{\mathcal{R}}$ 
receives---instead of the instantaneous contribution of the delta function---two integrated contributions (respectively, before and after the transition at $\tau_{\rm end}$) with opposite signs that almost cancel between each other (at the \text{\textperthousand} level).
\begin{figure}[!htb!]
\begin{center}
$$\includegraphics[width=.45\textwidth]{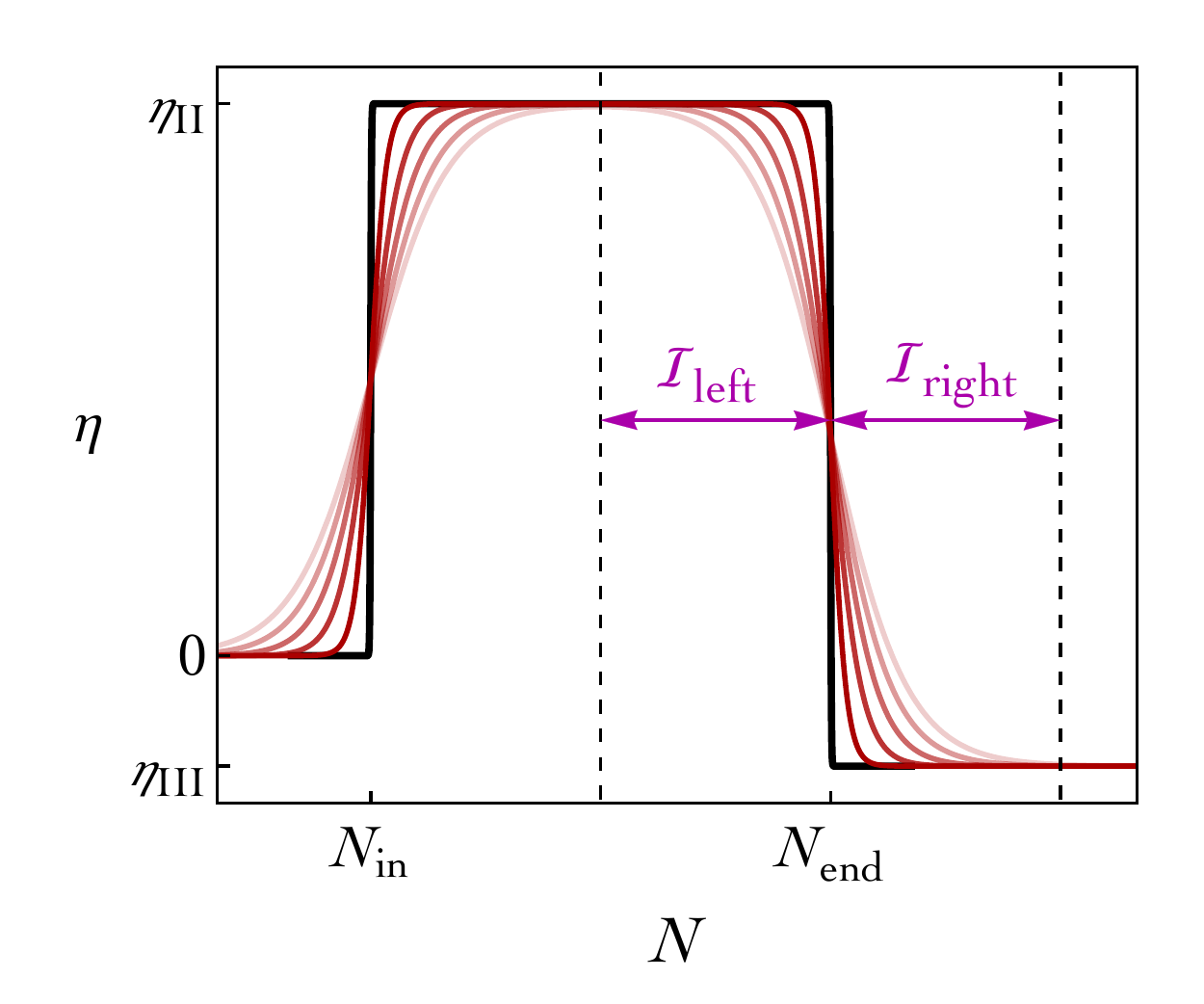}
\qquad\includegraphics[width=.45\textwidth]{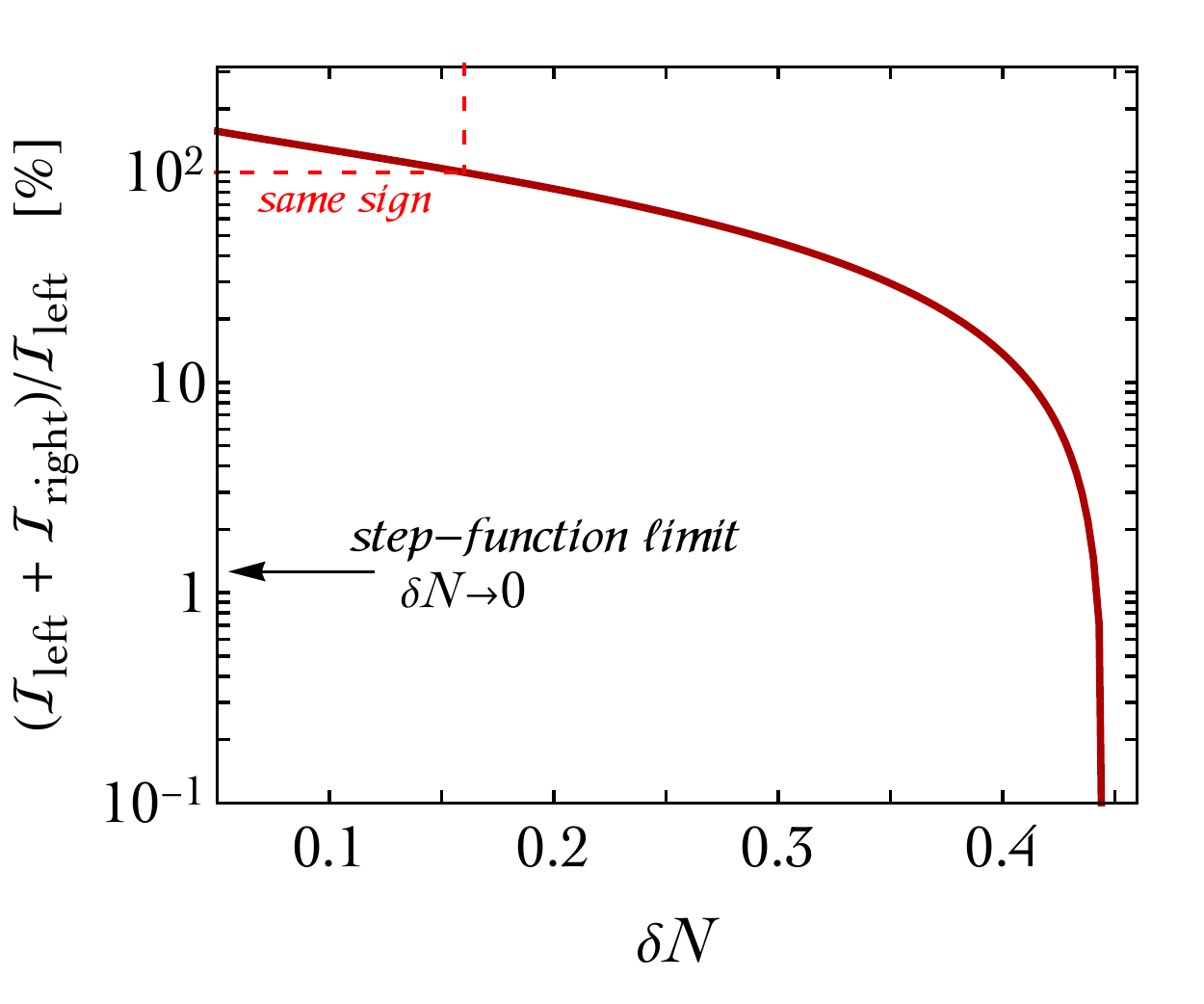}$$
\caption{\em \label{fig:SmoothEta}  
Left panel. Hyperbolic tangent parametrization of $\eta$ in eq.\,(\ref{eq:TanhEta}) 
for different $\delta N$ (from darker to lighter red, $\delta N = 0.1\div0.5$ in steps of $0.1$).
The black line is the step-function limit $\delta N\to 0$. 
The two integrals $\mathcal{I}_{\rm left}$ and $\mathcal{I}_{\rm right}$
refer to the intervals $[N_{\rm end}-1.5,N_{\rm end}]$ and $[N_{\rm end},N_{\rm end}+1.5]$, respectively.
Right panel. Sum of the two integrals $\mathcal{I}_{\rm left}+\mathcal{I}_{\rm right}$ (normalized to $\mathcal{I}_{\rm left}$, and
 expressed in percentage) as function of $\delta N$.
 When the ratio $(\mathcal{I}_{\rm left}+\mathcal{I}_{\rm right})/\mathcal{I}_{\rm left}$ is above 100\%, the two integrals 
 $\mathcal{I}_{\rm left}$ and $\mathcal{I}_{\rm right}$ have the same sign and they sum constructively. 
 Below this value, a cancellation occurs with increasing precision.
 }
\end{center}
\end{figure}
Let us give here just the idea of the proof (which is different compared to the one given in ref.\,\cite{Cai:2017bxr}).
Instead of the step-function approximation, we introduce the hyperbolic tangent parametrization
\begin{align}\label{eq:TanhEta}
\eta(N)  & = \frac{1}{2}\left[
-\eta_{\rm II}+ \eta_{\rm II}\tanh\left(\frac{N-N_{\rm in}}{\delta N}\right)
\right] + \frac{1}{2}\left[
\eta_{\rm II} + \eta_{\rm III} + (\eta_{\rm III}-\eta_{\rm II})\tanh\left(\frac{N-N_{\rm end}}{\delta N}\right)
\right]\,,
\end{align}
where the parameter $\delta N$ controls the width of the two transitions at $N_{\rm in}$ and $N_{\rm end}$. 
This parametrization can be considered as a proxy for the actual numerical result (see fig.\,\ref{fig:Pot}).
On the other hand, the limit $\delta N\to 0$ reproduces the step-function approximation. 
Notice that we also introduced an hyperbolic tangent to model the transition at $N = N_{\rm in}$, although not strictly needed 
for the validity of our considerations at $N = N_{\rm end}$.
The hyperbolic tangent parametrization in eq.\,(\ref{eq:TanhEta}) is shown in the left panel of fig.\,\ref{fig:SmoothEta} for different values of
$\delta N$. 
Consider the transition at $N = N_{\rm end}$. In the cubic action $-\int d\tau\,d^3x\, a^2\,\epsilon\,\eta^{\prime} \mathcal{R}^2\mathcal{R}^{\prime}$ we now have to integrate over time instead of just picking up
a delta function term at $N = N_{\rm end}$. 
One ends up, as anticipated before, with the sum of two integrals (respectively, $\mathcal{I}_{\rm left}$ and 
$\mathcal{I}_{\rm right}$) covering the two regions indicated in the left panel of fig.\,\ref{fig:SmoothEta}. 
For small $\delta N$, when the transition is very sharp, 
the two terms $\mathcal{I}_{\rm left}$ and  $\mathcal{I}_{\rm right}$ have the same sign, 
and their sum reconstructs the large bispectrum found in eq.\,(\ref{eq:Bi2}).
For larger $\delta N$, $\mathcal{I}_{\rm right}$ flips sign and the cancellation occurs with increasing precision. 
In the right panel of fig.\,\ref{fig:SmoothEta} we show the quantity $(\mathcal{I}_{\rm left}+\mathcal{I}_{\rm right})/\mathcal{I}_{\rm left}$
 as function of the width of the transition $\delta N$. 
 For $\delta N \gtrsim 0.4$, the sum $\mathcal{I}_{\rm left}+\mathcal{I}_{\rm right}$ quickly 
 gives a cancellation with the \text{\textperthousand} accuracy.
 
 In conclusion, we find that the analytical toy model in which the evolution of $\eta$ in the presence of USR is modeled by means of 
 a step-function approximation is not best-suited for the computation of non-gaussianities since it misses the important cancellation that is at work when the transition between adjacent regions is implemented in a more physical way. 
 For this reason, we switch from now on to the numerical model introduced in section\,\ref{sec:Abundance}. 
 By direct computation, we confirm 
 that  $B^{({\rm end})}_{\mathcal{R}}$ is negligible in the numerical model, where we verified the same cancellation discussed above for the hyperbolic tangent parametrization. This is because the evolution of $\eta$ in the numerical model can be indeed approximated very well 
 by eq.\,(\ref{eq:TanhEta}) with width parameter $\delta N \simeq 0.44$ (for which the cancellation is optimal, see 
 fig.\,\ref{fig:SmoothEta}). 
As a final comment before proceeding our analysis, notice that the cancellation of $B^{({\rm end})}_{\mathcal{R}}$ is not an exact mathematical identity but, as discussed before, its precision depends on the specific value of $\delta N$. 
In the numerical model introduced in section\,\ref{sec:Abundance}, the latter happens to be large enough that one can safely consider $B^{({\rm end})}_{\mathcal{R}}\simeq 0$ when compared to $B_{\mathcal{R}}^{({\rm red})}$.  
However, if one takes a model in which, say, $\delta N \simeq 0.3$ then our result shows that the cancellation works only at 
the $\sim 50$\% level; this means that $B^{({\rm end})}_{\mathcal{R}}$ gets reduced by, say, a factor of 2 but remains very sizable. 
Since the momentum structure of $B^{({\rm end})}_{\mathcal{R}}$ is not of local type, in such a situation it would become interesting to include it. 
The relevant question becomes the following: Is a model with $\delta N \simeq 0.3$ realistic? To answer this question, we can reverse-engineering the inflaton equation of motion and get
\begin{align}
V(N) &= V(N_{\rm ref})\exp\left\{
-2\int_{N_{\rm ref}}^{N}dN^{\prime}\left[\frac{\epsilon(3-\eta)}{3-\epsilon}\right]
\right\}\,,\label{eq:Master1}\\
\phi(N) &= \phi(N_{\rm ref}) \pm \int_{N_{\rm ref}}^N dN^{\prime}\sqrt{2\epsilon}\,,\label{eq:Master2}
\end{align}
as far as inflaton potential and field profile are concerned. 
These two equations are exact. We can now use the parametrization of $\eta$ given in eq.\,(\ref{eq:TanhEta}) and derive the time-evolution of $\epsilon$. Using $\eta \simeq  - (1/2)d\log\epsilon/dN$, we find the following (cumbersome but analytical) expression
\begin{align}
\epsilon(N) = &\epsilon_{\rm I}e^{-\eta_{\rm III}(N-N_{\rm ref})}\left[
\cosh\left(\frac{N-N_{\rm end}}{\delta N}\right)\cosh\left(\frac{N-N_{\rm in}}{\delta N}\right)
\right]^{-\frac{\delta N\eta_{\rm III}}{2}}
\left[
\cosh\left(\frac{N_{\rm ref}-N_{\rm end}}{\delta N}\right)\cosh\left(\frac{N_{\rm ref}-N_{\rm in}}{\delta N}\right)
\right]^{\frac{\delta N\eta_{\rm III}}{2}} \nn\\
&
\left[
\cosh\left(\frac{N-N_{\rm end}}{\delta N}\right){\rm sech}\left(\frac{N-N_{\rm in}}{\delta N}\right)
\right]^{\delta N\left(\eta_{\rm II} - \frac{\eta_{\rm III}}{2}\right)}
\left[
\cosh\left(\frac{N_{\rm ref}-N_{\rm end}}{\delta N}\right){\rm sech}\left(\frac{N_{\rm ref}-N_{\rm in}}{\delta N}\right)
\right]^{\frac{\delta N}{2}(-2\eta_{\rm II} + \eta_{\rm III})}\,,\label{eq:EspEvo}
\end{align}
where $\epsilon_{\rm I}\ll 1$ is the value of $\epsilon$ at some reference $N_{\rm ref}$ in region\,I.
We now integrate eqs.\,(\ref{eq:Master1},\,\ref{eq:Master2}) and get the inflaton potential as function of $\phi$ for different $\delta N$.
\begin{figure}[!htb!]
\begin{center}
%$$
\includegraphics[width=.38\textwidth]{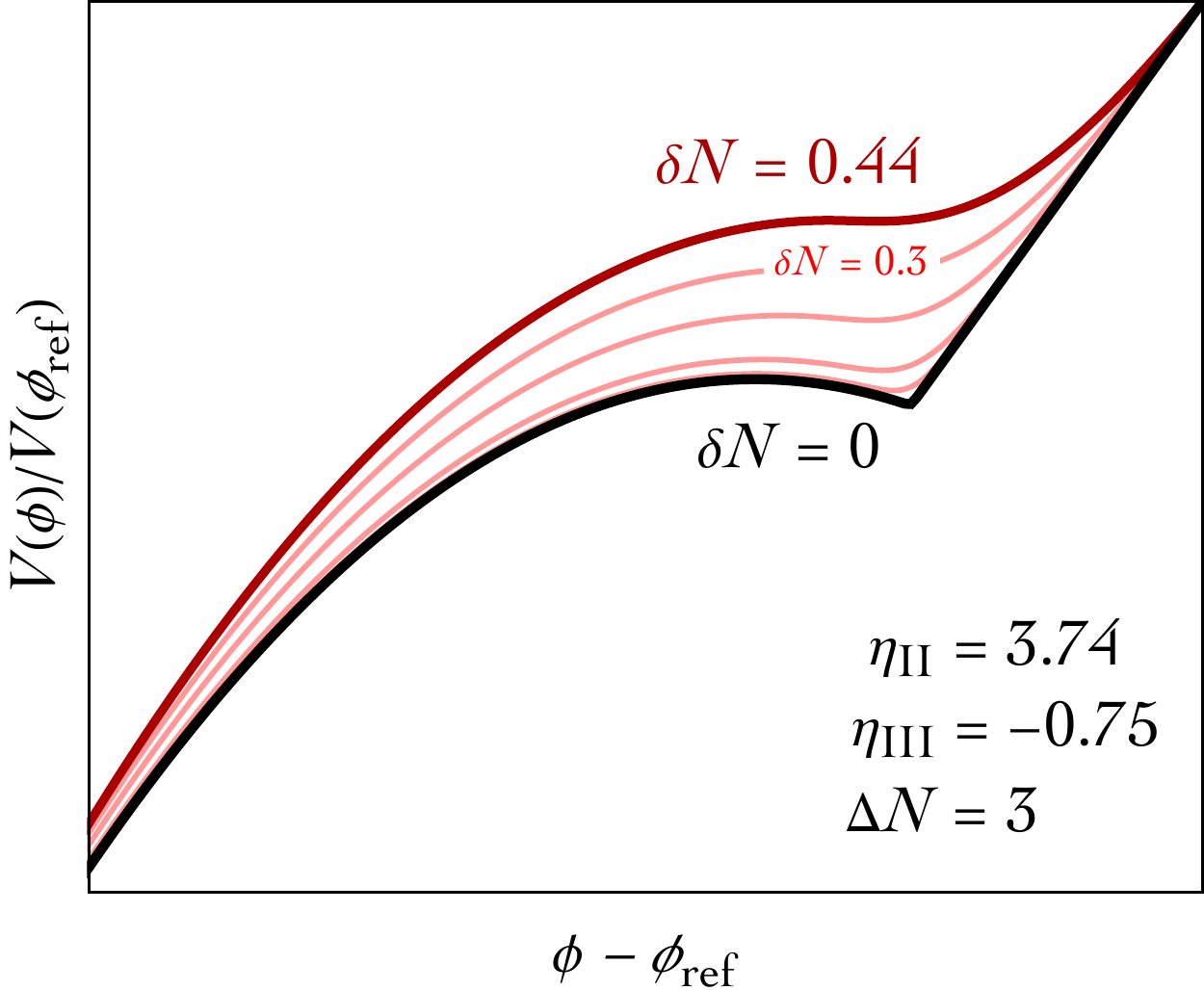}
\caption{\em \label{fig:PotentialProfile}  
Solution of eqs.\,(\ref{eq:Master1},\,\ref{eq:Master2}) with $\eta$-evolution given in eq.\,(\ref{eq:TanhEta}) and $\epsilon$-evolution
given in eq.\,(\ref{eq:EspEvo}) for different $\delta N$ with $\phi_{\rm ref}\equiv \phi(N_{\rm ref})$.
 }
\end{center}
\end{figure}
We show our result in fig.\,\ref{fig:PotentialProfile}.
We find that the width of the transition region of $\eta$ is related to the deformation of the 
stationary inflection point of the potential. Small values $\delta N \to 0$ deepen the local minimum and lead to a situation 
in which the inflaton risks being trapped into it (thus preventing a graceful exit). 
This is clear as far as the limit $\delta N \to 0$ is concerned. 
One is tempted to conclude that the same result holds true when comparing the cases with $\delta N = 0.44$ and $\delta N = 0.3$. 
However, any  intuition must be taken with a grain of salt since we know that small changes in the way one 
deforms the inflection point from its stationary configuration have a huge impact on the amplitude of the power spectrum (see discussion in ref.\,\cite{Ballesteros:2020qam}). For this reason, we postpone a quantitative estimate of the relation between the power spectrum and 
$\delta N$ to the end of this section, and we move on with our discussion of  non-gaussianities in the context of the numerical model in its benchmark configuration where $B_{\mathcal{R}}^{({\rm end})}$ can be neglected.

The smoothed skewness in the context of the numerical model is shown in the right panel of fig.\,\ref{fig:Skewness}.
The contribution $B_{\mathcal{R}}^{({\rm red})}$ survives also in the numerical model, and it is shown in yellow in 
the right panel of fig.\,\ref{fig:Skewness}. 
In the numerical model, the contribution $B_{\mathcal{R}}^{({\rm red})}$ is evaluated at some later time after the end of the USR phase 
when the relevant modes cease evolving (that is region\,III in the analytical model). 
We indicate with $\eta_0$ (which is {\it de facto} equivalent to $\eta_{\rm III}$) the (negative) value of $\eta$ at this transition time.
Since $\eta_0$ enters as an overall multiplicative factor in the computation of the local contribution of the bispectrum (the dominant one, as explained above),
in the right panel of fig.\,\ref{fig:Skewness} we plot the smoothed skewness in units of $-\eta_0$, that is $\mathcal{S}_R/(-\eta_0)$.
As far as this contribution is concerned, the numerical model gives a result that is in good agreement with the analytical model (despite the poor description of the latter at large scales).

In conclusion, we find that the presence of USR leaves a sizable imprint on the bispectrum of comoving curvature perturbations (and, consequently, on the density contrast) at the typical scales which are relevant for PBH formation.
Non-gaussianities are predominantly of local type, meaning that the bispectrum of comoving curvature perturbations in Fourier space has the form (see eq.\,(\ref{eq:Bi3}))\footnote{Locality here is just the statement that if the curvature is evaluated at some point $\vec{x}$ then the non-gaussianity is localized at same position. To be clear, a non-local type of non-gaussianity takes the form 
$\mathcal{R}(\vec{x}) = \mathcal{R}_{\mathcal{G}}(\vec{x}) + \int d^3\vec{y}d^{3}\vec{z}K(\vec{x};\vec{y},\vec{z})
[\mathcal{R}_{\mathcal{G}}(\vec{y})\mathcal{R}_{\mathcal{G}}(\vec{z}) - \langle\mathcal{R}_{\mathcal{G}}(\vec{y})\mathcal{R}_{\mathcal{G}}(\vec{z}) \rangle]$ with the local case corresponding to the kernel 
$K(\vec{x};\vec{y},\vec{z}) = 3f_{\rm NL}/5\delta^{(3)}(\vec{y}-\vec{x})\delta^{(3)}(\vec{z}-\vec{x})$.}
\begin{align}\label{eq:LocalBispectrumFourier}
B^{\rm \,loc}_{\mathcal{R}}(k_1,k_2,k_3) = (6f_{\rm NL}/5)[\Delta_{\mathcal{R}}(k_1)\Delta_{\mathcal{R}}(k_2) + 
\Delta_{\mathcal{R}}(k_1)\Delta_{\mathcal{R}}(k_3) +
\Delta_{\mathcal{R}}(k_2)\Delta_{\mathcal{R}}(k_3)]\,.
\end{align}
We introduced the amplitude parameter $f_{\rm NL}$ with its historical normalization that became conventional,  
which for the class of models that we are studying reads $f_{\rm NL}=-5/6\eta_0$.

 The presence of USR dynamics is crucial to generate potentially sizable non-gaussianities. 
 Conceptually, the reason is the following.
 The modes that contribute to the formation of the peak in the power spectrum of comoving curvature perturbations are those that are (in modulus)  enhanced (or suppressed, as far as the modes that contribute to the dip of the power spectrum are concerned) by the negative friction phase that takes place during USR. 
 These modes cease evolving only after the end of USR, a handful of $e$-folds before the end of inflation. 
 During this last part of the inflationary dynamics, the Hubble parameter $\eta$ takes $O(1)$ negative values which are needed  to connect the end of the USR phase with the end of inflation. In fact the inflaton almost stops on the top of the local maximum that forms when an approximate stationary inflection point is present in the potential, and accelerates afterwards to reach the subsequent absolute minimum where inflation ends.
 Since we are forced by the presence of the USR dynamics to evaluate correlators 
 for comoving scales $k \gtrsim 10^{12}$ Mpc$^{-1}$ in a region where $|\eta| \sim O(1)$, the usual slow-roll suppression 
 which affects non-gaussianities at CMB scales is not at work anymore, thus opening the possibility of having potentially large effects.

Let us now close this section with a final comment related to the cancellation of $B_{\mathcal{R}}^{({\rm end})}$. 
Consider the scalar potential in eq.\,(\ref{eq:TempHDO1}) that we write in the Einstein frame in the form\footnote{Notice that here $\phi$ is not canonically normalized but this fact does not alter the validity of the qualitative argument below. 
See ref.\,\cite{Ballesteros:2020qam} for details. We also neglect here for simplicity the presence of higher-dimensional operators.}
\begin{align}
V(\phi) = \frac{\lambda \phi^4}{4!(1+\xi \phi^2)^2}\left[
3 + \xi^2 \phi_0^4 -8(1+c_3)\frac{\phi_0}{\phi} +
2(1+c_2)(3 + \xi\phi_0^2)\frac{\phi_0^2}{\phi^2} 
\right]~~= \resizebox{50mm}{!}{
\parbox{25mm}{
\begin{tikzpicture}[]
\node (label) at (0,0)[draw=white]{ 
       {\fd{2.8cm}{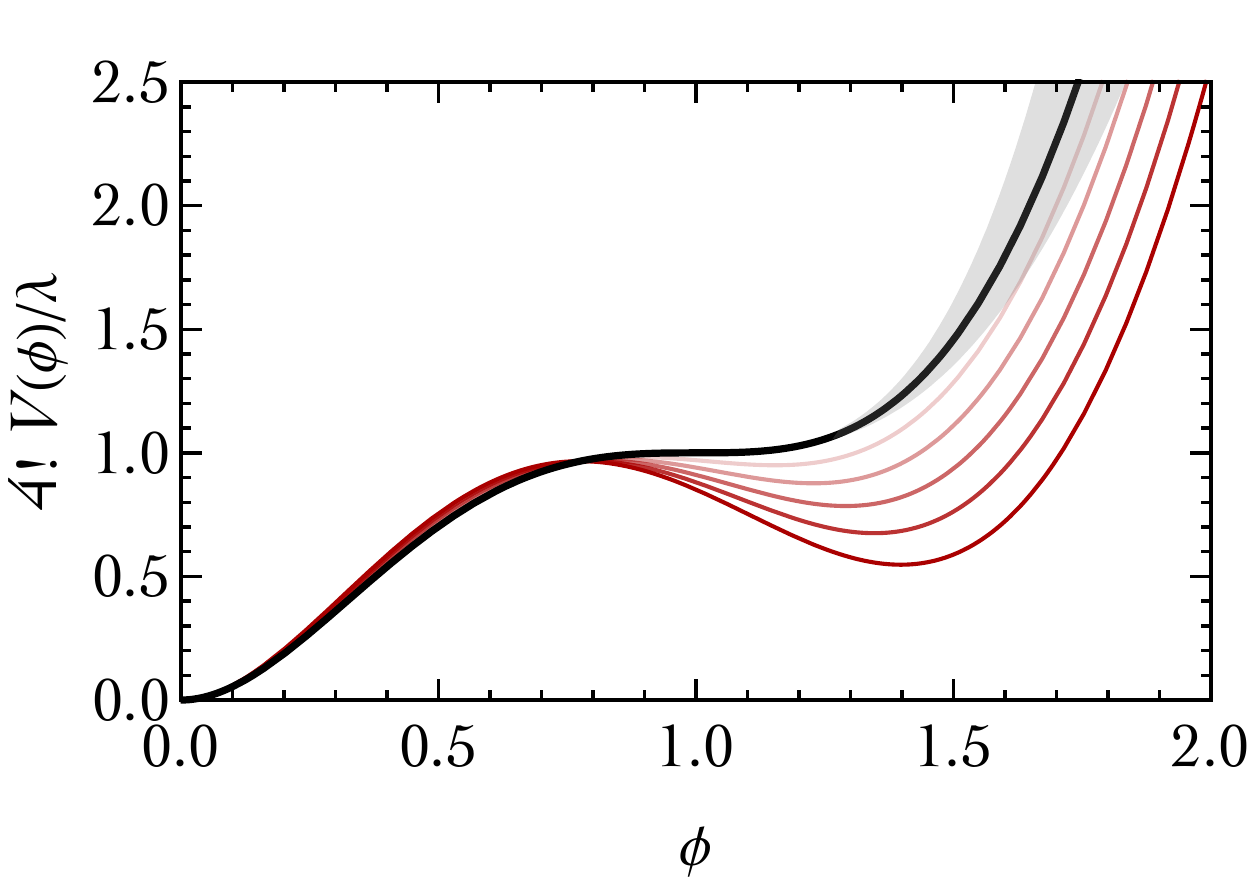}} 
      };
\end{tikzpicture}
}}\nonumber
\end{align}
where $\phi_0 = 1$ indicates the position of the inflection point and $c_{2,3}$ describe deformations from its stationary 
configurations (that corresponds to $c_{2,3} = 0$). 
Changing the value of the non-minimal coupling (gray region in the plot above) keeping $c_{2,3} \neq 0$ fixed does not alter the shape 
of the potential around the inflection point. This means that 
all these solutions will be described by a similar value of $\delta N.$
However, we find that a \text{\textperthousand}  variation of $\xi$ changes the peak of the power spectrum by many orders
of magnitude.  The smaller $\xi$ the lower $\mathcal{P}_{\mathcal{R}}(k_{\rm peak})$. 
In practice, in the context of the numerical model, the peak of the power spectrum is controlled by the size of the non-minimal coupling $\xi.$
A small tuning of $\xi$ leaves $\delta N $ almost unchanged but allows to select the desired value of $\mathcal{P}_{\mathcal{R}}(k_{\rm peak}).$
We will see this explicitly in section\,\ref{sec:Local}.

On the other hand, it is clear that changing the values of $c_{2,3}$ directly alters, by construction, the shape of the 
approximate stationary inflection point (red curves in the plot above, see ref.\,\cite{Ballesteros:2020qam}). 
For instance, we found good inflationary solutions (that is solutions of the inflaton equation of motion in which the inflaton successfully 
overcomes the inflection point and reaches the absolute minimum at the origin) with $\delta N \simeq 0.38$. 
According to the right panel of fig.\,\ref{fig:SmoothEta}, this is enough to get a cancellation that works only at the 
10\% level. In such a case, neglecting   $B_{\mathcal{R}}^{({\rm end})}$ is not justifiable.
 Of course, more work is needed to validate these solutions against CMB data and bound from Hawking evaporation 
(in particular, all solutions found with $\delta N \simeq 0.38$ tend to generate PBHs that are too light to be dark matter).
The point that we want to make here is that the condition $B_{\mathcal{R}}^{({\rm end})} \simeq 0$ is 
not guaranteed but one should always confront the $\eta$-evolution of a particular solution with our result in  
the right panel of our fig.\,\ref{fig:SmoothEta} before drawing any conclusion about the size of non-gaussianities generated by the 
transition at $N = N_{\rm end}$. 
The recipe to follow is simple. Compute the time-evolution of $\eta$ by solving the inflaton equation of motion and compare it with eq.\,(\ref{eq:TanhEta}) to extract the value of $\delta N$ that best suits the shape of the transition at $N = N_{\rm end}$. From 
the right panel of fig.\,\ref{fig:SmoothEta}, one then reads the accuracy of the cancellation.

In the rest of this work we will focus on cases with  $B_{\mathcal{R}}^{({\rm end})} \simeq 0$.

\section{The trispectrum and the kurtosis}\label{sec:Kurtosis}

The four-point correlation function in momentum space has the structure
\begin{align}
\langle \mathcal{R}_{k_1}\mathcal{R}_{k_2}\mathcal{R}_{k_3}\mathcal{R}_{k_4} \rangle 
 = &\underbrace{\left[(2\pi)^6\delta^{(3)}(\vec{k}_1+\vec{k}_2)\delta^{(3)}(\vec{k}_3+\vec{k}_4)
 \Delta_{\mathcal{R}}(k_1)
\Delta_{\mathcal{R}}(k_3) + {\rm two\,permutations}\right]}_{\rm disconnected\,contribution}  + \nonumber \\
& \underbrace{(2\pi)^3\delta^{(3)}(\vec{k}_1+\vec{k}_2+\vec{k}_3+\vec{k}_4)
T_{\mathcal{R}}(k_1, k_2, k_3,k_4,k_{12},k_{14})}_{\rm connected\,contribution}\,~~~~~~
\raisebox{-11mm}{\begin{tikzpicture}
	\draw [->,thick,draw=blue]  (-0.15,0.85)--(.55,-0.15);
	\draw [->,thick,draw=oucrimsonred]  (.6,-0.25)--(1,0.85);
	\draw [->,thick,draw=verdes]  (1,0.9)--(1.5,1.3);
	\draw [->,thick,draw=magenta]  (1.35,1.3)--(-0.15,0.95);
	\draw [thick,dashed]  (-0.,0.9)--(1,0.9);
	\draw [thick,dashed]  (.75,-0.05)--(1.45,1.15);
	\node[anchor=west] at (.5,1.45) {\scalebox{0.85}{{\color{magenta}{$\vec{k}_4$}}}};
	\node[anchor=west] at (-.5,.45) {\scalebox{0.85}{{\color{persianblue}{$\vec{k}_1$}}}};
	\node[anchor=west] at (0.25,.5) {\scalebox{0.85}{{\color{oucrimsonred}{$\vec{k}_2$}}}};
	\node[anchor=west] at (1.35,0.95) {\scalebox{0.85}{{\color{verdes}{$\vec{k}_3$}}}};
	\end{tikzpicture}}
\end{align}
The trispectrum $T_{\mathcal{R}}$ depends on the six scalars specifying the quadrilateral
formed by the wave-vectors, as illustrated in the schematic picture above where the two dashed lines correspond to 
$k_{12}\equiv |\vec{k}_1 - \vec{k}_2|$ and $k_{14}\equiv |\vec{k}_1 - \vec{k}_4|$.\footnote{In general, $T_{\mathcal{R}}$ depends on four three-momenta, namely twelve parameters;  however, as discussed before in the case of the bispectrum, assuming isotropy and homogeneity the number of parameters reduces to six.}
Notice that in the computation of $\langle \mathcal{R}_{k_1}\mathcal{R}_{k_2}\mathcal{R}_{k_3}\mathcal{R}_{k_4} \rangle $ we need to isolate the connected part since the disconnected part is already present at the gaussian level, see eq.\,(\ref{eq:Gaussian4pt}).
In analogy with the reduced bispectrum introduced in eq.\,(\ref{eq:ReducedBi}), it is useful to define the reduced trispectrum 
$\mathcal{T}_{\mathcal{R}}$
which is given by
\begin{align}\label{eq:RedcuedTrispectrum}
T_{\mathcal{R}}(k_1, k_2, k_3,k_4,k_{12},k_{14}) = \mathcal{T}_{\mathcal{R}}
\left\{
\Delta_{\mathcal{R}}(k_1)\Delta_{\mathcal{R}}(k_2)\left[
\Delta_{\mathcal{R}}(|\vec{k}_1 + \vec{k}_4|) + \Delta_{\mathcal{R}}(|\vec{k}_1 + \vec{k}_3|)\right] + 11\,{\rm permutations}
\right\}\,.
\end{align}

As done in the previous section, 
we define the smoothed kurtosis as
\begin{align}\label{}
\mathcal{K}_R = \frac{1}{(\sigma_R^2)^{2}}\underbrace{\int\prod_{i=1}^4\frac{d^3\vec{k}_i}{(2\pi)^3}
W(k_iR)\frac{16(1+\omega)^4}{(5+3\omega)^4}
\left(\frac{k_i}{aH}\right)^2
\langle \mathcal{R}_{k_1}\mathcal{R}_{k_2}\mathcal{R}_{k_3}\mathcal{R}_{k_4} \rangle}_{\equiv 
\langle \delta^4\rangle_R}\,.\label{eq:SmoothedKurt}
\end{align}
The exact computation of $\mathcal{K}_R$ is beyond the scope of this paper. 
However, an important remark is in order. 
When non-gaussianities are of local type, they generate---in addition to the bispectrum of the form 
defined by eq.\,(\ref{eq:LocalBispectrumFourier}) and obtained in eq.\,(\ref{eq:Bi3}) as a consequence of our explicit computation---also a contribution to the reduced trispectrum in eq.\,(\ref{eq:RedcuedTrispectrum}).
For the sake of simplicity, we ignore in the following the smoothing procedure and the relation between 
density contrast and curvature perturbations, and we just consider
the curvature perturbation $\mathcal{R}$.
It is a simple exercise to show that if we take $\mathcal{R} = \mathcal{R}_{\mathcal{G}} + (3f_{\rm NL}/5)
\left(\mathcal{R}_{\mathcal{G}}^2 - \langle \mathcal{R}_{\mathcal{G}}^2\rangle\right)$
we find a trispectrum of the form given in eq.\,(\ref{eq:RedcuedTrispectrum}) with 
$\mathcal{T}_{\mathcal{R}} = 18f_{\rm NL}^2/25$\,\cite{Komatsu:2010hc,Boubekeur:2005fj}.  
This can be understood in the context of the {\it in-in} formalism as shown in appendix\,\ref{app:ThreePoints2} (see derivation 
of eq.\,(\ref{eq:RedTri})).

We do not attempt here to compute $\mathcal{K}_R$ carefully but we just limit the discussion to the simple (and well-known) observation that 
a non-zero local-type bispectrum of order $O(f_{\rm NL})$ also implies a non-zero contribution 
to the reduced trispectrum of order $O(f_{\rm NL}^2)$. 
Statistically speaking, this means that we expect a non-zero fourth-order cumulant related to the third-order one. 
This comment actually applies also to higher-order cumulants as well.

Of course, we stress again that in addition to this $O(f_{\rm NL}^2)$ contribution other terms are expected which are generated, for 
instance,
by a pure quartic interaction.
In the following, we shall work under the assumption that all these additional terms are negligible, and that 
the dominant contribution to the four-point correlator is the one generated by local cubic interactions.

\section{On the PBH abundance in the presence of non-gaussianities: the local case.}\label{sec:Local}

We are now in the position, using eq.\,(\ref{eq:FundBetaNG}), to analyze the impact of non-gaussianities for the PBH abundance. 
To make our discussion more clear, let us start by considering the PDF for 
the fractional overdensity $\delta$ whose integral above the threshold gives the mass fraction $\beta$ (see eq.\,(\ref{eq:FundBeta})). We indicate with $P_{\mathcal{NG}}(\delta)$ the generic non-gaussian PDF, and with 
$P_{\mathcal{G}}(\delta)$ the PDF in the gaussian approximation.
As anticipated, eq.\,(\ref{eq:FundBetaNG}) follows from the Gram-Charlier series of the PDF 
\begin{align}
P_{\mathcal{NG}}(\delta) & = 
P_{\mathcal{G}}(\delta)\sum_{n=0}^{\infty}
\frac{1}{n!2^{n/2}}B_n[0,0,\mathcal{C}_R^{(3)},\dots,\mathcal{C}_R^{(n)}]H_n\left(
\frac{\nu}{\sqrt{2}}
\right) \label{eq:GC} \\
& = P_{\mathcal{G}}(\delta)\bigg\{
1 + \frac{\mathcal{C}_R^{(3)}}{3!2^{3/2}}H_3\left(\frac{\nu}{\sqrt{2}}\right)
+ 
\frac{\mathcal{C}_R^{(4)}}{4!2^{4/2}}H_4\left(\frac{\nu}{\sqrt{2}}\right) +
\frac{\mathcal{C}_R^{(5)}}{5!2^{5/2}}H_5\left(\frac{\nu}{\sqrt{2}}\right)  +
\frac{[10\mathcal{C}_R^{(3)\,2}+\mathcal{C}_R^{(6)}]}{6!2^{6/2}}H_6\left(\frac{\nu}{\sqrt{2}}\right) +\dots
\bigg\}\,,\nn
\end{align}
with $\nu\equiv \delta/\sigma$ and where we introduced the $n$-th complete exponential Bell polynomial 
\begin{align}\label{eq:BellPol}
B_n(x_1,\dots,x_n) = \sum_{k=1}^{n}B_{n,k}(x_1,\dots,x_{n-k+1}) 
\,,
\end{align}
with  $B_{n,k}$ the partial exponential Bell polynomials.
$\mathcal{C}_R^{(n)}$ is the $n$-th cumulant defined in eq.\,(\ref{eq:CumulantsR}).
In eq.\,(\ref{eq:GC}), we carried out explicitly the expansion up to the sixth order to show that, from this order on,
the coefficient of the $n$-th term is not just given by $\mathcal{C}_R^{(n)}$.
This is a well-known result in the theory of statistics\,\cite{Kendall,kolassa}.
Without entering to deep into the mathematical detail, eq.\,(\ref{eq:GC}) can be thought as 
a Taylor expansion of $P_{\mathcal{NG}}(\delta)$ around a reference PDF, taken to be the gaussian distribution $P_{\mathcal{G}}(\delta)$.\footnote{We remind that 
\begin{align}
\frac{d^n}{dx^n}e^{-x^2/2\sigma^2} = \frac{(-1)^n}{2^{n/2}\sigma^n}e^{-x^2/2\sigma^2}H_n\left(\frac{x}{\sqrt{2}\,\sigma}\right)\,,
\end{align}
so that the reader can immediately recognize in eq.\,(\ref{eq:GC}) a derivative expansion.
}
For completeness, if we integrate eq.\,(\ref{eq:GC}) to compute $\beta_{\mathcal{NG}} = \gamma \int_{\delta_{\rm th}}^{\infty}d\delta P_{\mathcal{NG}}(\delta)$ and use the property 
\begin{align}
\frac{1}{\sqrt{2\pi\sigma^2}}\int_{\delta_{\rm th}}^{\infty}d\delta
e^{-\delta^2/2\sigma^2}H_n\left(
\frac{\delta}{\sqrt{2}\sigma}
\right) = \frac{1}{\sqrt{\pi}}e^{-\delta_{\rm th}^2/2\sigma^2}H_{n-1}\left(
\frac{\delta_{\rm th}}{\sqrt{2}\sigma}
\right)\,,
\end{align}
we find precisely eq.\,(\ref{eq:FundBetaNG}).
From this simple discussion, we therefore learn that eq.\,(\ref{eq:FundBetaNG}) follows from a series expansion 
around the gaussian PDF.

\begin{figure}[!htb!]
\begin{center}
$$
\includegraphics[width=.48\textwidth]{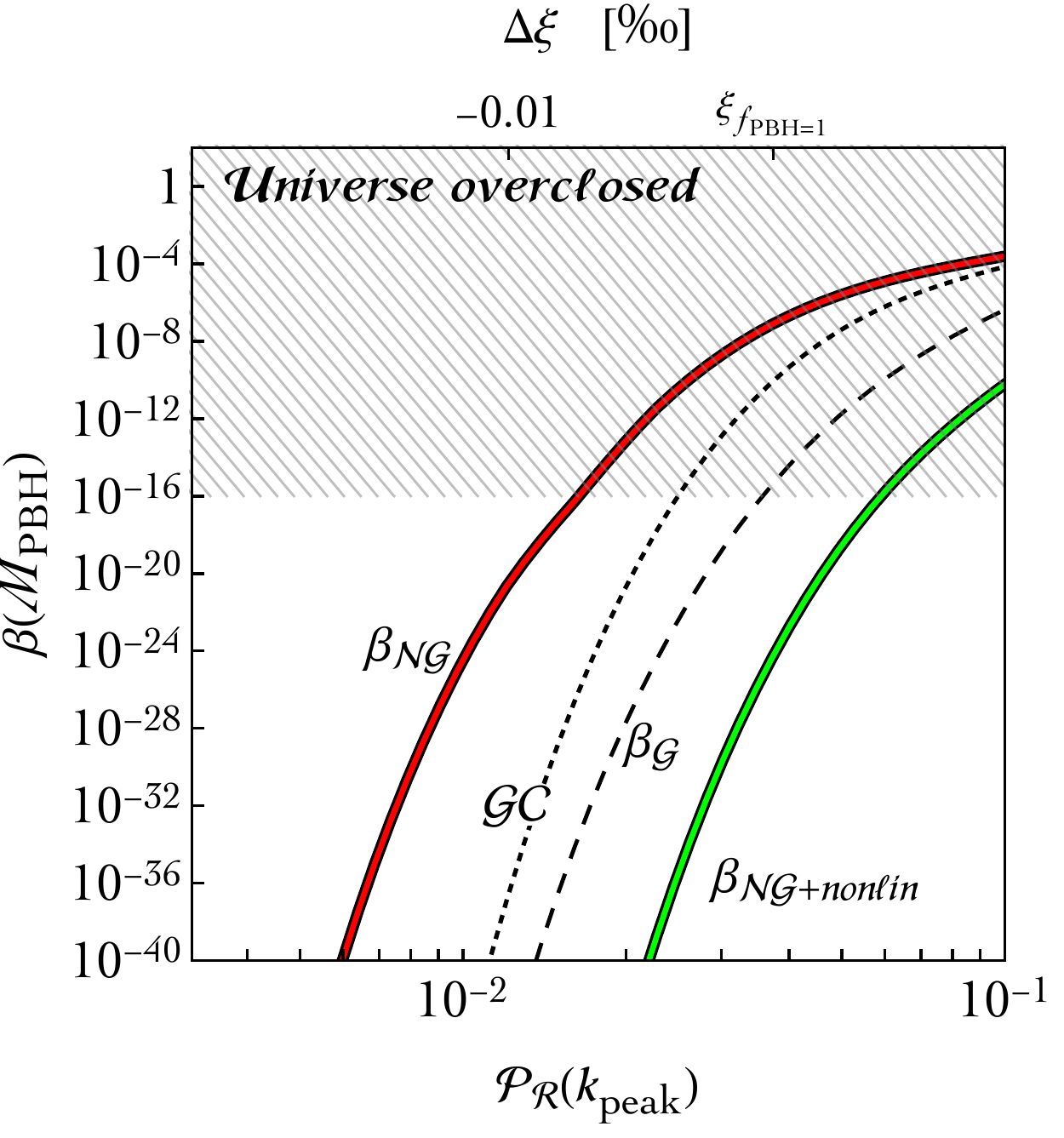}
\qquad\includegraphics[width=.48\textwidth]{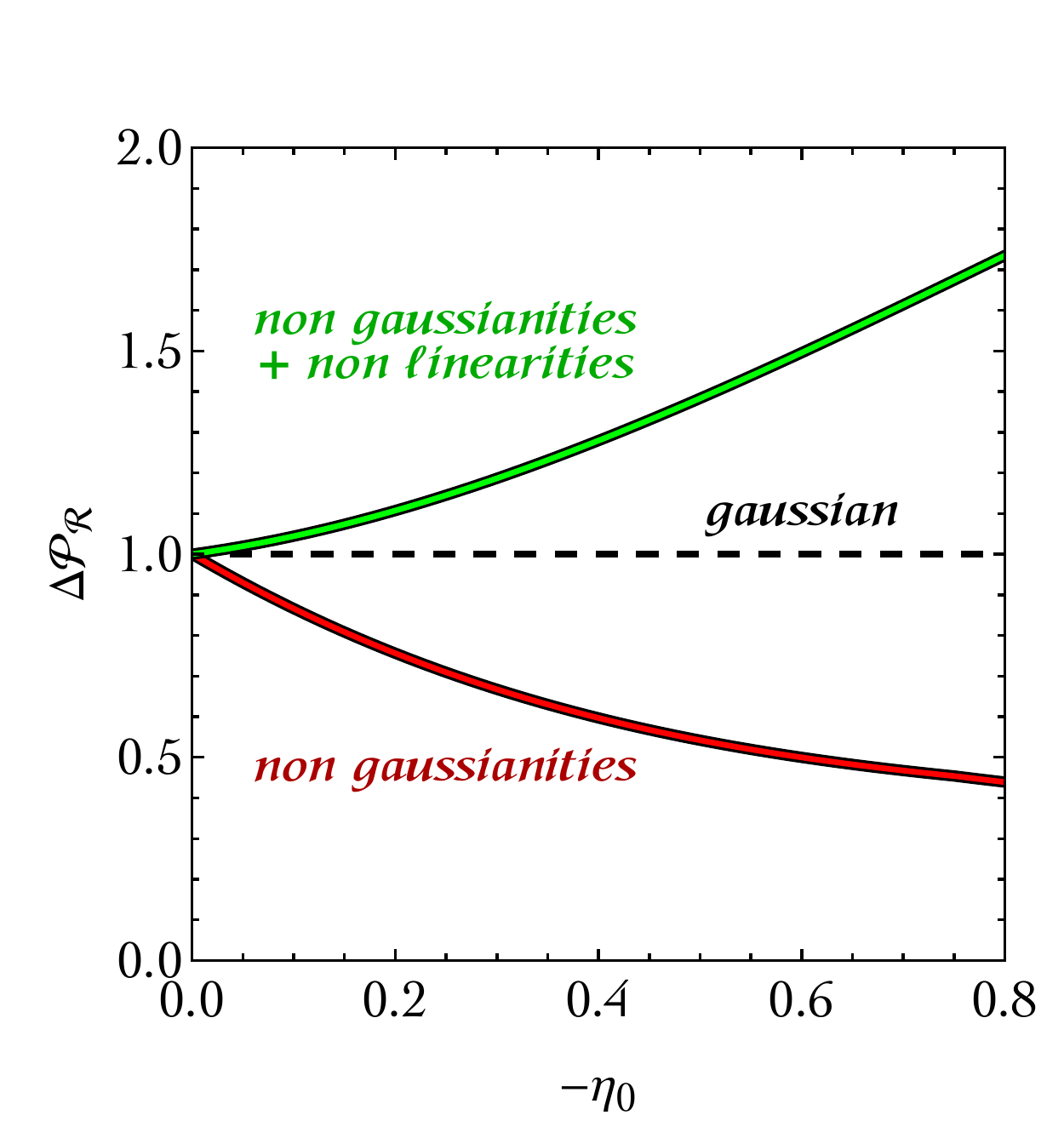}$$
\caption{\em \label{fig:FinalPlot}
Left panel: PBH abundance as a function of the the power spectrum of curvature perturbations at peak position.
The dashed black line is for the gaussian approximation, eq.\,(\ref{eq:FundBeta}). 
The dotted black line refers to the abundance computed integrating the Gram-Charlier series in eq.\,(\ref{eq:GC}), see text for details. 
The red line takes into account the effect of primordial non-gaussianities, and the green line further includes the non-linearities between density and curvature perturbations. 
Right panel: Rescaling factor of the power spectrum needed, in presence of non-gaussianities, in order to get the same PBH abundance predicted by the gaussian approximation. This quantity is shown as a function of $-\eta_0,$ the parameter controlling the amount of local non-gaussianities.
The color code of the lines is the same as in the left panel.
The calculations in both panels are for the numerical model used in section\,\ref{sec:SkewnessExact}.
 }
\end{center}
\end{figure}

In principle, all the cumulants are needed to compute the PBH abundance with eq.\,(\ref{eq:FundBetaNG}). 
However, to examine the relevance of non-gaussianities, we first start focusing only on the third order cumulant $\mathcal{C}_R^{(3)},$ which we computed in section\,\ref{sec:SkewnessExact} (see fig.\,\ref{fig:Skewness}), and set the remaining ones to zero. A complete calculation is performed later in this section.
We show our results in the left panel of fig.\,\ref{fig:FinalPlot}.
The analysis is performed in the context of the numerical model,
and we focus on the PBH mass corresponding to the peak of the mass distribution (i.e. the most abundant PBHs). 
The 
dotted line (labelled $_{\mathcal{G}\mathcal{C}}$) shows the PBH abundance computed integrating eq.\,(\ref{eq:GC}) above the threshold as a function of the power spectrum of curvature perturbations at the peak position.
Let us stress again that in eq.\,(\ref{eq:GC}) $\mathcal{C}_R^{(3)}$ does not enter only at third order in the expansion, but also affects higher ones.
For definiteness, we include terms up to $n=15$ in eq.\,(\ref{eq:GC}).
This means that $\mathcal{C}_R^{(3)}$ enters at $n=3,6,9,12,15$. 
We do not attempt here to include more terms and test the convergence of eq.\,(\ref{eq:GC}) since, as we shall discuss 
in a moment, we will eventually adopt a different approach for the computation of the abundance (see\,\cite{Riccardi:2021rlf} for a critical discussion on the truncation of the Gram-Charlier serie).
On the top $x$-axis we also show the relative change of the non-minimal coupling (with respect to the value $\xi_{f_{\rm PBH}=1}$ that gives the totality of dark matter in PBHs for the gaussian model) that is needed to modify the peak of the power spectrum 
according to the corresponding point on the bottom $x$-axis.
The dashed black line corresponds instead to the gaussian approximation in eq.\,\ref{eq:FundBeta}.
Two main messages can be extracted from this plot.
Non-gaussianities are relevant, and modify the abundance computed with gaussian statistics by several orders of magnitude.
On the other hand, the abundance is very sensitive to the amplitude of the power spectrum, and it is enough to change the latter quantity by a small amount to obtain the same abundance predicted by the gaussian approximation.

As mentioned above, this calculation is incomplete, because it misses the contributions from higher orders in the expansion and from the cumulants $\mathcal{C}_R^{(n)}$ with $n>3$ (which are non-zero because, as discussed in section\,\ref{sec:Kurtosis} for the specific case of the 
fourth-order cumulant, they at least receive a non-zero ``induced'' contribution from the local three-point function).
Baring cancellations among the different contributions, one would expect that, at qualitative level, the results presented above would still hold, once all the other terms are included.
In other words, the purpose of the estimate performed above is to highlight the relevance of non-gaussianities for the PBHs abundance.
On the other hand, for a proper calculation 
one could in principle use eq.\,(\ref{eq:FundBetaNG}), and compute all the cumulants needed to reach the convergence of the serie at the desired accuracy.
However, exploiting the results of section\,\ref{sec:SkewnessExact}, we follow a simpler strategy, which will also allow us to take into account the non-gaussianities arising from the non-linear relation between curvature and density perturbations.
The idea is the following.\footnote{A similar strategy has been presented in\,\cite{DeLuca:2019qsy} for the study of the non-gaussianities produced by the non-linear relation among density and curvature perturbations, see below.}
In section\,\ref{sec:SkewnessExact} we have shown that the three-point correlator in the presence of USR generates non-gaussianities that are predominantly of local type, since they give a bispectrum of the form given in eq.\,(\ref{eq:LocalBispectrumFourier}). This means that we can write in position space
\begin{align}\label{eq:LocalR}
\mathcal{R} = \mathcal{R}_{\mathcal{G}} 
+ \frac{3f_{\rm NL}}{5}
\left(\mathcal{R}_{\mathcal{G}}^2 - \langle \mathcal{R}_{\mathcal{G}}^2\rangle\right)\,,
\end{align}
where $\mathcal{R}_{\mathcal{G}}$ is gaussian and the term $\langle \mathcal{R}_{\mathcal{G}}^2\rangle$ is added to ensure that $\langle \mathcal{R} \rangle = 0$.
The statistical properties of the non-gaussian variable $\mathcal{R}$ are therefore completely specified in terms of those of the gaussian variable $\mathcal{R}_{\mathcal{G}}$ and the parameter $f_{\rm NL},$ which controls the amount of primordial non-gaussianities.
As explained in sec.\,\ref{sec:Abundance}, in threshold statistics the abundance of PBHs is obtained computing the probability that the density perturbation exceeds the threshold value $\delta_{\rm th}.$ 
Here, the strategy is to relate the density perturbation to the comoving curvature perturbation $\mathcal{R}$ and its spatial derivatives $\partial_i \mathcal{R}$ and $\triangle \mathcal{R},$ where $\partial_i $ denotes the derivative with respect to the spatial direction $i.$
Using eq.\,(\ref{eq:LocalR}), these non-gaussian variables can be expressed in terms of the analogous ones for the gaussian variable $\mathcal{R}_{\mathcal{G}},$ whose statistical properties are known.
In practice, exploiting the relation in eq.(\ref{eq:LocalR}), we bypass the need of computing all the cumulants in eq.\,(\ref{eq:FundBetaNG}).

To see explicitly how this procedure can be implemented, let us first define the following quantity:
\begin{align}\label{eq:sigmaj}
\sigma_j^2 = \int \frac{dk}{k}\,W(kR)^2\,\mathcal{P}_{\mathcal{R}}(k)\,
k^{2j}
\,,~~~~~\gamma=\frac{\sigma_1^2}{\sigma_0\,\sigma_2}.
\end{align}
For ease of notation, we also define the following gaussian variables:
\begin{align}\label{eq:derivatives}
\nu=\frac{\mathcal{R}_{\mathcal{G}}}{\sigma_0}\,,~~~~~\eta_i =
\partial_i \mathcal{R}_{\mathcal{G}}\,,~~~~~ x= -\frac{\triangle \mathcal{R}_{\mathcal{G}}}{\sigma_2},
\end{align}
The joint PDF of the gaussian variables $\nu$, $\vec{\eta}$ and $x$ is \,\cite{Bardeen:1985tr}:
\begin{align}\label{eq:PDGgauss}
P(\nu,\vec{\eta},x)\,d^3\nu\,d\vec{\eta}\,dx= A\, e^{-Q}\,d\nu\,d^3\vec{\eta}\,dx\,,~~~~~{\rm with}~Q=\frac{\nu^2}{2} + \frac{(x-x_*)^2}{2\,(1-\gamma^2)} +\frac{3\,\vec{\eta}\cdot\vec{\eta}}{2\sigma_1^2},
\end{align}
and
\begin{align}\label{eq:PDGgauss2}
A=\frac{6\sqrt{3}}{8\pi^{5/2}\sqrt{2(1-\gamma^2)}\sigma_1^3}\,,~~~~~x_*=\gamma\,\nu.
\end{align}

To proceed, we should relate the comoving curvature perturbation with the density perturbation. As anticipated, this map is non-linear, and therefore it hides some unavoidable non-gaussianities. 
The full relation can be written in the form\,\cite{Harada:2015yda}
\begin{align}\label{eq:NonLinearDelta}
\delta(\vec{x},t) = -\frac{4(1+\omega)}{5+3\omega}\left(
\frac{1}{aH}
\right)^2e^{-5\mathcal{R}(\vec{x})/2}\triangle e^{\mathcal{R}(\vec{x})/2} = 
-\frac{2(1+\omega)}{5+3\omega}\left(
\frac{1}{aH}
\right)^2 
\underbrace{e^{-2\mathcal{R}(\vec{x})}}_{i)}
\bigg[
\triangle\mathcal{R}(\vec{x}) + \underbrace{\frac{1}{2}\partial_i\mathcal{R}(\vec{x})
\partial_i\mathcal{R}(\vec{x})}_{ii)}
\bigg]\,,
\end{align}
We have two sources of non-linearities (and thus non-gaussianities) that come from {\it i)} terms of order higher than two in the expansion 
of the exponential and {\it ii)} the quadratic term in the square brackets.
Notice that eq.\,(\ref{eq:NonLinearDelta}) is written in comoving gauge (defined by requiring both comoving slicing and comoving threading)
whose choice seems appropriate since often 
used to perform simulations that describe the formation of PBHs in the context of numerical relativity.
Furthermore, eq.\,(\ref{eq:NonLinearDelta}) is not exact but includes terms up to the second order in the so-called spatial gradient expansion\,\cite{Harada:2015yda}. 
Notice that in eq.\,(\ref{eq:NonLinearDelta}) $\mathcal{R}$ denotes the non-gaussian comoving curvature perturbation. Using eq.\,(\ref{eq:LocalR}) one can recast the expression above in terms of the gaussian variables $\nu$, $\vec{\eta},$ $x$ in eq.\,(\ref{eq:derivatives}).
Then, solving for the variable $x$ one finds:
\begin{align}\label{eq:xdelta}
x_{\delta}\left(\delta,\nu,\vec{\eta}\right) = \frac{9 (a H)^2\,e^{2\sigma_0\left[\nu+\left(\nu^2-1\right)\sigma_0\alpha\right]}\delta+\left(2+8\alpha+8\nu\alpha\sigma_0+8\nu^2\alpha^2\sigma_0^2\right)\, \vec{\eta}\cdot\vec{\eta}}{4\left(1+2\nu\,\alpha\,\sigma_0\right)\sigma_2}.
\end{align}
Above we have defined $\alpha= 3f_{\rm NL}/5$ and taken $w=1/3$, corresponding to the radiation domination era. 

Finally, the fraction of the Universe collapsing into PBHs is obtained integrating the joint PDF in eq.\,(\ref{eq:PDGgauss}) rewritten in terms of $\nu$, $\vec{\eta}$ and $\delta,$
and imposing that density perturbation is above the threshold $\delta_{\rm th}:$ 
 \begin{align}\label{eq:betaNG} 
\beta_{\mathcal{NG}}= \gamma \int_{\delta_{\rm th}}^{\infty} d\delta\,\int d^3\vec{\eta}\, \int_{-\infty}^{\infty} d\nu\, J\, P(\nu,\vec{\eta},x_{\delta}),
\end{align}
where $J$ is the Jacobian of the transformation in eq.\,(\ref{eq:xdelta}):
\begin{align}\label{eq:Jacobian}
J= \left| 9 (aH)^2 \frac{e^{2\sigma_0\left[\nu+\left(\nu^2-1\right)\sigma_0\alpha\right]}}{4\left(1+2\nu\,\alpha\,\sigma_0\right)\sigma_2}\right|.
\end{align}

Eq.\,(\ref{eq:betaNG}) generalizes the gaussian expression in eq.\,(\ref{eq:FundBeta}) including the effect of primordial non-gaussianities of local type, and of non-gaussianities induced by the non-linear relation in eq.\,(\ref{eq:NonLinearDelta}).
For the latter source of non-gaussianities, an analogous expression has been obtained in\,\cite{DeLuca:2019qsy}.
One can check that in the limit of $f_{\rm NL}\rightarrow 0$ and using eq.\,(\ref{eq:NonLinearDelta}) at linear order in $\mathcal{R}$, eq.\,(\ref{eq:betaNG}) reduces to eq.\,(\ref{eq:FundBeta}).

We are now in position to quantify the impact of the non-gaussianities on the PBH abundance. 
We show in the left panel of fig.\,\ref{fig:FinalPlot} our results for $\beta(M_{\rm PBH})$ as a function of the power spectrum at peak position for 
both the gaussian and non-gaussian cases. As before, we consider the numerical model, 
for which $\eta_0\simeq-0.75$ and therefore $f_{\rm NL}=-5/6\, \eta_0 \simeq 0.625.$
To isolate the effect of primordial local non-gaussianites, we first perform the calculation employing the linear relation among density and the comoving curvature perturbations, eq.\,(\ref{eq:DensityCurvature}). The result is shown with a red line in fig.\,\ref{fig:FinalPlot}.
Local non-gaussianities (with positive $f_{\rm NL}$) tend to enhance the abundance of PBHs, and their effect can be reabsorbed by 
decreasing the amplitude of the power spectrum by a factor $2\div3$. This can be achieved at the prize of a small re-tuning of the parameter that controls the height of the peak of the power spectrum. In the case discussed in this paper, such parameter is the non-minimal coupling to gravity, and the re-tuning that one needs to gauge away the effect of non-gaussianities is of $\sim O(10^{-5})$. 

We notice that using the Gram-Charlier series with the only inclusion of the third order cumulant $\mathcal{C}_R^{(3)}$ and few terms in the expansion, leads to an increase of the PBH abundance and therefore, at the qualitative level, it captures the main effect of primordial non-gaussianities (in the limit in which non-linearities are neglected). 
However, we reiterate that some care is needed in the truncation of the serie, and a sufficient number of terms should be included in order to obtain a good convergence.
We do not discuss further the Gram-Charlier series, and we focus on the approach that led to eq.\,(\ref{eq:betaNG}). 
A critical discussion about the validity of the approach based on the Gram-Charlier series can be found in ref.\,\cite{Riccardi:2021rlf}.

Then, we introduce in the calculation also the non-linearities among density and comoving curvature perturbations. Using eq.\,(\ref{eq:betaNG}) we obtain the green line in fig.\,\ref{fig:FinalPlot}.
We find that their effect is opposite to the one of primordial local non-gaussianities, making the formation of PBHs harder, in agreement with what found in \,\cite{DeLuca:2019qsy} (but differently from this reference we also include primordial non-gaussianities).
The reason for that can be read from eq.\,(\ref{eq:xdelta}): non-linearities tends to increase the threshold $x_{\delta}$ relevant for PBH formation.
From fig.\,\ref{fig:FinalPlot} one can see that,
once again, the effect of non-gaussianities on the abundance can be reabsorbed by rescaling (in this case decreasing) the height of the power spectrum by a modest amount.

One could wonder how these results change for different values of $\eta_0.$
To address this question, we show in the right panel of fig.\,\ref{fig:FinalPlot}, as a function of $-\eta_0,$ the rescaling factor of the amplitude of power spectrum needed to get the same PBH abundance predicted by the gaussian approximation. 
For comparison, let us mention that the models of single field inflation with an USR phase in refs.\,\cite{Ballesteros:2020qam,Ballesteros:2017fsr,Dalianis:2018frf}, which are designed to obtained a sizable PBH abundance, predict $-\eta_0\simeq0.48-1.22$ (see also\,\cite{Atal:2018neu} for a survey of different models in the literature). 
Our results show that non-gaussianities do not play a dramatic role for the computation of the abundance in this class of models, meaning that their effect can be compensated by a small tuning of the model parameters.

Before concluding we shall mention a couple of additional ingredients that are relevant to quantify the impact of non-gaussianities.
In our analysis we focused on how non-gaussianties, primordial or from the non-linear relation, change the statistics of the density field. However, they also modify the threshold $\delta_{\rm th}.$ This is because $\delta_{\rm th}$ depends on the shape of the density perturbation, which is affected by the presence of non-gaussianties.
These modifications has been studied in the literature following different approaches, focusing on the effect of non-linearities\,\cite{Yoo:2018kvb,Germani:2019zez,Kawasaki:2019mbl,Yoo:2020dkz,Musco:2020jjb}, or including also primordial non-gaussianities\,\cite{Kehagias:2019eil} (for the latter case the power spectrum was assumed to be a Dirac-delta distribution; our power spectrum is significantly different).
A complete investigation of the role of non-gaussianities should include also these effects. We leave it for future work.

Finally, quantum fluctuations of the inflaton field can alter the classical trajectory, modifying therefore the inflationary dynamics. This effect, which can be accounted for in the framework of stochastic inflation\,\cite{Starobinsky:1986fx}, introduces non-gaussianities in the PDF of the comoving curvature perturbations, which can substantially alter the prediction for the PBHs abundance.
Investigation of these quantum effects for the PBHs phenomenology is underway in the literature, see e.g.\,\cite{Ballesteros:2020sre,Biagetti:2018pjj,Ezquiaga:2018gbw,Cruces:2018cvq,Firouzjahi_2019,Pattison:2019hef,Ezquiaga:2019ftu,Pattison:2021oen,De:2020hdo,Figueroa:2020jkf}.

\section{Summary and conclusions}
\label{sec:Conclusions}
Let us conclude with a brief summary of the main novelties and results of our work.
We have focused on the study of non-gaussianities which are generated by the presence of USR dynamics during inflation. 
The latter occur when the Hubble parameter $\eta$ takes values $>3/2.$
An USR phase is crucial for the production of a sizable number of PBHs, and previous studies often limited the computation of their abundance to the gaussian approximation. 

We have computed the bispectrum of the comoving curvature perturbations $\mathcal{R}$ in the context of the {\it in-in} formalism. From the operators appearing in the cubic action of $\mathcal{R}$ we have isolated two dominant terms. The contribution from one of those drastically depends on the transition between the USR phase and the conventional slow-roll phase. We have shown that unless this transition is very sharp, a cancellation occurs in the calculation, dramatically reducing the imprint of this operator on the bispectrum.
We have then offered a procedure, based on a simple analysis of the time evolution of $\eta$, to estimate how sharp the transition should be in order for this cancellation to apply. 
Our results extend the findings of ref.\,\cite{Cai:2017bxr}, which focused on a scenario of non-attractor inflation where the USR phase, taken with $\eta=3,$ is connected to a phase with negligible $\eta.$ On the contrary, in the context of models for PBHs production, the USR phase (generically with $\eta\neq3$) is followed by a phase where $\eta$ is negative and $O(1)$. This is a generic property in the presence of USR, required to connect the end of the latter with the end of inflation.

We have found that the remaining contributions to the bispectrum generate non-gaussianities predominantly of the local type, with 
\begin{align}
f_{\rm NL}=5\,/6\,(-\eta_0).
\end{align}
where $\eta_0$ is the value of the parameter $\eta$ after the end of the USR phase. 
We worked out the details of our analysis both semi-analytically and numerically, in the context of the model introduced in eqs.\,(\ref{act},\,\ref{eq:TempHDO1}), in which the inflaton field is a real scalar singlet equipped with a generic polynomial potential and non-minimally coupled to gravity in the Jordan frame. For this class of models, we extend the results presented in\,\cite{Atal:2018neu}.
Recently it has been claimed that at the end of an USR phase non-gaussianities are slow-roll suppressed and thus negligible\,\cite{Bravo:2020hde}. However, as we have shown with an explicit example, this is not necessarily true. On the contrary, in the class of models that we have studied, and which are motivated by the PBHs production, $\eta_0$ is generically negative and $O(1).$  

Then, using the formalism of threshold statistics, we have investigated the impact of non-gaussianities found here on the PBH abundance.
To evaluate their relevance, or in other words the validity of the gaussian approximation, we have considered eq.\,(\ref{eq:FundBetaNG}), which allows to compute the PBH abundance once the cumulants of non-gaussian distribution are known.
Focusing only on the first non-gaussian correction, i.e. the third-order cumulant $\mathcal{C}_R^{(3)},$ we have found that primordial non-gaussianities sizeably modify the estimation of the PBH abundance based on gaussian statistics.
Qualitatively, the same conclusion has been obtained in\,\cite{Franciolini:2018vbk,Atal:2018neu}, where an approximation of eq.\,(\ref{eq:FundBetaNG}) was adopted to quantify the impact of non-gaussianities.
A quantitative comparison with these analysis and the approximations adopted there will be presented in\,\cite{Riccardi:2021rlf}. 
Then, we have taken a step forward and we have shown how to compute the PBH abundance in presence of local  non-gaussianities, extending 
the gaussian threshold statistics in eq.\,(\ref{eq:FundBeta}). This leads to eq.\,(\ref{eq:betaNG}).
In our calculations we have kept into account the non-linear relation between the density and curvature perturbations, which introduces another source of non-gaussianities in the addition to the primordial ones. 
A similar formalism has been presented in\,\cite{DeLuca:2019qsy}, but focused only to the study of non-linearities.
Using eq.\,(\ref{eq:betaNG}), we have found that the impact of non-gaussianities can be reabsorbed by a change of the amplitude of the power spectrum of curvature perturbations of a factor $2\div3$ at most.
Namely, after such shift one recovers the same PBH abundance obtained with the gaussian approximation.
For the inflationary model that we have analyzed, this modification is achieved by a minor (below the \text{\textperthousand} level) tuning of one parameter (the non-minimal coupling to gravity).

Finally, let us stress that these results do not necessarily imply that non-gaussianities are irrelevant for the phenomenology of PBHs. A notable example of the contrary are the primordial gravitational wave induced by scalar curvature perturbations. This signal is an important route to test PBHs with future gravitational-wave interferometers. The energy density of the gravitational waves scales quadratically with the amplitude of the power spectrum of curvature perturbations, and
therefore there exists a correlation between the PBH abundance and the gravitational wave signal. 
One shall be interested in scenarios where PBHs account for all the dark matter, or such that their abundance is in agreement with observational constraints in the relevant mass range. The gravitational signal would then be a prediction, whose theoretical uncertainty is tied to the ones involved in the calculations of the PBHs abundance.
The effects discussed here are therefore relevant in this context.
The interplay between gravitational signal and PBH abundance will be studied in a future work.

\begin{acknowledgments}
The research of A.U. is supported in part by the MIUR under contract 2017\,FMJFMW (``{New Avenues in Strong Dynamics},'' PRIN\,2017) 
and by the INFN grant ``SESAMO -- SinergiE di SApore e Materia Oscura.'' 
M.T. acknowledges support from the INFN grant ``LINDARK,'' the research grant ``The Dark Universe: A Synergic Multimessenger Approach No. 2017X7X85'' funded by MIUR, and the project ``Theoretical Astroparticle Physics (TAsP)'' funded by the INFN.

\end{acknowledgments}

\appendix

\section{}\label{secAPP}

\subsection{Computation of the three-point correlator $\langle \mathcal{R}_{k_1}\mathcal{R}_{k_2}\mathcal{R}_{k_3}\rangle$}\label{app:ThreePoints}

In this appendix we sketch the derivation of eqs.\,(\ref{eq:Bi1},\,\ref{eq:Bi2},\,\ref{eq:Bi3}).  
Even though the formalism used is quite standard, we shall explain in detail our analysis in order to better highlight 
the difference between our result and the existing literature.

The obtain the power spectrum of the comoving curvature perturbation, computed in section\,\ref{sec:Abundance}, one has to solve the equation of motion obtained from the second-order action for the comoving curvature perturbation.

This is not enough for the computation of non-gaussianities (equivalently, for the computation of higher-order correlators) since the latter can be thought as an imprint left by interactions.  
More in detail, we can split the full Hamiltonian in three parts
$\mathcal{H}_{\rm full} = \mathcal{H}_{\rm cl} + \mathcal{H} + \mathcal{H}_{\rm int}$, 
where $\mathcal{H}_{\rm cl}$ is the background Hamiltonian constructed with the classical background field, 
$\mathcal{H}$ is the Hamiltonian constructed with scalar field perturbations up to quadratic order 
and $\mathcal{H}_{\rm int}$ is constructed with perturbations 
 at order equal or higher than three. 
 The formalism to compute correlation functions in cosmological settings is called the {\it in-in} 
 formalism (introduced in ref.\,\cite{Schwinger:1960qe,Keldysh:1964ud} for condensed
matter systems and the applied to cosmology in refs.\,\cite{Jordan:1986ug,Calzetta:1986ey,Maldacena:2002vr,Weinberg:2005vy}). The three-point correlator can be obtained by applying  the master 
 formula\footnote{For the sake of clarity, and to connect with section\,\ref{sec:SkewnessExact}, let us be clear with our notation. In the following we use
 \begin{align}
 \hat{\mathcal{R}}_I(\tau,\vec{x}) =\int\frac{d^3\vec{k}}{(2\pi)^{3}}
 \hat{\mathcal{R}}_I(\tau,\vec{k})e^{i\vec{x}\cdot\vec{k}}\,,~~~~~~~~~~
 \hat{\mathcal{R}}_I(\tau,\vec{k}) = R_k(\tau)a_{\vec{k}}(\tau_{\star}) + R_k^*(\tau)a_{-\vec{k}}^{\dag}(\tau_{\star})\,.
 \end{align}
The subscript $_I$ is
used here 
to distinguish between the left- and right-hand side of eq.\,(\ref{eq:MasterInIn3pt}).  
Furthermore, we have the contraction 
 \begin{align}
\langle 0|\hat{\mathcal{R}}_I(\tau_1,\vec{k}_1)\hat{\mathcal{R}}_I(\tau_2,\vec{k}_2)|0\rangle =  
(2\pi)^3\delta^{(3)}(\vec{k}_1+\vec{k}_2) R_{k_1}(\tau_1)R_{k_2}^*(\tau_2)\,.
 \end{align}
 }
 \begin{align}\label{eq:MasterInIn3pt}
 \langle \hat{\mathcal{R}}(\tau,\vec{k}_1)
 \hat{\mathcal{R}}(\tau,\vec{k}_2)
 \hat{\mathcal{R}}(\tau,\vec{k}_3)\rangle = 
 2\Im\left[
 \int_{-\infty(1-i\epsilon)}^{\tau}d\tau^{\prime}a(\tau^{\prime})\langle
 \hat{\mathcal{R}}_I(\tau,\vec{k}_1)
 \hat{\mathcal{R}}_I(\tau,\vec{k}_2)
 \hat{\mathcal{R}}_I(\tau,\vec{k}_3)
 \hat{\mathcal{H}}_{\rm int}(\tau^{\prime})
  \rangle
 \right]\,,
 \end{align}
 where the subscript $_{I}$ denotes interaction picture operators, 
and the interaction Hamiltonian $\hat{\mathcal{H}}_{\rm int}$ is also constructed with the interaction picture operators. 
The $i\epsilon$ prescription (where $\epsilon$, not to be confused with the Hubble parameter, 
 is a small real positive number) is introduced in order to regularize the
time integral in the early time limit. This prescription assures that the integrand vanishes at early
times, and hence the integral remains finite.
At this stage, it is important to keep in mind 
that the vacuum expectation value on the left-hand side is evaluated on the interaction-picture 
vacuum while the the right-hand side is evaluated on the vacuum of the free theory. 
The interaction Hamiltonian $\hat{\mathcal{H}}_{\rm int}$ evolves
the vacuum of the free theory to the interaction vacuum at the time $\tau$ at which we evaluate the 
three-point function. 
Eq.\,(\ref{eq:MasterInIn3pt}) makes clear that non-gaussianities, 
encoded in three- and higher-point correlators, are a manifestation of interactions. 
Notice also that eq.\,(\ref{eq:MasterInIn3pt}) is valid at the first order in $\hat{\mathcal{H}}_{\rm int}$. 
We remark that our goal is to compute, using eq.\,(\ref{eq:MasterInIn3pt}), the three-point correlator 
at some late time $\tau$ close to the end of inflation when we can safely take 
 the super-horizon curvature perturbations as constants.
 
The cubic action, in terms of the cosmic time, is given by 
$\mathcal{S}_3 = \int dtd^3\vec{x}\mathcal{L}_3$, with~\cite{Maldacena:2002vr}
\begin{align}
\mathcal{L}_3 & = 
a^3 \epsilon^2\mathcal{R}\dot{\mathcal{R}}^2 
+ a\epsilon^2\mathcal{R}(\partial \mathcal{R})^2
-2a\epsilon\dot{\mathcal{R}}(\partial \mathcal{R})(\partial \chi)  + 
\frac{a^3\epsilon\dot{\epsilon}_2}{2}\mathcal{R}^2\dot{\mathcal{R}} 
 +\frac{\epsilon}{2a}(\partial \mathcal{R})(\partial \chi)\partial^2\chi
 +\frac{\epsilon}{4a}(\partial^2\mathcal{R})(\partial\chi)^2
 +2f(\mathcal{R})\frac{\delta\mathcal{S}_2}{\delta\mathcal{R}}\,,\label{eq:CubicMalda}\\
f(\mathcal{R}) & = \frac{\epsilon_2}{4}\mathcal{R}^2+ \frac{1}{H}\mathcal{R}\dot{\mathcal{R}} 
+ {\rm terms\,quadratic\,in}\,\mathcal{R}\,{\rm with\,spatial\,derivatives\,acting\,on\,}\mathcal{R}\,,\label{eq:DefF}
\end{align}
where $\epsilon_2 = \dot{\epsilon}/H\epsilon = 2\epsilon - 2\eta$, $\partial^2\chi = a^2\epsilon \dot{\mathcal{R}}$ 
and $\delta\mathcal{S}_2/\delta\mathcal{R}$ the variation of the quadratic action with
respect to the perturbation.
Notice that the cubic action $\mathcal{S}_3$ is exact for arbitrary $\epsilon$ and $\eta$.
As shown in ref.\,\cite{Maldacena:2002vr}, 
the last term in eq.\,(\ref{eq:CubicMalda}) can be absorbed by the field redefinition 
$\mathcal{R} \to \mathcal{R}_n + f(\mathcal{R}_n)$ so that, after this redefinition, 
the interaction Hamiltonian in conformal time $\tau$ is
\begin{align}\label{eq:MasterHint}
\mathcal{H}_{\rm int}(\tau) =-\int d^3\vec{x}\bigg[&
a\epsilon^2\mathcal{R}_n\left(\mathcal{R}_n^{\prime}\right)^2 
+a\epsilon^2 \mathcal{R}_n (\partial \mathcal{R}_n)^2 - 
2\epsilon\mathcal{R}_n^{\prime}(\partial\mathcal{R}_n)(\partial\chi_n)
+ \frac{a\epsilon\epsilon_2^{\prime}}{2}\mathcal{R}_n^2\mathcal{R}_n^{\prime}\nonumber \\ &
+\frac{\epsilon}{2a}(\partial\mathcal{R}_n)(\partial\chi_n)(\partial^2\chi_n)
+\frac{\epsilon}{4a}(\partial^2\mathcal{R}_n)(\partial\chi_n)^2
\bigg]\,,
\end{align}
with $\partial^2\chi_n = a^2\epsilon \dot{\mathcal{R}}_n$ 
and $^{\prime}$ denotes the derivative with respect to $\tau$. 
This is not the only effect that is introduced by the the field redefinition since it also gives some extra terms in the three-point correlation function of $\mathcal{R}$. In Fourier space, we have
\begin{align}
\langle \hat{\mathcal{R}}&(\tau,\vec{k}_1)
 \hat{\mathcal{R}}(\tau,\vec{k}_2)
 \hat{\mathcal{R}}(\tau,\vec{k}_3)\rangle \label{eq:Red1} \\
& = \langle \hat{\mathcal{R}}_n(\tau,\vec{k}_1)
 \hat{\mathcal{R}}_n(\tau,\vec{k}_2)
 \hat{\mathcal{R}}_n(\tau,\vec{k}_3)\rangle
+\frac{\epsilon_2}{4}\left[
\langle \hat{\mathcal{R}}_n^2(\tau,\vec{k}_1)
 \hat{\mathcal{R}}_n(\tau,\vec{k}_2)
 \hat{\mathcal{R}}_n(\tau,\vec{k}_3) \rangle + {\rm two\,perms}
 \right] + {\rm higher\,order\,terms}\,, \nn
\end{align}
where one should keep in mind that in eq.\,(\ref{eq:Red1}) $\hat{\mathcal{R}}_n^2(\tau,\vec{k}_1)$ 
is not just the square of $\hat{\mathcal{R}}_n(\tau,\vec{k}_1)$ but it originates from the Fourier transform of the square of
 $\hat{\mathcal{R}}_n(\tau,\vec{x})$ (since the field redefinition is defined at the level of the action in position space) meaning that, by means of the convolution theorem, we actually have
\begin{align}
 \hat{\mathcal{R}}_n^2(\tau,\vec{k}_1) = \int \frac{d^3\vec{q}}{(2\pi)^3}~
 \hat{\mathcal{R}}_n(\tau,\vec{q})\hat{\mathcal{R}}_n(\tau,\vec{k}_1-\vec{q})\,,
\end{align}
and similarely for the other two permutations. 
In eq.\,(\ref{eq:Red1}) we neglected terms with spatial derivatives in eq.\,(\ref{eq:DefF}) since we are interested in super-horizon modes and the contribution from the second term in eq.\,(\ref{eq:DefF}) since 
we are interested in the computation of the three-point function at late time. 
All in all, the strategy of the computation is the following.
\begin{itemize}
\item [{\it i)}] We compute the three-point function $\langle \hat{\mathcal{R}}_n(\tau,\vec{k}_1)
 \hat{\mathcal{R}}_n(\tau,\vec{k}_2)
 \hat{\mathcal{R}}_n(\tau,\vec{k}_3)\rangle$, that is the first term on the right-hand side of eq.\,(\ref{eq:Red1}), 
 by means of 
 eq.\,(\ref{eq:MasterInIn3pt}) using the interaction Hamiltonian given in eq.\,(\ref{eq:MasterHint});
\item [{\it ii)}] we shift back from the bispectrum of $\mathcal{R}_n$ to that of $\mathcal{R}$ by 
including the leading correction given by the second term on the right-hand side of eq.\,(\ref{eq:Red1}).
\end{itemize}
For notation simplicity, we drop from now on the subscript $_n$.
We start from point {\it i)}. In the interaction Hamiltonian most of the terms are $\epsilon$-suppressed.
The only exception which is relevant for us is the term that contains $\epsilon_2^{\prime}$.
In our model, we have
\begin{align}\label{eq:DeerivativeEta}
\frac{d\eta}{d\tau} = \eta_{\rm II}\delta(\tau-\tau_{\rm in}) + (-\eta_{\rm II} + \eta_{\rm III})\delta(\tau-\tau_{\rm end})\,,
\end{align}
meaning that in the integral in eq.\,(\ref{eq:MasterInIn3pt}) we have two contributions corresponding to the transition at 
$\tau=\tau_{\rm in}$ and $\tau=\tau_{\rm end}$. 
Let us compute explicitly only the first contribution. The integral on the right-hand side of eq.\,(\ref{eq:MasterInIn3pt}) gives (after promoting classical fields to operators)
\begin{align}	\label{eq:Wick1}
 2\Im\eta_{\rm II}&a(\tau_{\rm in})^2\epsilon_{\rm I}
\int d^{3}\vec{x}\langle
\hat{\mathcal{R}}_I(\tau,\vec{k}_1)
 \hat{\mathcal{R}}_I(\tau,\vec{k}_2)
 \hat{\mathcal{R}}_I(\tau,\vec{k}_3)
 \hat{\mathcal{R}}_I^2(\tau_{\rm in},\vec{x})
 \hat{\mathcal{R}}_I(\tau_{\rm in},\vec{x})
\rangle = 2\Im\eta_{\rm II}a(\tau_{\rm in})^2\epsilon_{\rm I}\times \\
& 
\int d^{3}\vec{x}\int \frac{d^3\vec{q}_1}{(2\pi)^{3}}
\frac{d^3\vec{q}_2}{(2\pi)^{3}}\frac{d^3\vec{q}}{(2\pi)^{3}}
\langle
\hat{\mathcal{R}}_I(\tau,\vec{k}_1)
 \hat{\mathcal{R}}_I(\tau,\vec{k}_2)
 \hat{\mathcal{R}}_I(\tau,\vec{k}_3)
\hat{\mathcal{R}}_I(\tau_{\rm in},\vec{q}) 
\hat{\mathcal{R}}_I(\tau_{\rm in},\vec{q}_1-\vec{q})
\hat{\mathcal{R}}_I^{\prime}(\tau_{\rm in},\vec{q}_2)
\rangle
e^{i\vec{x}\cdot(\vec{q}_1 + \vec{q}_2)}\nonumber\,,
\end{align}
where in the last step we switched to Fourier space and we used the convolution theorem.
The expectation value on the vacuum $|0\rangle$ of the free theory can be computed by means of the Wick theorem.
We only have three relevant contractions,\footnote{Notice that in principle there are 15 ways 
of contracting pairs of operators (divided in three classes: 
pairings of external fields with fields at the interaction vertex, pairings of external fields among themselves and vertex fields among themselves). 
However, we are only interested in the connected contractions, i.e. contractions 
where one external operator in the string $\hat{\mathcal{R}}_I(\tau,\vec{k}_1)
 \hat{\mathcal{R}}_I(\tau,\vec{k}_2)
 \hat{\mathcal{R}}_I(\tau,\vec{k}_3)$
is contracted
with one operator of the interaction vertex $H_{\rm int}$;
 this reduces the number of contractions down to 6.
 Physically, the disconnected contractions which are not included are proportional to tadpole diagrams 
  but, since we are assuming that the background solution is stable (or, equivalently, that $\langle \hat{\mathcal{R}}\rangle = 0$) all such amplitudes vanish.
} each of them that counts twice since, for instance, the two contractions 
\begin{align}
\int d^{3}\vec{x} & \int \frac{d^3\vec{q}_1}{(2\pi)^{3}}
\frac{d^3\vec{q}_2}{(2\pi)^{3}}\frac{d^3\vec{q}}{(2\pi)^{3}}\langle
\begC1{\hat{\mathcal{R}}_I(\tau,\vec{k}_1)} 
\begC2{\hat{\mathcal{R}}_I(\tau,\vec{k}_2)}
 \begC3{\hat{\mathcal{R}}_I(\tau,\vec{k}_3)}
\endC1{\hat{\mathcal{R}}_I(\tau_{\rm in},\vec{q}) }
\endC2{\hat{\mathcal{R}}_I(\tau_{\rm in},\vec{q}_1-\vec{q})}
\endC3{\hat{\mathcal{R}}_I^{\prime}(\tau_{\rm in},\vec{q}_2)} 
\rangle e^{i\vec{x}\cdot(\vec{q}_1 + \vec{q}_2)} = \\
& \underbrace{\int d^{3}\vec{x}e^{-i\vec{x}\cdot(\vec{k}_1+\vec{k}_2+\vec{k}_3)}}_{=(2\pi)^3
\delta^{(3)}(\vec{k}_1+\vec{k}_2+\vec{k}_3)
}
\mathcal{R}_{k_1}(\tau)\mathcal{R}_{k_2}(\tau)\mathcal{R}_{k_3}(\tau)
\mathcal{R}^*_{k_1}(\tau_{\rm in})
\mathcal{R}^*_{k_2}(\tau_{\rm in})\mathcal{R}^{\prime\,*}_{k_3}(\tau_{\rm in})\,,
\end{align}
and
\begin{align}
\int d^{3}\vec{x}\int \frac{d^3\vec{q}_1}{(2\pi)^{3}}
\frac{d^3\vec{q}_2}{(2\pi)^{3}}\frac{d^3\vec{q}}{(2\pi)^{3}}
 \begC1{\hat{\mathcal{R}}_I(\tau,\vec{k}_1)} 
\begC2{\hat{\mathcal{R}}_I(\tau,\vec{k}_2)}
 \begC3{\hat{\mathcal{R}}_I(\tau,\vec{k}_3)}
\endC2{\hat{\mathcal{R}}_I(\tau_{\rm in},\vec{q}) }
\endC1{\hat{\mathcal{R}}_I(\tau_{\rm in},\vec{q}_1-\vec{q})}
\endC3{\hat{\mathcal{R}}_I^{\prime}(\tau_{\rm in},\vec{q}_2)}\rangle e^{i\vec{x}\cdot(\vec{q}_1 + \vec{q}_2)}\,,
\end{align}
give the same result. If we factor out $(2\pi)^3
\delta^{(3)}(\vec{k}_1+\vec{k}_2+\vec{k}_3)$, the contribution of eq.\,(\ref{eq:Wick1}) to the bispectrum becomes
\begin{align}
B^{({\rm in})}_{\mathcal{R}}(k_1,k_2,k_3)
= 4\Im\bigg\{& \eta_{\rm II}a(\tau_{\rm in})^2\epsilon_{\rm I}
\mathcal{R}_{k_1}(\tau)\mathcal{R}_{k_2}(\tau)\mathcal{R}_{k_3}(\tau)\times\\
& \left[
\mathcal{R}^*_{k_1}(\tau_{\rm in})
\mathcal{R}^*_{k_2}(\tau_{\rm in})\mathcal{R}^{\prime\,*}_{k_3}(\tau_{\rm in})+
\mathcal{R}^*_{k_1}(\tau_{\rm in})
\mathcal{R}^*_{k_3}(\tau_{\rm in})\mathcal{R}^{\prime\,*}_{k_2}(\tau_{\rm in})+
\mathcal{R}^*_{k_2}(\tau_{\rm in})
\mathcal{R}^*_{k_3}(\tau_{\rm in})\mathcal{R}^{\prime\,*}_{k_1}(\tau_{\rm in})
\right]\bigg\}\,,\nn
\end{align}
taken at late time $\tau\to 0^-$ (so that the term $\mathcal{R}_{k_1}(\tau)\mathcal{R}_{k_2}(\tau)\mathcal{R}_{k_3}(\tau)$ has to be computed in region\,III). 
Using eq.\,(\ref{eq:HankelGen}) and the expansion of the Hankel functions for small argument, we find eq.\,(\ref{eq:Bi1}) used in the text.

The computation of the contribution at $\tau=\tau_{\rm end}$ follows the same steps 
but now we pick up the second delta function in eq.\,(\ref{eq:DeerivativeEta}). 
We find the result
\begin{align}
B^{({\rm end})}_{\mathcal{R}}&(k_1,k_2,k_3)
=  4\Im\bigg\{(-\eta_{\rm II}+\eta_{\rm III})a(\tau_{\rm end})^2\epsilon_{\rm II}(\tau_{\rm end})
\mathcal{R}_{k_1}(\tau)\mathcal{R}_{k_2}(\tau)\mathcal{R}_{k_3}(\tau)\times\\
& \left[
\mathcal{R}^*_{k_1}(\tau_{\rm end})
\mathcal{R}^*_{k_2}(\tau_{\rm end})\mathcal{R}^{\prime\,*}_{k_3}(\tau_{\rm end})+
\mathcal{R}^*_{k_1}(\tau_{\rm end})
\mathcal{R}^*_{k_3}(\tau_{\rm end})\mathcal{R}^{\prime\,*}_{k_2}(\tau_{\rm end})+
\mathcal{R}^*_{k_2}(\tau_{\rm end})
\mathcal{R}^*_{k_3}(\tau_{\rm end})\mathcal{R}^{\prime\,*}_{k_1}(\tau_{\rm end})
\right]\bigg\}\,.\nn
\end{align}

Finally, we move to consider the contribution {\it ii)} that comes from the field redefinition in eq.\,(\ref{eq:Red1}). 
If we neglect $\epsilon$, we have $\epsilon_2 = -2\eta_{\rm III}$ since this term must be evaluated towards the end of 
inflation in region\,III. The correction in eq.\,(\ref{eq:Red1}) reads
\begin{align}
-\frac{\eta_{\rm III}}{2}\int\frac{d^3\vec{q}}{(2\pi)^3}
\left[
\langle 
\hat{\mathcal{R}}_I(\tau,\vec{q})
\hat{\mathcal{R}}_I(\tau,\vec{k}_1-\vec{q})
\hat{\mathcal{R}}_I(\tau,\vec{k}_2)
\hat{\mathcal{R}}_I(\tau,\vec{k}_3) \rangle +
{\rm two\,permutations}
\right]
\end{align}
Consider the first term in the square brackets. 
If we apply the Wick theorem, we have again that the contraction
\begin{align}
-\frac{\eta_{\rm III}}{2}\int\frac{d^3\vec{q}}{(2\pi)^3}
\langle 
\begC1{\hat{\mathcal{R}}_I(\tau,\vec{q})}
\begC2{\hat{\mathcal{R}}_I(\tau,\vec{k}_1-\vec{q})}
\endC1{\hat{\mathcal{R}}_I(\tau,\vec{k}_2)}
\endC2{\hat{\mathcal{R}}_I(\tau,\vec{k}_3)}
 \rangle = -\frac{\eta_{\rm III}}{2}(2\pi)^3\delta^{(3)}(\vec{k}_1 + \vec{k}_3 + \vec{k}_3)
 |\mathcal{R}_{k_2}(\tau)|^2|\mathcal{R}_{k_3}(\tau)|^2
\end{align}
counts twice since it gives the same result of
\begin{align}
-\frac{\eta_{\rm III}}{2}\int\frac{d^3\vec{q}}{(2\pi)^3}
\langle 
\begC1{\hat{\mathcal{R}}_I(\tau,\vec{q})}
\begC2{\hat{\mathcal{R}}_I(\tau,\vec{k}_1-\vec{q})}
\endC2{\hat{\mathcal{R}}_I(\tau,\vec{k}_2)}
\endC1{\hat{\mathcal{R}}_I(\tau,\vec{k}_3)}
 \rangle\,.
\end{align}
Repeating the same argument for the other two permutations in eq.\,(\ref{eq:Red1}) and
factoring out $(2\pi)^3
\delta^{(3)}(\vec{k}_1+\vec{k}_2+\vec{k}_3)$ from the definition of the bispectrum, we find
\begin{align}\label{eq:RedCon}
B^{({\rm red})}_{\mathcal{R}}(k_1,k_2,k_3) & = -\eta_{\rm III}\left(
|\mathcal{R}_{k_2}(\tau)|^2|\mathcal{R}_{k_3}(\tau)|^2 + 
 |\mathcal{R}_{k_1}(\tau)|^2|\mathcal{R}_{k_3}(\tau)|^2 + 
  |\mathcal{R}_{k_1}(\tau)|^2|\mathcal{R}_{k_2}(\tau)|^2
\right)\nn \\
& = -\frac{4\pi^4 \eta_{\rm III}}{(k_1 k_2 k_3)^3}\left[
k_1^3 \mathcal{P}_{\mathcal{R}}(k_2)\mathcal{P}_{\mathcal{R}}(k_3) +
k_2^3 \mathcal{P}_{\mathcal{R}}(k_1)\mathcal{P}_{\mathcal{R}}(k_3) +
k_3^3 \mathcal{P}_{\mathcal{R}}(k_1)\mathcal{P}_{\mathcal{R}}(k_2) 
\right]\,,
\end{align}
which reproduces eq.\,(\ref{eq:Bi3}) (and agrees, if one takes $\mathcal{P}_{\mathcal{R}}(k) = H^2/8\pi^2\epsilon$, with the standard result obtained 
in single-field slow-roll inflation, see e.g. ref.\,\cite{Chen:2006nt}).

Before moving to the comparison of our results with the literature, let us mention that ref.\,\cite{Atal:2019cdz} claims that the field redefinition in eq.\,(\ref{eq:Red1}) is the first order term of a more general expression, which leads to non-gaussianities of local form. Using the $\delta N$ formalism and working with a quadratic expansion of the inflaton potential around its local maximum, ref.\,\cite{Atal:2019cdz} presents a resummation of the higher order corrections. Since the potential studied here differs from the simple approximation mentioned above, these results can not be directly applied to our case. 
Still, it would be interesting to extend such calculation for our scenario. Moreover, as mentioned in section\,\ref{sec:Kurtosis}, it is important to investigate the role of higher order contributions leading to non-local non-gaussianities. 

We are now in the position to compare our result with refs.\,\cite{Namjoo:2012aa,Cai:2017bxr}.
Ref.\,\cite{Namjoo:2012aa} computes the amount of non-gaussianities in the case of non-attractor inflation models. 
In the specific realization considered, a canonically normalized scalar field rolls under a potential 
\begin{align}\label{eq:nonattractorinflation}
V(\phi) =
\left\{
\begin{array}{ccc}
 V_0 & {\rm for} & \phi<\phi_c    \\
 V_2(\phi) & {\rm for} & \phi>\phi_c     
\end{array}
\right.
\end{align}
which is constant below some critical field value $\phi_c$ and where $V_2(\phi)$ 
supports a second phase of slow-roll inflation as in conventional models.
In this model, which assumes the Minkowski vacuum deep inside the horizon, 
$\eta$ transits from $\eta= 3$ (dubbed non-attractor phase) to $\eta= 0$ during the second slow-roll phase. Since $\eta = 0$ during the second inflationary phase, the only contribution to 
the three-point function comes from the delta function at transition time for $\eta$.   
Let us indicate this transition time with $\tau_e$. 
Ref.\,\cite{Namjoo:2012aa} also assumes that, immediately after the end of the non-attractor phase, 
the super-horizon curvature perturbations cease evolving (meaning that the comoving time $\tau$ at which  
eq.\,(\ref{eq:MasterInIn3pt}) has to be evaluated is taken to be the transition time $\tau_e$). 
As a result, ref.\,\cite{Namjoo:2012aa} finds a local bispectrum (see eq.\,(\ref{eq:LocalR})) with amplitude which, in the squeezed limit, is 
given by $f_{\rm NL} = 5/2$. 
Notice that, based on this result, ref.\,\cite{Franciolini:2018vbk} evaluated the amount of non-gaussianities expected 
in the case of PBH formation in single-field inflationary models featuring a USR phase.
However, the model of ref.\,\cite{Namjoo:2012aa}, designed for the case of non-attractor inflation, 
cannot be applied to the case relevant for PBH formation. 

Ref.\,\cite{Cai:2017bxr}
 improves the analysis of ref.\,\cite{Namjoo:2012aa} by assuming that 
 the non-attractor inflation model consists of at least three different phases: 
 the non-attractor phase (with $\eta=3$), the transition phase (which is modeled in two ways dubbed smooth and sharp), and the final slow-roll phase (where $\eta$ eventually relaxes to $\eta = 0$).
 \begin{figure}[!htb!]
\begin{center}
$$\includegraphics[width=.48\textwidth]{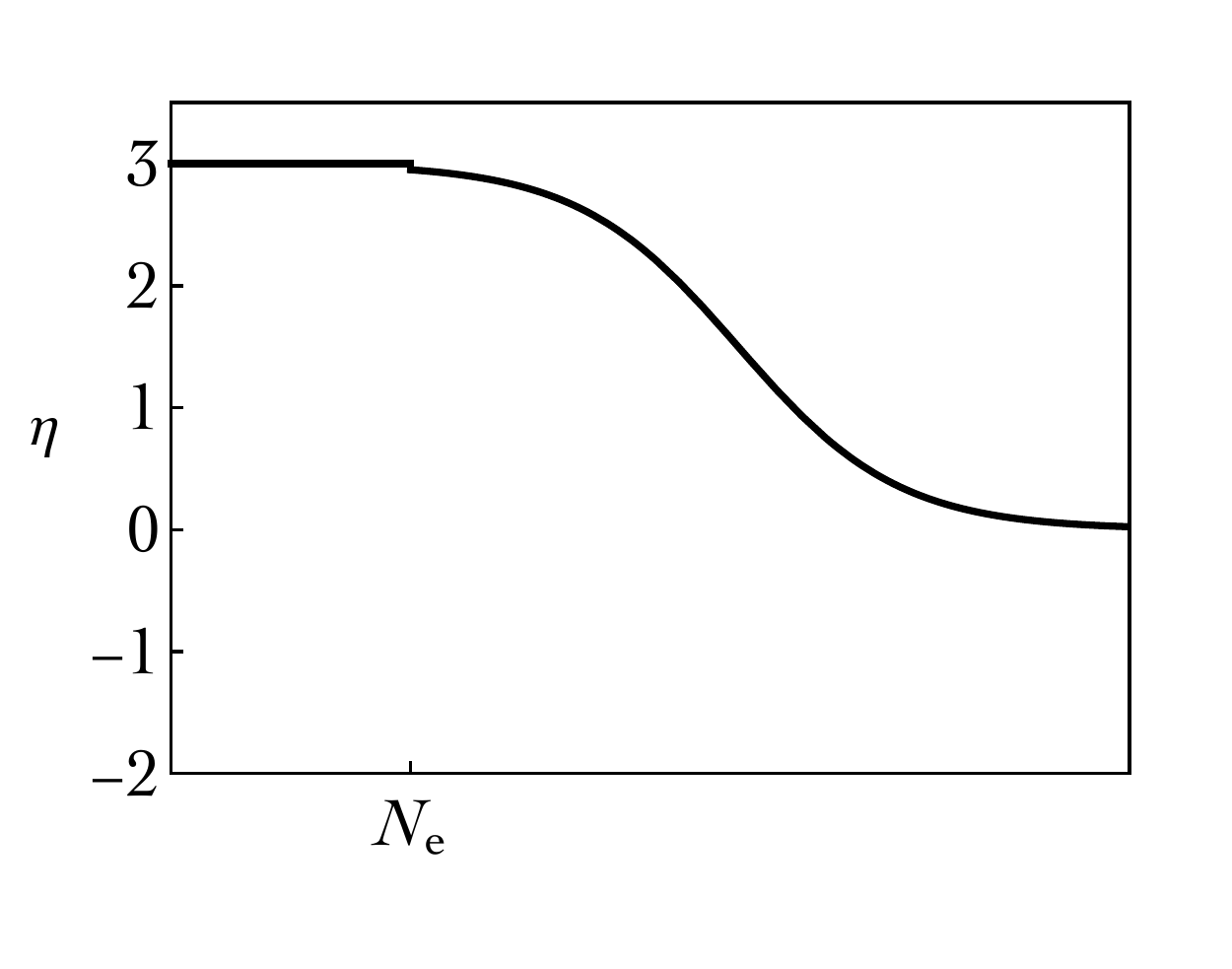}
\qquad\includegraphics[width=.48\textwidth]{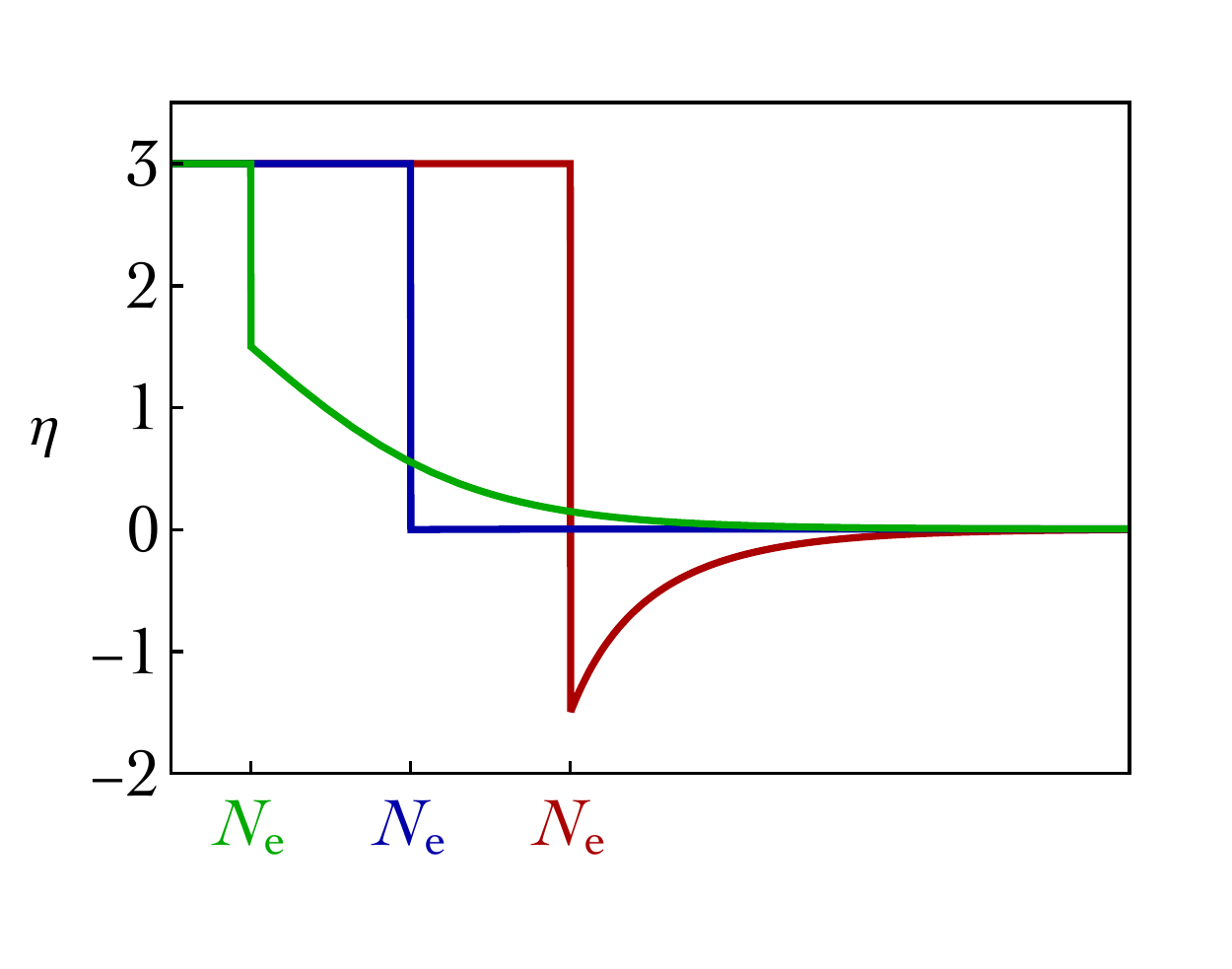}$$
\caption{\em \label{fig:Compa} 
Time evolution of $\eta$ for the two models discussed in ref.\,\cite{Cai:2017bxr}.
The left panel shows a typical smooth transition from the non-attractor to the slow-roll phase.
We indicate with $N_e$ the $e$-fold time at which the non-attractor phase ends.
The right panel shows three typical sharp transitions from the non-attractor to the slow-roll phase (the three lines 
have different inflaton velocity at the end of the non-attractor phase, see ref.\,\cite{Cai:2017bxr} for details; 
furthermore we chose, for the sake of clarity, three different values of $N_e$).}
\end{center}
\end{figure}
We reproduce, for ease of reading, in 
fig.\,\ref{fig:Compa} the qualitative behavior of $\eta$ as function of time (with $N_e$ the $e$-fold time at which the non-attractor phase ends) for the two cases discussed in 
ref.\,\cite{Cai:2017bxr} (smooth and sharp transition).  
In the case of a smooth transition, ref.\,\cite{Cai:2017bxr} finds that the non-gaussianity computed in 
ref.\,\cite{Namjoo:2012aa} by means of the instantaneous approximation is almost cancelled. 
In the case of a sharp transition, ref.\,\cite{Cai:2017bxr} finds that the amplitude of local non-Gaussianity is mainly determined by 
the value of the inflaton velocity at the end of the non-attractor phase.

Without going into detail, we remark here that none of these models, designed 
for the case of non-attractor inflation,
can be considered as an optimal proxy for the case of PBH production. 
This is evident if we compare the time evolution of $\eta$ shown in fig.\,\ref{fig:Pot} with 
those in fig.\,\ref{fig:Compa}. 
In realistic single-field inflationary models of PBH production which are compatible with observations, 
the field starts from slow-roll conditions ($\eta \approx 0$), goes through a USR phase ($\eta > 3$) and 
approaches the end of inflation via a third phase during which $\eta < 0$. 
This dynamics leaves a specific imprints on the three-point correlator which departs, both 
qualitatively and quantitatively, from the previous works, as discussed in this appendix.

\subsection{Computation of the three-point correlator $\langle \mathcal{R}_{k_1}\mathcal{R}_{k_2}\mathcal{R}_{k_3}\rangle$: 
further remarks}\label{app:ThreePoints2} 

Let us add few additional comments to the computation performed in appendix\,\ref{app:ThreePoints}.
\begin{itemize}

\item [{\it a)}] We use the cubic action in eq.\,(\ref{eq:CubicMalda}) as opposed to the expression\,\cite{Maldacena:2002vr,Chen:2006nt}
\begin{align} 
\mathcal{S}_3 = \int dtd^3\vec{x}\bigg\{&  -a\epsilon\mathcal{R}(\partial\mathcal{R})^2 - \frac{a^3\epsilon}{H}\dot{\mathcal{R}}^3 
+3a^3\epsilon\mathcal{R}\dot{\mathcal{R}}^2 \nn \\
& +\frac{1}{2a}\bigg(
3\mathcal{R} - \frac{\dot{\mathcal{R}}}{H}
\bigg)\left[
(\partial_i\partial_j\psi)(\partial_i\partial_j\psi) - (\partial^2\psi)(\partial^2\psi)
\right] -\frac{2}{a}(\partial_i\psi)(\partial_i\mathcal{R})(\partial^2\psi)\bigg\}\,,\label{eq:CubicMalda2}
\end{align}
with $\psi = -\mathcal{R}/H + a^2\epsilon\partial^{-2}\mathcal{R}$
because all terms in eq.\,(\ref{eq:CubicMalda}) are explicitly proportional to the Hubble parameters and, therefore, 
more easy to handle. 
Furthermore, in eq.\,(\ref{eq:CubicMalda}) the cubic vertex is  simpler since the last term can be eliminated 
via a field redefinition\,\cite{Maldacena:2002vr}.
The action obtained by means of the Lagrangian density in eq.\,(\ref{eq:CubicMalda}) can be derived from the one in
eq.\,(\ref{eq:CubicMalda2}) after integrating by parts, discarding total
derivatives, and using the field equation that follows from the quadratic action\,\cite{Maldacena:2002vr}. 
Notice that eq.\,(\ref{eq:CubicMalda2}) was used in ref.\,\cite{Saito:2008em} to compute 
non-gaussianities related to PBH production in the context of the chaotic new inflation model.
 \begin{figure}[!htb!]
\begin{center}
$$\includegraphics[width=.42\textwidth]{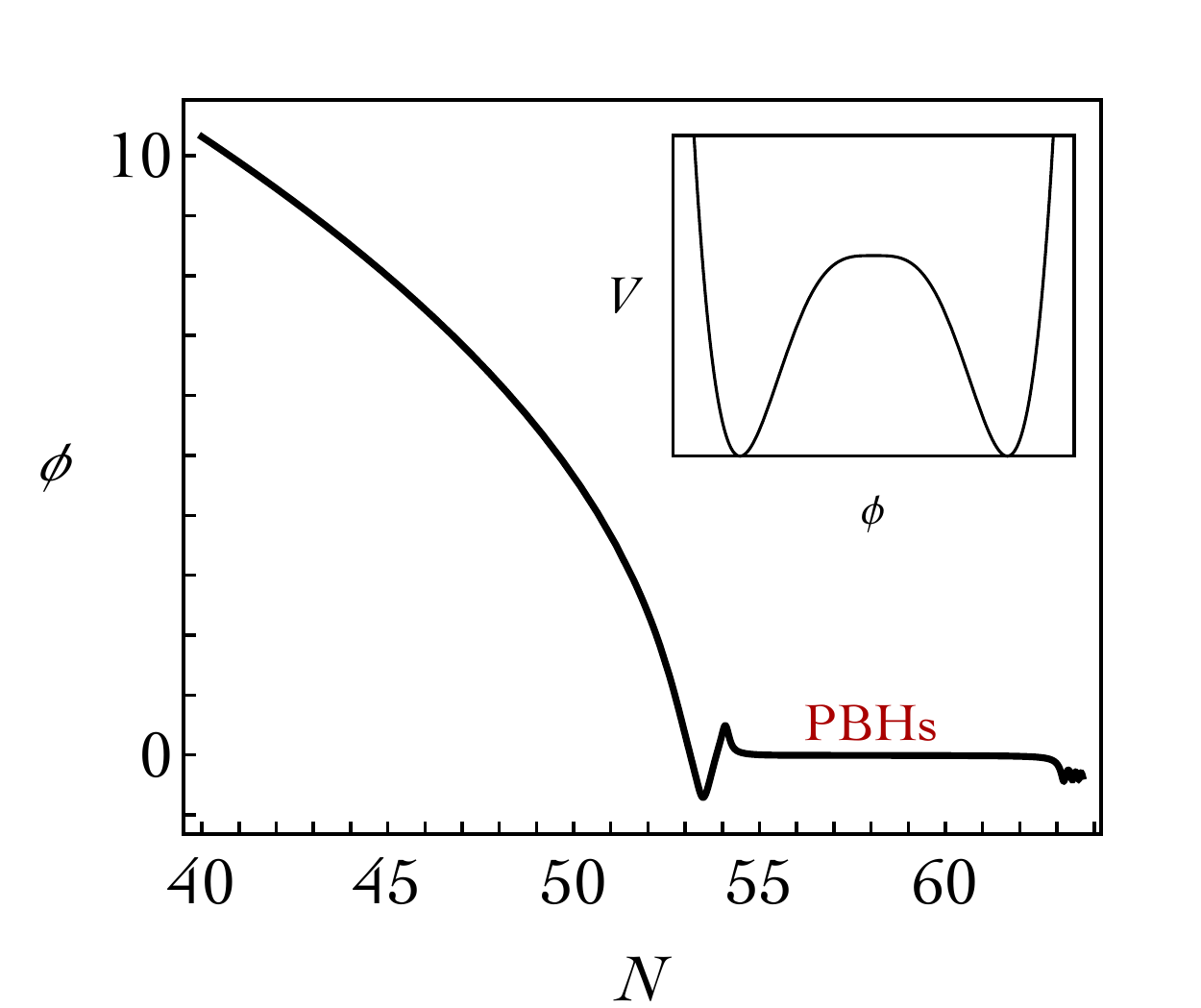}
\qquad\includegraphics[width=.42\textwidth]{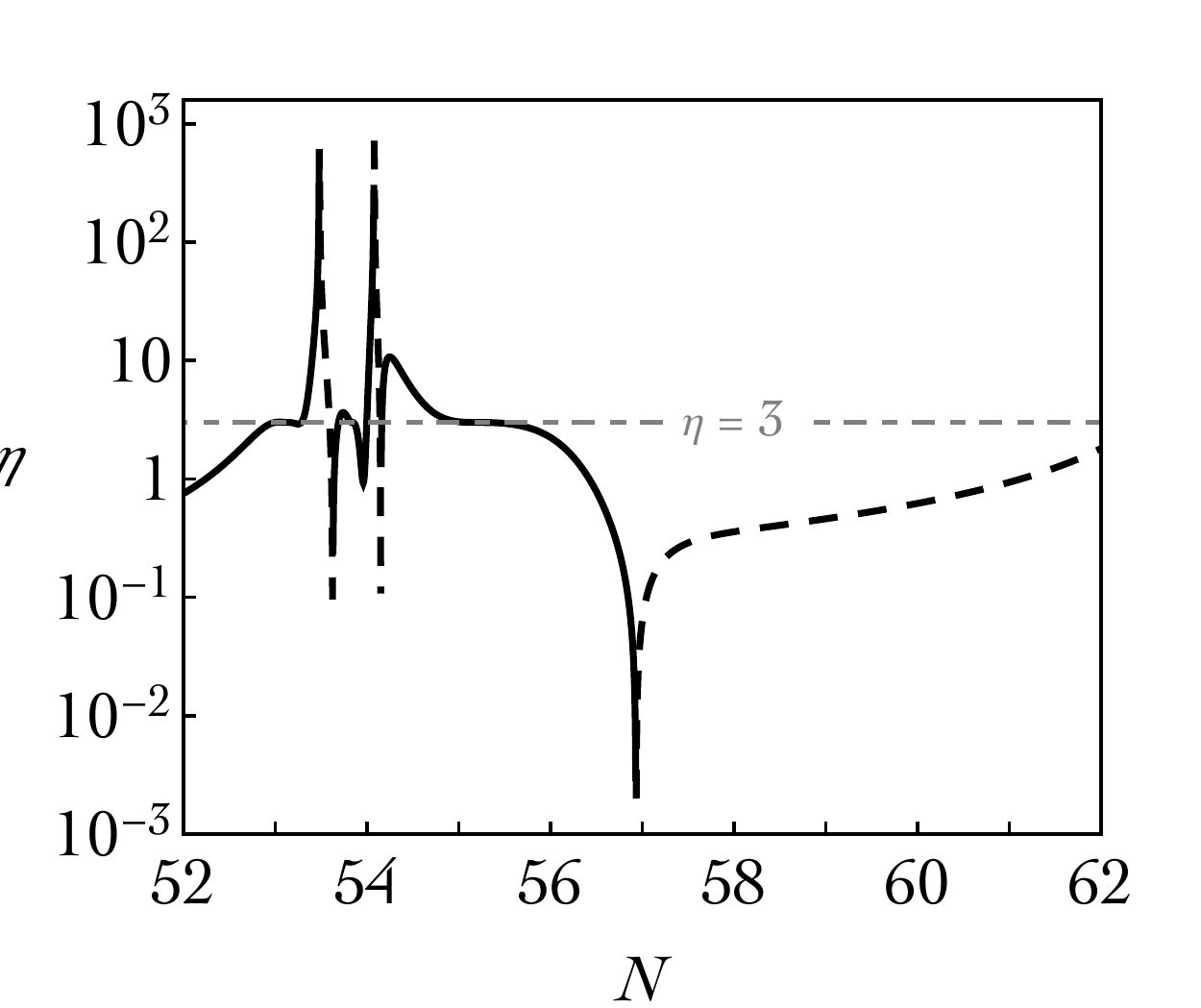}$$
\caption{\em \label{fig:Chaotic} 
Inflaton dynamics for the Coleman-Weinberg model studied in ref.\,\cite{Saito:2008em}. 
In the left panel we show the time evolution of the inflaton field and in the right panel the time evolution of the Hubble parameter $\eta$.
 }
\end{center}
\end{figure}
Ref.\,\cite{Saito:2008em} finds that non-gaussianities give small corrections to the PBH abundance. 
However, the model presented in ref.\,\cite{Saito:2008em} differs---both qualitatively and quantitatively---from the class of models studied in this paper. The inflaton follows a Coleman-Weinberg potential (see inset plot in the left panel of fig.\,\ref{fig:Chaotic}) starting from large field values and rolling down towards the origin. Initial conditions and model parameters are tuned in such a way that the inflaton 
crosses the positive potential minimum and the barrier at the origin, falls in the negative 
potential minimum, bounces back, crosses a second time the origin, bounces back again (this time on the positive side of the potential), climbs again the central barrier and it has just enough inertia to spend $N\gtrsim 10$ $e$-folds 
 moving slowly in the neighborhood of the origin after the oscillation. 
 For ease of reading, we reproduce the classical dynamics of the inflaton in the left panel of fig.\,\ref{fig:Chaotic} where 
 the oscillation and the subsequent second phase of inflation is evident. 
 The latter ends when the inflaton falls again in the negative 
potential minimum, where it finally settles down.  
In this model, PBHs are produced during the second phase of inflation when the inflaton moves 
slowly close to the origin.
In the right panel of fig.\,\ref{fig:Chaotic}, we computed the time evolution of the $\eta$ for the same model parameters and 
initial conditions used in the left panel. The solid (dashed) line indicates positive (negative) values. 
The time profile of $\eta$ is considerably more complicated than the one expected in our class of models mostly 
because of the presence of the oscillating phase during which $\eta$ changes twice from large positive to large negative values 
before taking for $\Delta N \simeq 1$ $e$-folds the constant value $\eta \simeq 3$ close to the origin (where the potential is flat). 
This is a clear example of a model that differs from the class of models that we analyze in this paper.
However, we do not proceed further in the explicit analysis of this setup since one gets 
 $n_s \simeq 0.94$ and $r\simeq 0.32$ for, respectively, the spectral index and the tensor-to-scalar ratio.
These values are now grossly excluded by the most recent CMB measurements.

\item [{\it b)}] Consider the field redefinition 
$\mathcal{R} \to \mathcal{R}_n + f(\mathcal{R}_n)$ and the quadratic 
action for $\mathcal{R}$ 
\begin{align}
\mathcal{S}_2[\mathcal{R}] = \int d\tau d^3\vec{x}\,a^2\epsilon\left[
(\mathcal{R}^{\prime})^2 - (\partial\mathcal{R})^2 
\right]\,.
\end{align}
Since $f(\mathcal{R}_n)$ is a quadratic function in $\mathcal{R}_n$, the shift in the quadratic action 
generates cubic and quartic terms for $\mathcal{R}_n$ but not additional quadratic ones; schematically, we have
$\mathcal{S}_2[\mathcal{R}] \to \mathcal{S}_2[\mathcal{R}_n] + \tilde{\mathcal{S}}_3[\mathcal{R}_n] + O(\mathcal{R}_n^4)$, and  
the quadratic action for $\mathcal{R}_n$ has, therefore, exactly the same form as for $\mathcal{R}$. 
This is the reason why even if we formally work with $\mathcal{R}_n$ in eqs.\,(\ref{eq:MasterHint},\,\ref{eq:Red1}) 
we can still use the free-field solutions.
For the very same reason, the shift does not change $\mathcal{L}_3$ in eq.\,(\ref{eq:CubicMalda}) but it just relabels 
$\mathcal{R}\to \mathcal{R}_n$; however, the utility of the field redefinition introduced in\,\cite{Maldacena:2002vr} is that 
what cancels the last term in eq.\,(\ref{eq:CubicMalda}) is
the cubic correction $\tilde{\mathcal{S}}_3[\mathcal{R}_n]$ introduced before and originating  
from the quadratic action, as one can explicitly check.

\item [{\it c)}] Consider now some of the interaction terms in eq.\,(\ref{eq:MasterHint}) that we neglected. 
Let us focus on the interaction term
\begin{align}
\mathcal{H}^{(c_1)}_{\rm int}(\tau) \equiv -\int d^3\vec{x}a\epsilon^2\mathcal{R}\left(\mathcal{R}^{\prime}\right)^2\,,
\end{align}
which generates a contribution to the bispectrum of the form
\begin{align}\label{eq:FirstNeg}
B_{\mathcal{R}}^{(c_1)} = -4\Im
\mathcal{R}_{k_1}(\tau)
\mathcal{R}_{k_2}(\tau)
\mathcal{R}_{k_3}(\tau)
\int_{-\infty(1-i\epsilon)}^{\tau}d\bar{\tau}\,a(\bar{\tau})^2\epsilon(\bar{\tau})^2
\left[
\mathcal{R}^{*}_{k_1}(\bar{\tau})
\mathcal{R}^{\prime\,*}_{k_2}(\bar{\tau})\mathcal{R}^{\prime\,*}_{k_3}(\bar{\tau}) 
+ {\rm two\,perms}
\right]\,.
\end{align}
At late time the integrand eventually goes to zero because the perturbations settle to constant values. 
However, the integral is computed along the whole inflationary history, and, in particular, 
it crosses the USR phase during which the perturbations grow exponentially.
It is simple to see that the contribution to the bispectrum from eq.\,(\ref{eq:FirstNeg}) scales according to
\begin{align}\label{eq:FirstNeg2}
B_{\mathcal{R}}^{(c_1)} = \epsilon_I\left(\frac{H^4}{\epsilon_I^2 k_{\rm in}^6}\right)
\mathcal{F}_{c_1}(x_1,x_2,x_3)\,,
\end{align}
where $\mathcal{F}_{c_1}$ is a shape function which depends on the integral in eq.\,(\ref{eq:FirstNeg}).
As expected, eq.\,(\ref{eq:FirstNeg2}) pays a suppression factor $\epsilon_I \ll 1$ if compared to the 
contributions we included in our computation. For this reason, we neglected this contribution.
Of course, in more realistic models in which it is well possible to reach $\epsilon_I = O(1)$ during the pre-USR dynamics,
 the same argument can be invalidated.
To proceed further, one has to evaluate the time integral and compute the shape function explicitly.
In our model, we can split the integral in three parts.
In region\,I, the integral is analytical. The contribution at $\tau = -\infty(1-i\epsilon)$ vanishes thanks to the 
$\epsilon$ prescription (physically, this means that interaction are switched off in the Bunch-Davis vacuum), and only the contribution at $\tau=\tau_{\rm in}$ survives. 
In region\,II and region\,III, the integral can be computed numerically. 
In region\,III perturbations with $k \lesssim k_{\rm peak}$ settle exponentially fast to their final constant value, and 
the integral gives a small contribution since the integrand quickly vanishes. 
On the contrary, as discussed in ref.\,\cite{Ballesteros:2020qam}, 
perturbations with $k \gtrsim k_{\rm peak}$  
evolves in region\,III until their horizon crossing time $N_k > N_{\rm end}$ (after which they become constant) and they 
 contribute to the integral in eq.\,(\ref{eq:FirstNeg}). 
 We checked numerically that the contribution in eq.\,(\ref{eq:FirstNeg2}) to the bispectrum is largely subdominant.

Consider now the interaction term
\begin{align}
\mathcal{H}^{(c_2)}_{\rm int}(\tau) \equiv 2\int d^3\vec{x}\epsilon\mathcal{R}^{\prime}
(\partial_i\mathcal{R})(\partial_i\chi)\,,~~~~~\chi = \epsilon a^2\partial^{-2}\dot{\mathcal{R}}\,,
\end{align}
with $\partial^{-2}$ the inverse laplacian. The corresponding contribution to the bispectrum reads
\begin{align}\label{eq:SecondNeg}
B_{\mathcal{R}}^{(c_2)} = 4\Im
\mathcal{R}_{k_1}(\tau)
\mathcal{R}_{k_2}(\tau)
\mathcal{R}_{k_3}(\tau)
\int_{-\infty(1-i\epsilon)}^{\tau}d\bar{\tau}\,a(\bar{\tau})^2\epsilon(\bar{\tau})^2
\left[\frac{\vec{k}_2\cdot\vec{k}_3}{k_3^2}
\mathcal{R}^{\prime\,*}_{k_1}(\bar{\tau})
\mathcal{R}^{*}_{k_2}(\bar{\tau})\mathcal{R}^{\prime\,*}_{k_3}(\bar{\tau}) 
+ {\rm six\,perms}
\right]\,,
\end{align}
where the scalar products can be expressed in terms of the magnitudes $k_i$ 
using the conservation enforced by the triangle identity, for instance
$\vec{k}_2 + \vec{k}_3 = -\vec{k}_1~~~\Longrightarrow~~~ \vec{k}_2\cdot\vec{k}_3 = (k_1^2 - k_2^2 - k_3^2)/2$.
As before, eq.\,(\ref{eq:SecondNeg}) features the scaling
\begin{align}\label{eq:SecondNeg2}
B_{\mathcal{R}}^{(c_2)} = \epsilon_I\left(\frac{H^4}{\epsilon_I^2 k_{\rm in}^6}\right)
\mathcal{F}_{c_2}(x_1,x_2,x_3)\,,
\end{align}
with the extra suppression given by $\epsilon_I \ll 1$. The explicit computation of $\mathcal{F}_{c_2}(x_1,x_2,x_3)$ is not
conceptually different from that of $\mathcal{F}_{c_1}(x_1,x_2,x_3)$ and leads to comparable results.

The few examples considered here confirm that the bispectrum computed from the time variation 
of $\eta$ captures the leading contribution coming from the interaction Hamiltonian.
We also showed that organizing cubic interactions according to eq.\,(\ref{eq:MasterHint}) 
is indeed useful since it allows to 
exploit a power counting in terms of the Hubble parameter $\epsilon$ (assuming the latter 
remains small during the inflationary dynamics).   

\item [{\it d)}] A comment on eq.\,(\ref{eq:Red1}) is needed. 
As anticipated, the three-point function is meant to be evaluated in the vacuum state of the full interacting theory. 
More formally, one should write
\begin{align}
\langle \hat{\mathcal{R}}&(\tau,\vec{k}_1)
 \hat{\mathcal{R}}(\tau,\vec{k}_2)
 \hat{\mathcal{R}}(\tau,\vec{k}_3)\rangle = 
 \langle\Omega(\tau)| \hat{\mathcal{R}}_I(\tau,\vec{k}_1)
 \hat{\mathcal{R}}_I(\tau,\vec{k}_2)
 \hat{\mathcal{R}}_I(\tau,\vec{k}_3)|\Omega(\tau)\rangle\,,
\end{align}
with the vacuum of the interacting theory given by $|\Omega(\tau)\rangle = {\rm T}e^{-i\int_{-\infty(1-i\epsilon)}^{\tau}d\bar{\tau}a(\bar{\tau})H_{\rm int}(\bar{\tau})}|0\rangle$
where T is the time-ordering operator.
 At initial time, when interactions are turned off (thanks to the $i\epsilon$ prescription) the initial vacuum reduces to the Bunch-Davis vacuum $|0\rangle$. 
Notice that the three-point function
$\langle \hat{\mathcal{R}}(\tau,\vec{k}_1)
 \hat{\mathcal{R}}(\tau,\vec{k}_2)
 \hat{\mathcal{R}}(\tau,\vec{k}_3)\rangle$ 
 would vanish if evaluated in $|0\rangle$ (since it contains an odd numbers of operators) so that its non-zero value is a genuine imprint 
 left by interactions.
 
 The same comment applies to all terms on the right-hand side of eq.\,(\ref{eq:Red1}) which should be 
 evaluated in the vacuum state of the full interacting theory. 
 This is indeed the way in which we computed the first contribution on the right-hand side, $\langle \hat{\mathcal{R}}_n(\tau,\vec{k}_1)
 \hat{\mathcal{R}}_n(\tau,\vec{k}_2)
 \hat{\mathcal{R}}_n(\tau,\vec{k}_3)\rangle$. Consider now the other term
\begin{align}
\frac{\epsilon_2}{4}\left[
\langle \hat{\mathcal{R}}_n^2(\tau,\vec{k}_1)
 \hat{\mathcal{R}}_n(\tau,\vec{k}_2)
 \hat{\mathcal{R}}_n(\tau,\vec{k}_3) \rangle + {\rm two\,permutations}
 \right]\,.
\end{align}
We computed this term in the vacuum of the free theory.
The difference is that in this case the operator is quartic in $\hat{\mathcal{R}}_n$, and 
we have a non-zero leading contribution already in the vacuum $|0\rangle$ (the first term in the Dyson series 
of $|\Omega(\tau)\rangle$). This is the contribution that has been computed in eq.\,(\ref{eq:RedCon}). 
Further corrections to this contribution can be discussed (dubbed ``higher order terms'' in eq.\,(\ref{eq:Red1})). 
One possibility is given by the term
\begin{align}
\frac{\epsilon_2^3}{64}\left[
\langle \hat{\mathcal{R}}_n^2(\tau,\vec{k}_1)
 \hat{\mathcal{R}}_n^2(\tau,\vec{k}_2)
 \hat{\mathcal{R}}_n^2(\tau,\vec{k}_3) \rangle + {\rm permutations}
 \right]\,,
\end{align}
which is also non-zero evaluated on the free-field vacuum. 
This term generates a contribution to the bispectrum which has the form
\begin{align}\label{eq:d1}
B_{\mathcal{R}}^{(d_1)} =  -\frac{\eta_{\rm III}^3}{8}\int\frac{d\Omega_q}{(4\pi)}
\int\frac{dq}{q}\frac{(2\pi^2)^2}{(k_1+q)^3(k_3-q)^3}\mathcal{P}_{\mathcal{R}}(k_1+q)
\mathcal{P}_{\mathcal{R}}(q)
\mathcal{P}_{\mathcal{R}}(k_3-q) + {\rm permutations}\,.
\end{align}
If we approximate $\mathcal{P}_{\mathcal{R}}(q)$ as a spiky power spectrum so that the integration over $q$ picks up 
only the contribution at $k_{\rm peak}$, a quick estimate of the previous term compared with the 
one in eq.\,(\ref{eq:Bi3}) would be
\begin{align}\label{eq:Suppr}
\frac{B_{\mathcal{R}}^{(d_1)}}{B^{({\rm red})}_{\mathcal{R}} } = O(\eta_{\rm III}^2\mathcal{P}_{\mathcal{R}}(k_{\rm peak}))\,,
~~~~{\rm with}~~\mathcal{P}_{\mathcal{R}}(k_{\rm peak}) \sim 10^{-2}\,.
\end{align}
The point to make here is that terms like the one in eq.\,(\ref{eq:d1}) are usually tossed away in slow-roll inflation models 
since highly suppressed by $\mathcal{P}_{\mathcal{R}} \sim 10^{-9}$. In the presence of USR, 
one can not take lightly the same conclusion 
since the power spectrum can be much larger than the value suggested by CMB measurements. However, 
the suppression in eq.\,(\ref{eq:Suppr}) might still allow to neglect these contributions.
We can compute numerically the ratio in eq.\,(\ref{eq:Suppr}) and check our estimate.
\begin{figure}[!htb!]
\begin{center}
$$\includegraphics[width=.42\textwidth]{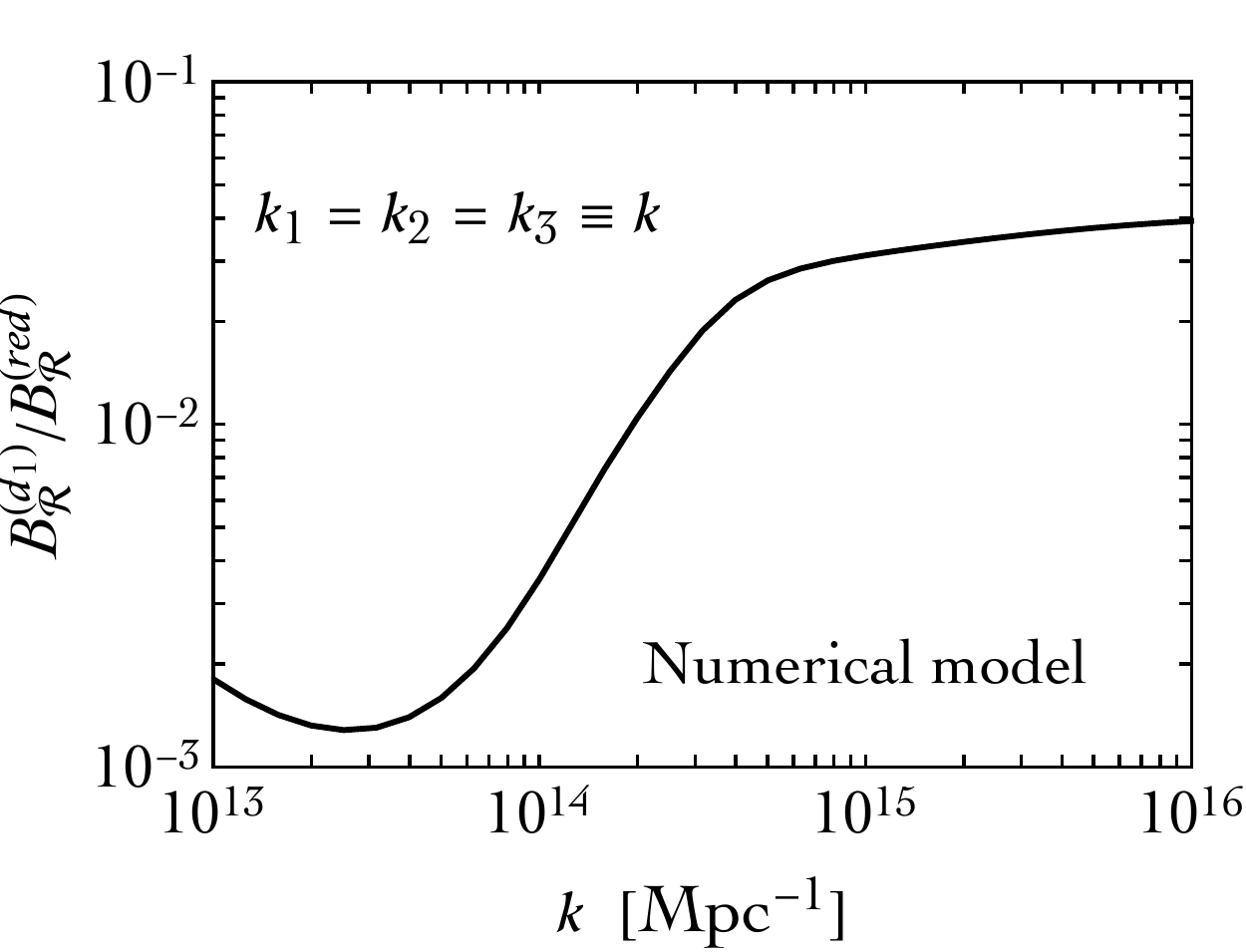}
\qquad\includegraphics[width=.42\textwidth]{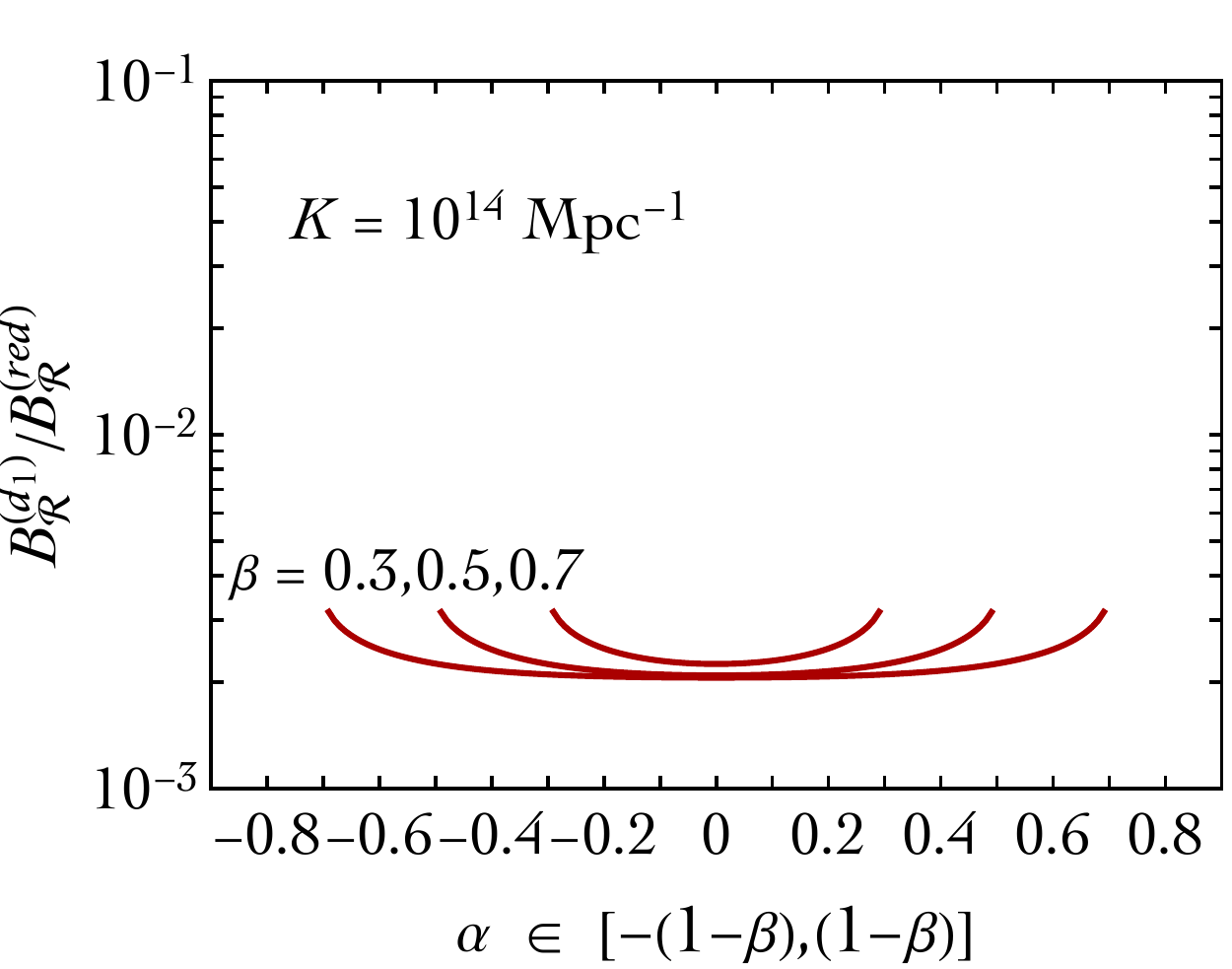}$$
\caption{\em \label{fig:BCompa} 
Ratio in eq.\,(\ref{eq:Suppr}) computed numerically. 
We consider the numerical model, and we fix $\eta_{\rm III} = \eta_0 = -0.75$. 
In the left panel, we show the ratio in the equilateral configuration. 
In the right panel, we use the parametrization in eq.\,(\ref{eq:IntegrationBispectrum}) for fixed $K$ and three different choices 
for $\beta$.
 }
\end{center}
\end{figure}
It is worth doing this check explicitly since this is the largest correction to our approximation. 
We show our results, computed in the context of the numerical model, in fig.\,\ref{fig:BCompa}. 
The suppression estimated in eq.\,(\ref{eq:Suppr}) is confirmed.

\item [{\it e)}]  Consider the effect of the field redefinition on the four-point correlator
 $\langle \hat{\mathcal{R}}(\tau,\vec{k}_1)
 \hat{\mathcal{R}}(\tau,\vec{k}_2)
 \hat{\mathcal{R}}(\tau,\vec{k}_3)\hat{\mathcal{R}}(\tau,\vec{k}_4)\rangle$.
In Fourier space, the field redefinition generates, at order $O(\epsilon_2^2)$, the following 6 terms 
\begin{align}
\frac{\epsilon_2^2}{16}\bigg[&
\langle \hat{\mathcal{R}}_n(\tau,\vec{k}_1)
 \hat{\mathcal{R}}_n(\tau,\vec{k}_2)
 \hat{\mathcal{R}}^2_n(\tau,\vec{k}_3)\hat{\mathcal{R}}^2_n(\tau,\vec{k}_4) \rangle
 +
 \langle \hat{\mathcal{R}}^2_n(\tau,\vec{k}_1)
 \hat{\mathcal{R}}^2_n(\tau,\vec{k}_2)
 \hat{\mathcal{R}}_n(\tau,\vec{k}_3)\hat{\mathcal{R}}_n(\tau,\vec{k}_4) \rangle \nn \\&+
 \langle \hat{\mathcal{R}}^2_n(\tau,\vec{k}_1)
 \hat{\mathcal{R}}_n(\tau,\vec{k}_2)
 \hat{\mathcal{R}}^2_n(\tau,\vec{k}_3)\hat{\mathcal{R}}_n(\tau,\vec{k}_4) \rangle+
 \langle \hat{\mathcal{R}}^2_n(\tau,\vec{k}_1)
 \hat{\mathcal{R}}_n(\tau,\vec{k}_2)
 \hat{\mathcal{R}}_n(\tau,\vec{k}_3)\hat{\mathcal{R}}^2_n(\tau,\vec{k}_4) \rangle \nn \\&+
 \langle \hat{\mathcal{R}}_n(\tau,\vec{k}_1)
 \hat{\mathcal{R}}^2_n(\tau,\vec{k}_2)
 \hat{\mathcal{R}}^2_n(\tau,\vec{k}_3)\hat{\mathcal{R}}_n(\tau,\vec{k}_4) \rangle+
 \langle \hat{\mathcal{R}}_n(\tau,\vec{k}_1)
 \hat{\mathcal{R}}^2_n(\tau,\vec{k}_2)
 \hat{\mathcal{R}}_n(\tau,\vec{k}_3)\hat{\mathcal{R}}^2_n(\tau,\vec{k}_4) \rangle
 \bigg]\,.\label{eq:RedQuartic}
\end{align}
Each one of them can be computed in the vacuum of the free theory. 
Consider for simplicity only the first one, which we write in the form
\begin{align}
\frac{\epsilon_2^2}{16}\int\frac{d^3\vec{q}_1}{(2\pi)^3}\frac{d^3\vec{q}_2}{(2\pi)^3}
\langle 
\hat{\mathcal{R}}_I(\tau,\vec{k}_1)
\hat{\mathcal{R}}_I(\tau,\vec{k}_2)
\hat{\mathcal{R}}_I(\tau,\vec{q}_1)
\hat{\mathcal{R}}_I(\tau,\vec{k}_3-\vec{q}_1)
\hat{\mathcal{R}}_I(\tau,\vec{q}_2)
\hat{\mathcal{R}}_I(\tau,\vec{k}_4 -\vec{q}_2) \rangle\,.
\end{align}
It is easy to see that there are 8 possible contractions, and one obtains
\begin{align}
\frac{\epsilon_2^2}{8}\left\{
\Delta(k_1)\Delta(k_2)\left[
\Delta(|\vec{k}_1+\vec{k}_3|)+\Delta(|\vec{k}_1+\vec{k}_4|) +
\Delta(|\vec{k}_2+\vec{k}_3|) + \Delta(|\vec{k}_2+\vec{k}_4|)
\right]
\right\}\,,
\end{align}
where the four terms in the curly brackets counted twice. 
Putting all together, eq.\,(\ref{eq:RedQuartic}) reads 
\begin{align}\label{eq:RedTri}
\frac{\epsilon_2^2}{8}\left[
\Delta(k_1)\Delta(k_2)\Delta(|\vec{k}_1+\vec{k}_4|) + {\rm 23\,permutations}
\right]\,,
\end{align}
where the $4!=24$ terms in the square brackets counts the permutations of the indices $_{1234}$. 
This is the same structure that we introduced for the reduced trispectrum in eq.\,(\ref{eq:RedcuedTrispectrum}). 
If we identify $\epsilon_2/4 = 3f_{\rm NL}/5$, the coefficient in front of eq.\,(\ref{eq:RedTri}) becomes 
$18f_{\rm NL}^2/25$.

\end{itemize}

\newpage

%%%%%%%%%%%%%%%%%%%%%%%%%%%%
%%%%%%%%%%%%%%%%%%%%%%%%%%%%%%%%%%%%%%

%%%%%%%%%%%%%%%%%%%%%%%%%%%%%%%%%%%%%%%%%%%%%%


\begin{thebibliography}{99}

\bibitem{Zelda}
Ya.~B.~Zel'dovich and I.~D.~Novikov,
``{\it The Hypothesis of Cores Retarded during Expansion and the Hot Cosmological Model,}''
Astronomicheskii Zhurnal, \textbf{43} (1966), 758.
     
%\cite{Hawking:1971ei,Carr:1974nx}
\bibitem{Hawking:1971ei}
S.~Hawking,
``{\it Gravitationally collapsed objects of very low mass,}''
Mon. Not. Roy. Astron. Soc. \textbf{152} (1971), 75
%612 citations counted in INSPIRE as of 26 May 2020   

%\cite{Carr:1974nx}
\bibitem{Carr:1974nx}
B.~J.~Carr and S.~Hawking,
``{\it Black holes in the early Universe,}''
Mon. Not. Roy. Astron. Soc. \textbf{168} (1974), 399-415
%710 citations counted in INSPIRE as of 26 May 2020  
     
 %\cite{Maldacena:2002vr,Weinberg:2005vy}
\bibitem{Maldacena:2002vr}
J.~M.~Maldacena,
``{\it Non-Gaussian features of primordial fluctuations in single field inflationary models,}''
JHEP \textbf{05} (2003), 013
%doi:10.1088/1126-6708/2003/05/013
[\hhref{arXiv:astro-ph/0210603} [astro-ph]].
%2055 citations counted in INSPIRE as of 12 Apr 2020    
  
%\cite{Leach:2000yw,Leach:2001zf}
\bibitem{Leach:2000yw}
S.~M.~Leach and A.~R.~Liddle,
``{\it Inflationary perturbations near horizon crossing,}''
Phys. Rev. D \textbf{63} (2001), 043508
%doi:10.1103/PhysRevD.63.043508
[\hhref{arXiv:astro-ph/0010082}].
%101 citations counted in INSPIRE as of 16 Jul 2020

%\cite{Leach:2001zf}
\bibitem{Leach:2001zf}
S.~M.~Leach, M.~Sasaki, D.~Wands and A.~R.~Liddle,
``{\it Enhancement of superhorizon scale inflationary curvature perturbations,}''
Phys. Rev. D \textbf{64} (2001), 023512
%doi:10.1103/PhysRevD.64.023512
[\hhref{arXiv:astro-ph/0101406}].
%142 citations counted in INSPIRE as of 16 Jul 2020  

%\cite{Tsamis:2003px}
\bibitem{Tsamis:2003px}
N.~C.~Tsamis and R.~P.~Woodard,
``{\it Improved estimates of cosmological perturbations,}''
Phys. Rev. D \textbf{69} (2004), 084005
%doi:10.1103/PhysRevD.69.084005
[\hhref{arXiv:astro-ph/0307463} [astro-ph]].
%88 citations counted in INSPIRE as of 23 Feb 2021

%\cite{Kinney:2005vj}
\bibitem{Kinney:2005vj}
W.~H.~Kinney,
``{\it Horizon crossing and inflation with large eta,}''
Phys. Rev. D \textbf{72} (2005), 023515
%doi:10.1103/PhysRevD.72.023515
[\hhref{arXiv:gr-qc/0503017} [gr-qc]].
%203 citations counted in INSPIRE as of 23 Feb 2021

%\cite{Kinney:1997ne}
\bibitem{Kinney:1997ne}
W.~H.~Kinney,
``{\it A Hamilton-Jacobi approach to nonslow roll inflation,}''
Phys. Rev. D \textbf{56} (1997), 2002-2009
%doi:10.1103/PhysRevD.56.2002
[\hhref{arXiv:hep-ph/9702427} [hep-ph]].
%94 citations counted in INSPIRE as of 23 Feb 2021

     
%\cite{Ivanov:1994pa,Saito:2008em}
\bibitem{Ivanov:1994pa}
P.~Ivanov, P.~Naselsky and I.~Novikov,
``{\it Inflation and primordial black holes as dark matter,}''
Phys.\ Rev.\ D {\bf 50} (1994) 7173.
% doi:10.1103/PhysRevD.50.7173
%%CITATION = doi:10.1103/PhysRevD.50.7173;%%
%161 citations counted in INSPIRE as of 19 Dec 2019
  
%\cite{Saito:2008em}
\bibitem{Saito:2008em}
R.~Saito, J.~Yokoyama and R.~Nagata,
``{\it Single-field inflation, anomalous enhancement of superhorizon fluctuations, and non-Gaussianity in primordial black hole formation,}''
JCAP \textbf{06} (2008), 024
%doi:10.1088/1475-7516/2008/06/024
[\hhref{arXiv:0804.3470} [astro-ph]].
%73 citations counted in INSPIRE as of 17 Nov 2020 

\bibitem{Starobinsky:1992ts}
A.~A.~Starobinsky,
%``Spectrum of adiabatic perturbations in the universe when there are singularities in the inflation potential,''
JETP Lett. \textbf{55} (1992), 489-494

  
     
%\cite{Green:2004wb}
\bibitem{Green:2004wb}
A.~M.~Green, A.~R.~Liddle, K.~A.~Malik and M.~Sasaki,
``{\it A New calculation of the mass fraction of primordial black holes,}''
Phys.\ Rev.\ D \textbf{70} (2004), 041502
%doi:10.1103/PhysRevD.70.041502
[\hhref{arXiv:astro-ph/0403181} [astro-ph]].
%91 citations counted in INSPIRE as of 06 Apr 2020  

%\cite{Germani:2019zez}
\bibitem{Germani:2019zez}
C.~Germani and R.~K.~Sheth,
``{\it Nonlinear statistics of primordial black holes from Gaussian curvature perturbations,}''
Phys. Rev. D \textbf{101} (2020) no.6, 063520
%doi:10.1103/PhysRevD.101.063520
[\hhref{arXiv:1912.07072} [astro-ph.CO]].
%12 citations counted in INSPIRE as of 21 May 2020

%\cite{DeLuca:2019qsy}
\bibitem{DeLuca:2019qsy}
V.~De Luca, G.~Franciolini, A.~Kehagias, M.~Peloso, A.~Riotto and C.~\"Unal,
``{\it The Ineludible non-Gaussianity of the Primordial Black Hole Abundance,}''
JCAP \textbf{07} (2019), 048
%doi:10.1088/1475-7516/2019/07/048
[\hhref{arXiv:1904.00970} [astro-ph.CO]].
%32 citations counted in INSPIRE as of 19 May 2020

%\cite{Yoo:2018kvb,Kawasaki:2019mbl,Young:2019yug}
\bibitem{Yoo:2018kvb}
C.~M.~Yoo, T.~Harada, J.~Garriga and K.~Kohri,
``{\it Primordial black hole abundance from random Gaussian curvature perturbations and a local density threshold,}''
PTEP \textbf{2018} (2018) no.12, 123E01
%doi:10.1093/ptep/pty120
[\hhref{arXiv:1805.03946} [astro-ph.CO]].
%55 citations counted in INSPIRE as of 21 May 2020

%\cite{Kawasaki:2019mbl,Young:2019yug}
\bibitem{Kawasaki:2019mbl}
M.~Kawasaki and H.~Nakatsuka,
``{\it Effect of nonlinearity between density and curvature perturbations on the primordial black hole formation,}''
Phys. Rev. D \textbf{99} (2019) no.12, 123501
%doi:10.1103/PhysRevD.99.123501
[\hhref{arXiv:1903.02994} [astro-ph.CO]].
%23 citations counted in INSPIRE as of 21 May 2020

%\cite{Young:2019yug}
\bibitem{Young:2019yug}
S.~Young, I.~Musco and C.~T.~Byrnes,
``{\it Primordial black hole formation and abundance: contribution from the non-linear relation between the density and curvature perturbation,}''
JCAP \textbf{11} (2019), 012
%doi:10.1088/1475-7516/2019/11/012
[\hhref{arXiv:1904.00984} [astro-ph.CO]].
%32 citations counted in INSPIRE as of 21 May 2020

%\cite{Kehagias:2019eil}
\bibitem{Kehagias:2019eil}
A.~Kehagias, I.~Musco and A.~Riotto,
``{\it Non-Gaussian Formation of Primordial Black Holes: Effects on the Threshold,}''
JCAP \textbf{12} (2019), 029
%doi:10.1088/1475-7516/2019/12/029
[\hhref{arXiv:1906.07135} [astro-ph.CO]].
%14 citations counted in INSPIRE as of 21 May 2020

%\cite{Yoo:2019pma}
\bibitem{Yoo:2019pma}
C.~M.~Yoo, J.~O.~Gong and S.~Yokoyama,
``{\it Abundance of primordial black holes with local non-Gaussianity in peak theory,}''
JCAP \textbf{09} (2019), 033
%doi:10.1088/1475-7516/2019/09/033
[\hhref{arXiv:1906.06790} [astro-ph.CO]].
%12 citations counted in INSPIRE as of 28 May 2020

%\cite{Yoo:2020dkz}
\bibitem{Yoo:2020dkz}
C.~M.~Yoo, T.~Harada, S.~Hirano and K.~Kohri,
``{\it Abundance of Primordial Black Holes in Peak Theory for an Arbitrary Power Spectrum,}''
[arXiv:2008.02425 [astro-ph.CO]].
%5 citations counted in INSPIRE as of 10 Dec 2020

%\cite{Musco:2020jjb}
\bibitem{Musco:2020jjb}
I.~Musco, V.~De Luca, G.~Franciolini and A.~Riotto,
%``The Threshold for Primordial Black Hole Formation: a Simple Analytic Prescription,''
[arXiv:2011.03014 [astro-ph.CO]].
%3 citations counted in INSPIRE as of 10 Dec 2020

\bibitem{Young:2019osy}
S.~Young,
``{\it The primordial black hole formation criterion re-examined: Parametrisation, timing and the choice of window function,}''
Int. J. Mod. Phys. D \textbf{29} (2019) no.02, 2030002
%doi:10.1142/S0218271820300025
[\hhref{arXiv:1905.01230} [astro-ph.CO]].
%41 citations counted in INSPIRE as of 21 Jun 2021

LaTeX (EU)
  
%\cite{Musco:2004ak}
\bibitem{Musco:2004ak}
I.~Musco, J.~C.~Miller and L.~Rezzolla,
``{\it Computations of primordial black hole formation,}''
Class. Quant. Grav. \textbf{22} (2005), 1405-1424
%doi:10.1088/0264-9381/22/7/013
[\hhref{arXiv:gr-qc/0412063} [gr-qc]].
%134 citations counted in INSPIRE as of 21 May 2020


%\cite{Young:2014ana,Germani:2018jgr,DeLuca:2019qsy}
\bibitem{Young:2014ana}
S.~Young, C.~T.~Byrnes and M.~Sasaki,
``{\it Calculating the mass fraction of primordial black holes,}''
JCAP \textbf{07} (2014), 045
%doi:10.1088/1475-7516/2014/07/045
[\hhref{arXiv:1405.7023} [gr-qc]].
%89 citations counted in INSPIRE as of 21 May 2020 

%\cite{Bardeen:1985tr}
\bibitem{Bardeen:1985tr}
J.~M.~Bardeen, J.~Bond, N.~Kaiser and A.~Szalay,
``{\it The Statistics of Peaks of Gaussian Random Fields,}''
Astrophys. J. \textbf{304} (1986), 15-61
%doi:10.1086/164143
%2396 citations counted in INSPIRE as of 21 May 2020

%\cite{Germani:2018jgr,Musco:2018rwt}
\bibitem{Germani:2018jgr}
C.~Germani and I.~Musco,
``{\it Abundance of Primordial Black Holes Depends on the Shape of the Inflationary Power Spectrum,}''
Phys. Rev. Lett. \textbf{122} (2019) no.14, 141302
%doi:10.1103/PhysRevLett.122.141302
[\hhref{arXiv:1805.04087} [astro-ph.CO]].
%75 citations counted in INSPIRE as of 21 May 2020 

%\cite{Musco:2018rwt}
\bibitem{Musco:2018rwt}
I.~Musco,
``{\it Threshold for primordial black holes: Dependence on the shape of the cosmological perturbations,}''
Phys. Rev. D \textbf{100} (2019) no.12, 123524
%doi:10.1103/PhysRevD.100.123524
[\hhref{arXiv:1809.02127} [gr-qc]].
%41 citations counted in INSPIRE as of 23 May 2020


%\cite{Neilsen:1998qc}
\bibitem{Neilsen:1998qc}
D.~W.~Neilsen and M.~W.~Choptuik,
``{\it Critical phenomena in perfect fluids,}''
Class. Quant. Grav. \textbf{17} (2000), 761-782
%doi:10.1088/0264-9381/17/4/303
[\hhref{arXiv:gr-qc/9812053} [gr-qc]].
%65 citations counted in INSPIRE as of 05 Jun 2020

%\cite{Kalaja:2019uju}
\bibitem{Kalaja:2019uju}
A.~Kalaja, N.~Bellomo, N.~Bartolo, D.~Bertacca, S.~Matarrese, I.~Musco, A.~Raccanelli and L.~Verde,
``{\it From Primordial Black Holes Abundance to Primordial Curvature Power Spectrum (and back),}''
JCAP \textbf{10} (2019), 031
%doi:10.1088/1475-7516/2019/10/031
[\hhref{arXiv:1908.03596} [astro-ph.CO]].
%22 citations counted in INSPIRE as of 05 Jun 2020

%\cite{inprep}
%\bibitem{inprep}
%\cite{Riccardi:2021rlf}
\bibitem{Riccardi:2021rlf}
F.~Riccardi, M.~Taoso and A.~Urbano,
``{\it Solving peak theory in the presence of local non-gaussianities,}''
[arXiv:2102.04084 [astro-ph.CO]].
%6 citations counted in INSPIRE as of 27 Jul 2021
%F.~Riccardi, M.~Taoso, A.~Urbano,
%``{\it Solving peak theory in the presence of local non-gaussianities,}''
%to appear.


%\cite{Ballesteros:2020qam}
\bibitem{Ballesteros:2020qam} 
  G.~Ballesteros, J.~Rey, M.~Taoso and A.~Urbano,
  ``{\it Primordial black holes as dark matter and gravitational waves from single-field polynomial inflation,}''
  JCAP \textbf{07}, 025 (2020)
  \hhref{arXiv:2001.08220} [astro-ph.CO].
  %%CITATION = ARXIV:2001.08220;%%
  
%\cite{Ballesteros:2020sre}
\bibitem{Ballesteros:2020sre}
G.~Ballesteros, J.~Rey, M.~Taoso and A.~Urbano,
`{\it`Stochastic inflationary dynamics beyond slow-roll and consequences for primordial black hole formation,}''
JCAP \textbf{08}, 043 (2020)
%doi:10.1088/1475-7516/2020/08/043
[\hhref{arXiv:2006.14597} [astro-ph.CO]].


%\cite{Carr:2009jm}
\bibitem{Carr:2009jm}
B.~Carr, K.~Kohri, Y.~Sendouda and J.~Yokoyama,
``{\it New cosmological constraints on primordial black holes,}''
Phys. Rev. D \textbf{81} (2010), 104019
%doi:10.1103/PhysRevD.81.104019
[\hhref{arXiv:0912.5297} [astro-ph.CO]].
%580 citations counted in INSPIRE as of 07 Jun 2020

%\cite{Laha:2019ssq}
\bibitem{Laha:2019ssq}
R.~Laha,
``{\it Primordial Black Holes as a Dark Matter Candidate Are Severely Constrained by the Galactic Center 511 keV $\gamma$ -Ray Line,}''
Phys. Rev. Lett. \textbf{123} (2019) no.25, 251101
%doi:10.1103/PhysRevLett.123.251101
[\hhref{arXiv:1906.09994} [astro-ph.HE]].
%28 citations counted in INSPIRE as of 07 Jun 2020

%\cite{Niikura:2017zjd}
\bibitem{Niikura:2017zjd}
H.~Niikura, M.~Takada, N.~Yasuda, R.~H.~Lupton, T.~Sumi, S.~More, T.~Kurita, S.~Sugiyama, A.~More, M.~Oguri and M.~Chiba,
``{\it Microlensing constraints on primordial black holes with Subaru/HSC Andromeda observations,}''
Nature Astron. \textbf{3} (2019) no.6, 524-534
%doi:10.1038/s41550-019-0723-1
[\hhref{arXiv:1701.02151} [astro-ph.CO]].
%195 citations counted in INSPIRE as of 07 Jun 2020

%\cite{Katz:2018zrn}
\bibitem{Katz:2018zrn}
A.~Katz, J.~Kopp, S.~Sibiryakov and W.~Xue,
``{\it Femtolensing by Dark Matter Revisited,}''
JCAP \textbf{12} (2018), 005
%doi:10.1088/1475-7516/2018/12/005
[\hhref{arXiv:1807.11495} [astro-ph.CO]].
%84 citations counted in INSPIRE as of 07 Jun 2020

%\cite{Capela:2013yf}
\bibitem{Capela:2013yf}
F.~Capela, M.~Pshirkov and P.~Tinyakov,
``{\it Constraints on primordial black holes as dark matter candidates from capture by neutron stars,}''
Phys. Rev. D \textbf{87} (2013) no.12, 123524
%doi:10.1103/PhysRevD.87.123524
[\hhref{arXiv:1301.4984} [astro-ph.CO]].
%122 citations counted in INSPIRE as of 06 Jun 2020

%\cite{Pani:2014rca,Capela:2014qea,Defillon:2014wla}
\bibitem{Pani:2014rca}
P.~Pani and A.~Loeb,
``{\it Tidal capture of a primordial black hole by a neutron star: implications for constraints on dark matter,}''
JCAP \textbf{06} (2014), 026
%doi:10.1088/1475-7516/2014/06/026
[\hhref{arXiv:1401.3025} [astro-ph.CO]].
%43 citations counted in INSPIRE as of 07 Jun 2020

%\cite{Capela:2014qea,Defillon:2014wla}
\bibitem{Capela:2014qea}
F.~Capela, M.~Pshirkov and P.~Tinyakov,
``{\it A comment on ``Exclusion of the remaining mass window for primordial black holes ...'', arXiv:1401.3025,}''
[\hhref{arXiv:1402.4671} [astro-ph.CO]].
%13 citations counted in INSPIRE as of 07 Jun 2020

%\cite{Defillon:2014wla}
\bibitem{Defillon:2014wla}
G.~Defillon, E.~Granet, P.~Tinyakov and M.~H.~Tytgat,
``{\it Tidal capture of primordial black holes by neutron stars,}''
Phys. Rev. D \textbf{90} (2014) no.10, 103522
%doi:10.1103/PhysRevD.90.103522
[\hhref{arXiv:1409.0469} [gr-qc]].
%18 citations counted in INSPIRE as of 07 Jun 2020

%\cite{Montero-Camacho:2019jte}
\bibitem{Montero-Camacho:2019jte}
P.~Montero-Camacho, X.~Fang, G.~Vasquez, M.~Silva and C.~M.~Hirata,
``{\it Revisiting constraints on asteroid-mass primordial black holes as dark matter candidates,}''
JCAP \textbf{08} (2019), 031
%doi:10.1088/1475-7516/2019/08/031
[\hhref{arXiv:1906.05950} [astro-ph.CO]].
%80 citations counted in INSPIRE as of 23 Feb 2021

%\cite{Graham:2015apa}
\bibitem{Graham:2015apa}
P.~W.~Graham, S.~Rajendran and J.~Varela,
``{\it Dark Matter Triggers of Supernovae,}''
Phys. Rev. D \textbf{92} (2015) no.6, 063007
%doi:10.1103/PhysRevD.92.063007
[\hhref{arXiv:1505.04444} [hep-ph]].
%120 citations counted in INSPIRE as of 06 Jun 2020

%\cite{Akrami:2019izv}
\bibitem{Akrami:2019izv}
Y.~Akrami \textit{et al.} [Planck],
``{\it Planck 2018 results. IX. Constraints on primordial non-Gaussianity,}''
[\hhref{arXiv:1905.05697} [astro-ph.CO]].
%95 citations counted in INSPIRE as of 05 Apr 2020


%\cite{Matarrese:1986et}
\bibitem{Matarrese:1986et}
S.~Matarrese, F.~Lucchin and S.~A.~Bonometto,
``{\it A path integral approach to large scale matter distribution originated by non-gaussian fluctuations,}''
Astrophys.\ J.\  \textbf{310} (1986), L21-L26
%doi:10.1086/184774
%100 citations counted in INSPIRE as of 05 Apr 2020

%\cite{Franciolini:2018vbk}
\bibitem{Franciolini:2018vbk}
G.~Franciolini, A.~Kehagias, S.~Matarrese and A.~Riotto,
``{\it Primordial Black Holes from Inflation and non-Gaussianity,}''
JCAP \textbf{03} (2018), 016
%doi:10.1088/1475-7516/2018/03/016
[arXiv:1801.09415 [astro-ph.CO]].


%\cite{Cai:2017bxr}
\bibitem{Cai:2017bxr}
Y.~Cai, X.~Chen, M.~H.~Namjoo, M.~Sasaki, D.~Wang and Z.~Wang,
``{\it Revisiting non-Gaussianity from non-attractor inflation models,}''
JCAP \textbf{05} (2018), 012
%doi:10.1088/1475-7516/2018/05/012
[\hhref{arXiv:1712.09998} [astro-ph.CO]].
%28 citations counted in INSPIRE as of 13 Apr 2020

%\cite{Passaglia:2018ixg}
\bibitem{Passaglia:2018ixg}
S.~Passaglia, W.~Hu and H.~Motohashi,
``{\it Primordial black holes and local non-Gaussianity in canonical inflation,}''
Phys. Rev. D \textbf{99} (2019) no.4, 043536
%doi:10.1103/PhysRevD.99.043536
[\hhref{arXiv:1812.08243} [astro-ph.CO]].
%24 citations counted in INSPIRE as of 08 May 2020

%\cite{Komatsu:2010hc}
\bibitem{Komatsu:2010hc}
E.~Komatsu,
``{\it Hunting for Primordial Non-Gaussianity in the Cosmic Microwave Background,}''
Class. Quant. Grav. \textbf{27} (2010), 124010
%doi:10.1088/0264-9381/27/12/124010
[\hhref{arXiv:1003.6097} [astro-ph.CO]].
%207 citations counted in INSPIRE as of 06 May 2020

%\cite{Boubekeur:2005fj}
\bibitem{Boubekeur:2005fj}
L.~Boubekeur and D.~Lyth,
``{\it Detecting a small perturbation through its non-Gaussianity,}''
Phys. Rev. D \textbf{73} (2006), 021301
%doi:10.1103/PhysRevD.73.021301
[\hhref{arXiv:astro-ph/0504046} [astro-ph]].
%185 citations counted in INSPIRE as of 06 May 2020


\bibitem{Kendall}
A.~Stuart and K.~Ord,
``{\it Kendall's Advanced Theory of Statistics. Volume I: Distribution Theory,}''
Oxford University Press, 1987.

\bibitem{kolassa}
 J.~E.~Kolassa,
``{\it Series Approximation Methods in Statistics,}''
Lecture Notes in Statistics, Springer, 2006.


%\cite{Harada:2015yda}
\bibitem{Harada:2015yda}
T.~Harada, C.~M.~Yoo, T.~Nakama and Y.~Koga,
``{\it Cosmological long-wavelength solutions and primordial black hole formation,}''
Phys. Rev. D \textbf{91} (2015) no.8, 084057
%doi:10.1103/PhysRevD.91.084057
[\hhref{arXiv:1503.03934} [gr-qc]].

%\cite{Ballesteros:2017fsr}
\bibitem{Ballesteros:2017fsr}
G.~Ballesteros and M.~Taoso,
``{\it Primordial black hole dark matter from single field inflation,}''
Phys. Rev. D \textbf{97} (2018) no.2, 023501
%doi:10.1103/PhysRevD.97.023501
[\hhref{arXiv:1709.05565} [hep-ph]].
%89 citations counted in INSPIRE as of 19 Aug 2020

%\cite{Dalianis:2018frf}
\bibitem{Dalianis:2018frf}
I.~Dalianis, A.~Kehagias and G.~Tringas,
``{\it Primordial black holes from $\alpha$-attractors,}''
JCAP \textbf{01} (2019), 037
%doi:10.1088/1475-7516/2019/01/037
[\hhref{arXiv:1805.09483} [astro-ph.CO]].
%41 citations counted in INSPIRE as of 19 Aug 2020

%\cite{Atal:2018neu}
\bibitem{Atal:2018neu}
V.~Atal and C.~Germani,
%``The role of non-gaussianities in Primordial Black Hole formation,''
Phys. Dark Univ. \textbf{24}, 100275 (2019)
doi:10.1016/j.dark.2019.100275
[arXiv:1811.07857 [astro-ph.CO]].
%41 citations counted in INSPIRE as of 17 Sep 2020

%\cite{Starobinsky:1986fx}
\bibitem{Starobinsky:1986fx} 
A.~A.~Starobinsky,
``{\it Stochastic De Sitter (inflationary) Stage In The Early Universe,}''
Lect.\ Notes Phys.\  {\bf 246}, 107 (1986).
%doi:10.1007/3-540-16452-9_6
%%CITATION = doi:10.1007/3-540-16452-9_6;%%
%213 citations counted in INSPIRE as of 24 Feb 2020

\bibitem{Biagetti:2018pjj}
M.~Biagetti, G.~Franciolini, A.~Kehagias and A.~Riotto,
``{\it Primordial Black Holes from Inflation and Quantum Diffusion,}''
JCAP \textbf{07} (2018), 032
%doi:10.1088/1475-7516/2018/07/032
[\hhref{arXiv:1804.07124} [astro-ph.CO]].
%37 citations counted in INSPIRE as of 12 May 2020

%\cite{Ezquiaga:2018gbw}
\bibitem{Ezquiaga:2018gbw} 
J.~M.~Ezquiaga and J.~García-Bellido,
``{\it Quantum diffusion beyond slow-roll: implications for primordial black-hole production,}''
JCAP {\bf 1808}, 018 (2018)
%doi:10.1088/1475-7516/2018/08/018
[\hhref{arXiv:1805.06731} [astro-ph.CO]].
%%CITATION = doi:10.1088/1475-7516/2018/08/018;%%
%38 citations counted in INSPIRE as of 25 Feb 2020
  
%\cite{Cruces:2018cvq}
\bibitem{Cruces:2018cvq}
D.~Cruces, C.~Germani and T.~Prokopec,
{\it ``Failure of the stochastic approach to inflation beyond slow-roll,}''
JCAP \textbf{03} (2019), 048
%doi:10.1088/1475-7516/2019/03/048
[\hhref{arXiv:1807.09057} [gr-qc]].

%\cite{Firouzjahi_2019}
\bibitem{Firouzjahi_2019}
H. Firouzjahi, A. Nassiri-Rad, and M. Noorbala,
{\it ``Stochastic ultra slow roll inflation,}'' 
JCAP \textbf{01} (2019), 040
[\hhref{arXiv:1811.02175} [hep-th]]

%\cite{Pattison:2019hef}
\bibitem{Pattison:2019hef}
C.~Pattison, V.~Vennin, H.~Assadullahi and D.~Wands,
{\it ``Stochastic inflation beyond slow roll,}''
JCAP \textbf{07}, 031 (2019)
%doi:10.1088/1475-7516/2019/07/031
[\hhref{arXiv:1905.06300} [astro-ph.CO]].
%14 citations counted in INSPIRE as of 08 Jul 2020

%\cite{Ezquiaga:2019ftu}
\bibitem{Ezquiaga:2019ftu}
J.~M.~Ezquiaga, J.~Garcia-Bellido and V.~Vennin,
``{\it The exponential tail of inflationary fluctuations: consequences for primordial black holes,}''
JCAP \textbf{03} (2020) no.03, 029
%doi:10.1088/1475-7516/2020/03/029
[\hhref{arXiv:1912.05399} [astro-ph.CO]].
%2 citations counted in INSPIRE as of 11 Apr 2020

\bibitem{Pattison:2021oen}
C.~Pattison, V.~Vennin, D.~Wands and H.~Assadullahi,
``{\it Ultra-slow-roll inflation with quantum diffusion,}''
[\hhref{arXiv:2101.05741} [astro-ph.CO]].
%7 citations counted in INSPIRE as of 21 Jun 2021

\bibitem{De:2020hdo}
A.~De and R.~Mahbub,
``{\it Numerically modeling stochastic inflation in slow-roll and beyond,}''
Phys. Rev. D \textbf{102} (2020) no.12, 123509
%doi:10.1103/PhysRevD.102.123509
[\hhref{arXiv:2010.12685} [astro-ph.CO]].
%1 citations counted in INSPIRE as of 21 Jun 2021

\bibitem{Figueroa:2020jkf}
D.~G.~Figueroa, S.~Raatikainen, S.~Rasanen and E.~Tomberg,
``{\it Non-Gaussian tail of the curvature perturbation in stochastic ultra-slow-roll inflation: implications for primordial black hole production,}''
[\hhref{arXiv:2012.06551} [astro-ph.CO]].
%5 citations counted in INSPIRE as of 21 Jun 2021
  
%\cite{Bravo:2020hde}
\bibitem{Bravo:2020hde}
R.~Bravo and G.~A.~Palma,
%``Unifying attractor and non-attractor models of inflation under a single soft theorem,''
[arXiv:2009.03369 [hep-th]].
%0 citations counted in INSPIRE as of 17 Sep 2020  

%\cite{Schwinger:1960qe}
\bibitem{Schwinger:1960qe}
J.~S.~Schwinger,
``{\it Brownian motion of a quantum oscillator,}''
J.\ Math.\ Phys.\  \textbf{2} (1961), 407-432.
%doi:10.1063/1.1703727
%1493 citations counted in INSPIRE as of 12 Apr 2020

%\cite{Keldysh:1964ud}
\bibitem{Keldysh:1964ud}
L.~Keldysh,
``{\it Diagram technique for non-equilibrium processes,}''
Zh.\ Eksp.\ Teor.\ Fiz.\  \textbf{47} (1964), 1515-1527.
%1481 citations counted in INSPIRE as of 12 Apr 2020

%\cite{Jordan:1986ug,Calzetta:1986ey,Maldacena:2002vr,Weinberg:2005vy}
\bibitem{Jordan:1986ug}
R.~Jordan,
``{\it Effective Field Equations for Expectation Values,}''
Phys.\ Rev.\ D \textbf{33} (1986), 444-454.
%doi:10.1103/PhysRevD.33.444
%376 citations counted in INSPIRE as of 12 Apr 2020

%\cite{Calzetta:1986ey,Maldacena:2002vr,Weinberg:2005vy}
\bibitem{Calzetta:1986ey}
E.~Calzetta and B.~Hu,
``{\it Closed Time Path Functional Formalism in Curved Space-Time: Application to Cosmological Back Reaction Problems,}''
Phys.\ Rev.\ D \textbf{35} (1987), 495
%doi:10.1103/PhysRevD.35.495
%427 citations counted in INSPIRE as of 12 Apr 2020

  
 %\cite{Weinberg:2005vy}
\bibitem{Weinberg:2005vy}
S.~Weinberg,
``{\it Quantum contributions to cosmological correlations,}''
Phys.\ Rev.\ D \textbf{72} (2005), 043514
%doi:10.1103/PhysRevD.72.043514
[\hhref{arXiv:hep-th/0506236} [hep-th]].
%628 citations counted in INSPIRE as of 12 Apr 2020

%\cite{Chen:2006nt}
\bibitem{Chen:2006nt}
X.~Chen, M.~Huang, S.~Kachru and G.~Shiu,
``{\it Observational signatures and non-Gaussianities of general single field inflation,}''
JCAP \textbf{01} (2007), 002
%doi:10.1088/1475-7516/2007/01/002
[\hhref{arXiv:hep-th/0605045} [hep-th]].
%763 citations counted in INSPIRE as of 13 Apr 2020  

%\cite{Atal:2019cdz}
\bibitem{Atal:2019cdz}
V.~Atal, J.~Garriga and A.~Marcos-Caballero,
``{\it Primordial black hole formation with non-Gaussian curvature perturbations,}''
JCAP \textbf{09} (2019), 073
%doi:10.1088/1475-7516/2019/09/073
[\hhref{arXiv:1905.13202} [astro-ph.CO]].
%19 citations counted in INSPIRE as of 23 Feb 2021

%\cite{Namjoo:2012aa}\cite{Cai:2017bxr}
\bibitem{Namjoo:2012aa}
M.~H.~Namjoo, H.~Firouzjahi and M.~Sasaki,
``{\it Violation of non-Gaussianity consistency relation in a single field inflationary model,}''
EPL \textbf{101} (2013) no.3, 39001
%doi:10.1209/0295-5075/101/39001
[\hhref{arXiv:1210.3692} [astro-ph.CO]].
%154 citations counted in INSPIRE as of 13 Apr 2020

  
\end{thebibliography}
\end{document}